\pdfoutput=1
\documentclass[twocolumn,times]{aastex62} 
\usepackage{graphicx}
\usepackage{mathtools}
\usepackage{natbib}
\usepackage{amsmath}
\usepackage{appendix}
\usepackage{multirow,tabularx}
\usepackage[version=4]{mhchem}
\usepackage{newtxtext,newtxmath}

\newcommand\reactionnonumber[1]%
{\begin{equation*}\ce{#1}\end{equation*}}

\newif\iftwocol
\twocoltrue

\received{February 27, 2018}
\revised{September 17, 2018}
\accepted{September 19, 2018}

\submitjournal{ApJ}

\newcommand{\um}{$\upmu$m}

\begin{document}
\title{Evolved Climates and Observational Discriminants for the TRAPPIST-1 Planetary System}

\correspondingauthor{Andrew P. Lincowski}
\email{alinc@uw.edu}

\author[0000-0003-0429-9487]{Andrew P. Lincowski}
\affiliation{Department of Astronomy and Astrobiology Program, University of Washington, Box 351580, Seattle, Washington 98195, USA}
\affiliation{NASA Astrobiology Institute's Virtual Planetary Laboratory, Box 351580, University of Washington, Seattle, Washington 98195, USA}

\author{Victoria S. Meadows}
\affiliation{Department of Astronomy and Astrobiology Program, University of Washington, Box 351580, Seattle, Washington 98195, USA}
\affiliation{NASA Astrobiology Institute's Virtual Planetary Laboratory, Box 351580, University of Washington, Seattle, Washington 98195, USA}

\author{David Crisp}
\affiliation{NASA Astrobiology Institute's Virtual Planetary Laboratory, Box 351580, University of Washington, Seattle, Washington 98195, USA}
\affiliation{Jet Propulsion Laboratory, California Institute of Technology, Earth and Space Sciences Division, Pasadena, California 91011, USA}

\author{Tyler D. Robinson}
\affiliation{NASA Astrobiology Institute's Virtual Planetary Laboratory, Box 351580, University of Washington, Seattle, Washington 98195, USA}
\affiliation{Department of Physics and Astronomy, Northern Arizona University, Flagstaff, AZ 86011, USA}

\author{Rodrigo Luger}
\affiliation{Department of Astronomy and Astrobiology Program, University of Washington, Box 351580, Seattle, Washington 98195, USA}
\affiliation{NASA Astrobiology Institute's Virtual Planetary Laboratory, Box 351580, University of Washington, Seattle, Washington 98195, USA}

\author[0000-0002-0746-1980]{Jacob Lustig-Yaeger}
\affiliation{Department of Astronomy and Astrobiology Program, University of Washington, Box 351580, Seattle, Washington 98195, USA}
\affiliation{NASA Astrobiology Institute's Virtual Planetary Laboratory, Box 351580, University of Washington, Seattle, Washington 98195, USA}

\author{Giada~N.~Arney}
\affiliation{NASA Astrobiology Institute's Virtual Planetary Laboratory, Box 351580, University of Washington, Seattle, Washington 98195, USA}
\affiliation{NASA/Goddard Space Flight Center, Greenbelt, MD 20771, USA}

\shorttitle{Climates \& Spectra for TRAPPIST-1 Planets}
\shortauthors{Lincowski et al.}

\begin{abstract}

 The TRAPPIST-1 planetary system provides an unprecedented opportunity to study terrestrial exoplanet evolution with the \textit{James Webb Space Telescope} (JWST) and ground-based observatories. Since M~dwarf planets likely experience extreme volatile loss, the TRAPPIST-1 planets may have highly-evolved, possibly uninhabitable atmospheres.  We used a versatile, 1D terrestrial-planet climate model with line-by-line radiative transfer and mixing length convection (VPL~Climate) coupled to a terrestrial photochemistry model to simulate environmental states for the TRAPPIST-1 planets.  We present equilibrium climates with self-consistent atmospheric compositions, and observational discriminants of post-runaway, desiccated, 10--100 bar \ce{O2}- and \ce{CO2}-dominated atmospheres, including interior outgassing, as well as for water-rich compositions. Our simulations show a range of surface temperatures, most of which are not habitable, although an aqua-planet TRAPPIST-1~e could maintain a temperate surface given Earth-like geological outgassing and \ce{CO2}. We find that a desiccated TRAPPIST-1~h may produce habitable surface temperatures beyond the maximum greenhouse distance. Potential observational discriminants for these atmospheres in transmission and emission spectra are influenced by photochemical processes and aerosol formation, and include collision-induced oxygen absorption (\ce{O2}-\ce{O2}), and \ce{O3}, CO, \ce{SO2}, \ce{H2O}, and \ce{CH4} absorption features, with transit signals of up to 200~ppm.
 Our simulated transmission spectra are consistent with K2, HST, and \textit{Spitzer} observations of the TRAPPIST-1 planets. For several terrestrial atmospheric compositions, we find that TRAPPIST-1~b is unlikely to produce aerosols.
 These results can inform JWST observation planning and data interpretation for the TRAPPIST-1 system and other M~dwarf terrestrial planets. 

\end{abstract}

\keywords{planets and satellites: atmospheres --- planets and satellites: detection --- planets and satellites: terrestrial planets --- planets and satellites: individual (TRAPPIST-1)}

\section{Introduction}

M dwarfs are the most common type of star \citep{Henry:2006}, with a high occurrence rate of both multi-planet systems \citep[e.g.][]{Ballard:2016,Gillon:2017}, and Earth-sized planets in the habitable zone \citep{Dressing:2015}. Here, we define the classical habitable zone (HZ) as that region around the star where a planet with an Earth-like environment could maintain liquid water on its surface \citep{Kasting:1993}. Consequently, M~dwarfs host the most HZ planets of any stellar type by a large margin, and are therefore key to understanding the broader evolution of HZ terrestrial planets and the distribution of life beyond the Solar System.  Although numerous, M~dwarf HZ planets likely face significant challenges to habitability, as their parent stars exhibit high stellar activity \citep{Tarter:2007} and intense, protracted stellar evolution \citep{Baraffe:2015}. These stellar characteristics could strip the primordial atmosphere and cause significant volatile loss \citep{Khodachenko:2007,Lammer:2007,Lammer:2011,Luger:2015a,Luger:2015b,Tian:2015,Ribas:2016,Airapetian:2017,Dong:2017,Garcia:2017} or destroy ozone, which can protect exposed life on the surface from UV radiation \citep{Segura:2010}.  In particular, the super-luminous pre-main-sequence phase of an M~dwarf could drive an ocean-bearing terrestrial planet into a runaway greenhouse state for hundreds of millions of years until the planet enters the habitable zone, as the star dims and enters the main sequence.   \citep{Luger:2015b,Meadows:2018}. 

Determining whether these abundant M~dwarf HZ planets are in fact habitable is therefore a key focus in astrobiology and exoplanet science \citep[see e.g.][for eviews]{Shields:2016,Meadows:2018}. M~dwarf terrestrial planets are likely to experience very different evolutionary paths compared to those in our Solar System. Their HZ planets may therefore have atmospheres that have evolved considerably from their primordial composition. Reflecting their extreme evolution, these atmospheres may not be Earth-like or \ce{N2}-dominated, and the planets may not be habitable \citep{Meadows:2018}.   

Soon the targets and technical capabilities will be available that are needed to observe M~dwarf planets, detect and characterize their atmospheric compositions, and search for signs of planetary evolution and habitability.  Several planets that are likely to be terrestrial have been found orbiting M~dwarfs, including within the habitable zone \citep{Berta-Thompson:2015,Anglada-Escude:2016,Gillon:2017,Dittmann:2017,Shvartzvald:2017}.  Facilities available in the near future, such as the \emph{James Webb Space Telescope} (JWST) and ground-based extremely large telescopes (ELTs), may be capable of probing M~dwarf planetary atmospheres, including a handful of HZ planets \citep{Rodler:2014,Cowan:2015,Snellen:2015,Quanz:2015,Greene:2016,Lovis:2017}. For the best targets, these facilities may have the precision required to undertake the first spectroscopic search for atmospheric water vapor and biosignature gases, such as \ce{O2} and \ce{CH4} \citep{DesMarais:2002,Rodler:2014,Meadows:2017a,Meadows:2017b,Schwieterman:2017,Catling:2017b,Fujii:2017,Walker:2017}.

The seven transiting planets of the nearby TRAPPIST-1 system provide a natural laboratory to study planetary atmospheric evolution and the associated impact on habitability.  The semi-major axes of these planets' orbits encompass and extend beyond both the inner and outer boundaries of the habitable zone, enabling a test of evolutionary processes as a function of distance from the star. TRAPPIST-1 is a very late-type M~dwarf star (M8V; \citealt{Liebert:2006}), and so is an ideal target for characterization with JWST because its planets have very short period orbits, facilitating repeated observations, and comparatively small star-to-planet size ratios, which increase the signal-to-noise of transit observations. To plan and support the interpretation of these JWST observations, numerical models are being used to study the possible evolutionary pathways and resultant climates of these planets
\citep{Wolf:2017,Turbet:2018}, and to generate simulated thermal emission and transit transmission spectra \citep{Morley:2017}.  These models and simulated observations can be used to predict observable phenomena that may help discriminate between habitable and uninhabitable planetary environmental states, and help identify the evolutionary processes that generated them.  

Exoplanet models for \ce{H2}-dominated worlds  \citep[e.g.][]{Miller-Ricci:2009,Hu:2014} may not be applicable for small M~dwarf planets, which may be more terrestrial in composition, and are therefore likely to have solid or liquid surfaces and secondary outgassed (i.e. high molecular weight) atmospheres. There is growing evidence that the bulk properties of planets with radii $\lesssim1.5R_\oplus$ are more similar to our Solar System terrestrials than to sub-Neptune planets with \ce{H2}-dominated atmospheres \citep[e.g.][]{Rogers:2015,DeWit:2016,Fulton:2017,DeWit:2018}.  The densities of nearby M~dwarf HZ planets (e.g. GJ1132~b, 6.0$\pm$2.5~g~cm$^{-3}$, \citealp{Berta-Thompson:2015}, and LHS1140~b, 12.5$\pm$3.4~g~cm$^{-3}$, \citealp{Dittmann:2017}) are similar to or greater than Earth's (5.5~g~cm$^{-3}$) and Venus' (5.3~g~cm$^{-3}$) densities, consistent with mixtures of silicate rock and iron. The TRAPPIST-1 planets have currently measured densities that span 0.6 to 1.0 times Earth's density (i.e. 3.3--5.5~g~cm$^{-3}$; see Table~\ref{table:planets}). The generally lower densities of the TRAPPIST-1 planets \citep{Gillon:2017,Grimm:2018}, along with their resonant orbits, suggest migration \citep{Luger:2017b,Unterborn:2018}. If this is the case, they may have formed at larger distances from the star with a more volatile-rich composition.

If these small HZ M~dwarf planets are terrestrial, considerable evolution of the host star and migration processes may have caused changes in the bulk and atmospheric compositions and evolutionary paths unlike those seen in our Solar System. For example, \citet{Ribas:2016}, \citet{Barnes:2018}, and \citet{Coleman:2017} assessed evolutionary outcomes for a single planet in the habitable zone of an M5.5 dwarf star, Proxima Centauri b.
Their work suggests that these evolutionary paths could produce desiccated planets that rapidly lost their oceans by evaporation, photolysis, and subsequent hydrogen loss to space, leaving massive \ce{O2}-dominated atmospheres \citep{Kasting:1995,Luger:2015b,Barnes:2018}, or runaway greenhouse conditions with \ce{CO2}-dominated atmospheres  \citep[i.e. Venus-like;][]{Kasting:1995,Meadows:2018}.
The former hypothesized planetary environment assumes \ce{O2} build-up as a result of the super-luminous pre-main-sequence phase and subsequent water loss, with inefficient atmospheric loss or surface sinks for oxygen.
Processes other than XUV-driven hydrodynamic escape may cause \ce{O2} loss, such as surface interactions \citep{Hamano:2013,Schaefer:2016,Wordsworth:2018} and non-thermal processes favoring ion escape \citep{Collinson:2016,Airapetian:2017,Dong:2017,Garcia:2017}. Our current understanding of the history of Venus suggests that oxygen loss may be efficient.  If so, a \ce{CO2}-dominated atmosphere is also a likely case that may evolve from the loss processes that the TRAPPIST-1 planets may have experienced, since these planets could continue to outgas volatiles \citep{Donahue:1983}, eventually producing Venus-like atmospheric conditions.  These planets could also have migrated inwards to their current positions \citep{Luger:2017a,Unterborn:2018}, which would have allowed primordial \ce{H2} envelopes to be stripped by the young star to reveal a terrestrial core \citep{Lammer:2011,Luger:2015a} that may be volatile-rich and result in a more Earth-like \ce{H2O}- or \ce{N2}-rich atmosphere. 

The stellar spectra of M-dwarfs are redder and far more variable at UV wavelengths than those of G-type stars like the Sun.  The combined effects of these spectral differences and planetary semi-major axis on the climate and chemistry for M~dwarf terrestrial atmospheres has been previously explored in work that focused mostly on Earth-like planets in the habitable zone.
\citet{Segura:2005} modeled the effects of M~dwarf stellar spectra on Earth-like atmospheres with incoming host-star stellar radiation (instellation) levels similar to Earth, and found robust \ce{O3} production and increased \ce{CH4} levels. \citet{Grenfell:2007} explored the effect of orbital semi-major axis on the atmospheric lifetime of biosignature gases and their detectability for Earth-like HZ planets around F-G-K stars, finding increased \ce{O3} and \ce{CH4} for larger semi-major axes. This trend is due both to colder temperatures at larger semi-major axes, as well as the reduction in radicals derived from \ce{H2O} photolysis, which destroy both \ce{O3} and \ce{CH4}. More recently, comprehensive studies of Proxima Centauri b were undertaken by \citet{Turbet:2016} using 3D General Circulation Models (GCMs) and by \citet{Meadows:2018} using 1D coupled climate-photochemical models. These studies provided a range of plausible atmospheric compositions and climates, including non-Earth-like cases, for planets in the HZ of M~dwarfs.

There have been a number of efforts to model aspects of the TRAPPIST-1 planetary system that might affect habitability. These have included the effects of stellar evolution \citep{Bolmont:2017} and planet formation \citep{Quarles:2017}. Planetary atmospheric models have been used to assess climate and atmospheric composition.  Simple energy balance climate models (EBMs) were used to assess the climate response to globally parameterized variables such as albedo and vegetation coverage  \citep{Alberti:2017}, and identified TRAPPIST-1~d as the planet most likely to maintain surface water over a large range of albedo/vegetation parameters, with f, g, and h unlikely to host surface liquid water.  To explore potential atmospheric states of the TRAPPIST-1 planets as a function of semi-major axis, \citet{Morley:2017} distilled the bulk atmospheric compositions of Earth, Venus, and Titan into their lowest Gibbs free energy components using thermochemistry. They used this approximation to explore a broad parameter space of atmospheric pressures spanning 0.0001--100~bars, with prescribed atmospheric thermal profiles and surface temperatures adjusted to be consistent with planetary equilibrium temperatures for different assumed values of planetary Bond albedo. \citet{Morley:2017} assessed the observational signatures of these atmospheres in transmission and emission, and concluded that planet b would require fewer than 10 eclipse observations with JWST/MIRI to detect thermal emission and constrain the planetary emission temperature. They suggested that planets d--f could have surface liquid water and that their environments may be probed using JWST photometry. \citet{Wolf:2017} and \citet{Turbet:2018} used 3D GCMs to model climates for HZ and outer TRAPPIST-1 planets with \ce{N2} and pure \ce{CO2} atmospheres and surface pressures up to 10~bars.  In contrast with the EBM study \citep{Alberti:2017}, \citet{Wolf:2017} concluded that only planet e resided within the classical habitable zone, and that it would require substantially more \ce{CO2} than Earth to maintain an Earth-like surface temperature. \citet{Turbet:2018} assessed the outer planets, and found that atmospheres consisting of \ce{N2}, CO, or \ce{O2} are resistant to collapse, even for synchronously rotating planets. They argued that greenhouse gases like \ce{CO2}, \ce{NH3}, and \ce{CH4} would be difficult to maintain in the atmospheres of the TRAPPIST-1 planets after outgassing, due to easy condensation of \ce{CO2} on the night side and the likely photochemical destruction of \ce{NH3} and \ce{CH4}.  They noted that if the planet started with larger amounts of \ce{CO2} ($>10$ bar), then the greenhouse effect and heat redistribution would allow \ce{CO2} to remain stable in the atmosphere. \citet{Turbet:2018} agreed with \citeauthor{Wolf:2017} that TRAPPIST-1~e has the highest potential for habitability, although they concluded that several bars of \ce{CO2} would be sufficient for TRAPPIST-1~f and g to also support liquid surface water. 

The above studies do not include photochemistry, and are largely insensitive to the effect of TRAPPIST-1's redder, more variable stellar spectrum on planetary atmospheric composition, which can be significant for terrestrial atmospheres at the range of instellation levels received by the TRAPPIST-1 planets \citep[e.g.][]{Segura:2005,Segura:2007,Grenfell:2014,Meadows:2018}. Without coupled climate-photochemistry models forced by the incoming stellar spectrum, it is not possible to obtain self-consistent solutions for planetary atmospheric composition and corresponding temperature structures.  This could have large effects on predicted spectra, especially for emission spectra at MIR wavelengths, which are strongly dependent on the vertical temperature structure of the atmosphere.   

To help guide and interpret current \citep[e.g.][]{DeWit:2016,DeWit:2018,Delrez:2018} and upcoming observations that will attempt to determine the nature of the TRAPPIST-1 exoplanetary environments, here we use self-consistent 1D coupled climate-chemistry models for terrestrial planets to explore the effects of stellar distance on atmospheric properties for a selection of possible environmental states.  Given the stellar evolution of the host star, we suggest that evolved, post ocean- or atmospheric-loss states may be more likely for the TRAPPIST-1 planets than Earth-like atmospheres.  Specifically, we model the atmospheric composition and climate of all seven TRAPPIST-1 planets for post-runaway \ce{O2}- and \ce{CO2}-dominated atmospheres, with and without continuous outgassing, and for surface pressures up to 100~bars. For comparison, we also model an Earth-like, ocean-covered environment (aqua planet) for TRAPPIST-1~e only. 

Our climate, photochemistry, and radiative transfer models were developed specifically for terrestrial planets with secondary outgassed atmospheres. They have been validated against Solar System planets \citep{Meadows:1996,Tinetti:2005,Robinson:2011,Arney:2014,Robinson:2018} and have been used to model \ce{O2}- and \ce{CO2}-based atmospheric compositions for Proxima Centauri b \citep{Meadows:2018}. Together, these models explicitly account for the photochemical and climate impacts of the redder, more active stellar spectrum.  We consider initial atmospheric compositions that may plausibly result from evolutionary processes, with Earth-like outgassing rates and a range of volatile inventory levels. These atmospheres are then iterated forward in time until they reach coupled climate-photochemical equilibrium. We then use a line-by-line radiative transfer model to generate high-spectral-resolution transmission, emission, and reflectance spectra of the resulting planetary states, which may be used as forward-model  predictions for upcoming observations with JWST, ground-based telescopes, and a future direct-imaging mission such as the Habitable Exoplanet Imaging Mission (HabEx) or Large UV/Optical/IR surveyor (LUVOIR) concepts. From these synthetic spectra, we identify observable properties of these highly-evolved atmospheres that may be used to discriminate among these environments.  For TRAPPIST-1~e, we determine whether a potentially habitable atmosphere could be detected and discriminated from possible desiccated atmospheres.  These evolutionary and spectral studies may also inform target selection for other late M~dwarf systems, including planets yet to be discovered by the Transiting Exoplanet Survey Satellite, TESS \citep{Sullivan:2015}.

In \S\ref{sec:models}, we give a description of our coupled climate-photochemical and atmospheric escape modeling suite, along with necessary input data. In \S\ref{sec:results}, we present results of our atmospheric escape calculations, which inform results of coupled climate-chemistry simulations of several different atmospheric compositions, and the associated simulated direct imaging and transmission spectra. In \S\ref{sec:discussion}, we discuss the significance of the atmospheric escape results and associated oxygen produced, the climate-photochemistry results, identify observational discriminants for the modeled environments, compare our results with other recent work, and assess recent observations.  We summarize our findings in \S\ref{sec:conclusions}. Detailed descriptions of our climate-photochemical models and their validations are given in the Appendices.

\section{Model and Input Descriptions} \label{sec:models}
 
At the core of our simulations of the TRAPPIST-1 planetary environments and spectra is a 1D coupled-climate-chemistry model, which we use to calculate climate-photochemical states that are self-consistent with the spectrum of the host star.  As input to this model, we use an atmospheric escape and evolution model to determine likely surface volatile inventories and initial \ce{O2} abundance as a function of semi-major axis. These models inform the initial assumptions for the present-day atmospheric composition used by the coupled-climate-photochemical model. The self-consistent environments generated by the coupled-climate-photochemical model are used by our radiative transfer model to generate  synthetic spectra as an aid to predicting observational discriminants for these different environmental outcomes.

Specifically, we use a 1D terrestrial radiative-convective-equilibrium climate model, VPL~Climate, capable of modeling the wide variety of potential atmospheres for these planets and can implement stellar types very different from the Sun. VPL~Climate has been validated against Venus, Earth, and Mars \citep{Robinson:2018}, and integrates SMART \citep{Meadows:1996,Crisp:1997} as its radiative transfer core, the same line-by-line radiative transfer code we use to generate high-resolution spectra.
Here, we couple this new climate model to a 1D atmospheric chemistry model that has been used in a variety of terrestrial exoplanet studies \citep[e.g.][]{Segura:2005,Arney:2016,Meadows:2018}, which we update for this study (see Appendix~\ref{app:atmos}).
We use an updated version of the atmospheric evolution model of \citet{Luger:2015b} to produce plausible inventories of oxygen accumulation that justify some of the posited environmental states we use in our climate-photochemical modeling. 
These models and their inputs are discussed in more detail below and in the Appendices.

\subsection{Radiative Transfer}

We use the Spectral Mapping and Atmospheric Radiative Transfer code (SMART), originally developed by David Crisp \citep{Meadows:1996,Crisp:1997}, for both high-resolution spectra and within the climate model to compute radiative fluxes and heating rates. SMART is capable of producing accurate synthetic planetary spectra \citep[e.g.][]{Robinson:2011}. SMART uses the Discrete Ordinate Radiative Transfer code (DISORT, \citealt{Stamnes:1988,Stamnes:2000}) to numerically compute the radiation field. SMART can combine the radiative fluxes from overlapping stellar and thermal sources, and includes the extinction in each layer from absorbing gases (due to visible-to-far-infrared vibrational-rotational transitions, collision-induced absorption, and UV pre-dissociation bands), Rayleigh scattering, cloud and aerosol scattering, a wavelength-dependent surface albedo, and a stellar source spectrum. We also use SMART to produce transit transmission spectra, including refraction \citep{Robinson:2017a}.

\subsection{The Climate Model}

We use VPL Climate, a general-purpose, 1D radiative-convective-equilibrium, terrestrial planet climate model \citep{Robinson:2018,Meadows:2018}. This model uses SMART for rigorous, spectrum-resolving (line-by-line) radiative transfer, coupled to a mixing length convection parameterization. As input, VPL Climate requires an initial atmospheric state (consisting of a temperature profile and gas mixing ratios), physical parameters, and fundamental laboratory and experimental data.  

We use a 1D coupled climate-photochemical model in this work, rather than a 3D~GCM, to more rigorously and self-consistently explore both the radiative and photochemical effects of the TRAPPIST-1 stellar spectrum on atmospheric composition and climate for non-Earth-like terrestrial planet atmospheres.  For Proxima Centauri b, \citet{Meadows:2018} performed a 1D climate comparison with the 3D~GCM results of \citet{Turbet:2016}, and demonstrated that differences in globally-averaged temperatures between 1D and 3D models were small. For \ce{N2}- and \ce{CO2}-dominated atmospheres, the difference was less than 20~K when the planet was synchronously rotating, and 5~K or less when the planet was in a 3:2 spin-orbit resonance state.  In slower, synchronously rotating planets, enhanced day-night temperature contrasts (which cannot be spatially resolved by 1D models) may induce atmospheric collapse \citep{Turbet:2016,Turbet:2018}. However, even if synchronously rotating, the TRAPPIST-1 planets, with orbital periods of 1.5--19~days, are in an orbital regime closer to rapid rotators, with the possibility of sufficient Coriolis force to drive zonal circulation and help prevent atmospheric collapse \citep{Yang:2014b,Kopparapu:2016}. For the desiccated planets that we model here, we also assume dense ($\geq10$ bar) atmospheres that will have smaller day-night contrast and are more likely to be stable against collapse \citep{Hu-Yang:2014,Meadows:2018,Turbet:2018}. Although our habitable planet example for TRAPPIST-1~e has only a 1~bar atmosphere, \citet{Hu-Yang:2014} showed that day-night temperature contrasts for synchronously-rotating planets with oceans were much smaller when a full dynamical ocean model, which strongly augments atmospheric heat transport, was used in a 3D GCM. With a dynamical ocean, even a relatively thin atmosphere was more stable against collapse.  

The VPL Climate model components are summarized below and described in more detail in Appendix~\ref{app:vplc}. The inputs are described in \S\ref{sec:inputs}.

\subsection{Atmospheric Chemistry Model}

To self-consistently model the effects of M~dwarf host star UV radiation on our modeled atmospheres, we couple {VPL Climate} to a publicly available atmospheric chemistry model\footnote{\url{https://github.com/VirtualPlanetaryLaboratory/atmos}}. This code is based on \citet{Kasting:1979} and \citet{Zahnle:2006}, and can simulate a range of planetary redox states, including high-oxygen atmospheres \citep{Arney:2016,Schwieterman:2016,Meadows:2018}.
This photochemical model is described in detail in \citet{Meadows:2018}. Briefly, it divides the nominal model atmosphere into plane-parallel layers and assumes hydrostatic equilibrium. A vertical transport scheme includes molecular and eddy diffusion. Boundary conditions can be set for all species at the top and bottom of the atmosphere, including initial gas mixing ratios, outgassing fluxes, and surface deposition velocities. We describe the specific boundary conditions for each type of atmosphere in \S\ref{sec:results}.

Here, we updated the photochemistry model by extending the photochemically active wavelength range considered by the model to include Lyman-$\alpha$ for a wide range of species. Updates to the model are described in Appendix~\ref{app:atmos}, and validations for Earth and Venus are given in Appendix~\ref{app:validation}.

\subsection{Coupled Climate-Chemistry Model Convergence}

The climate and chemistry models can be run independently or as a coupled model. When the climate model is run independently, the convergence criterion is met when all atmospheric layers are flux-balanced within 1~W~m$^{-2}$ and the heating rate for each layer is less than 10$^{-4}$~K per day. When run independently, the atmospheric chemistry model tracks changes in gas concentrations in each time step and computes the relative error. It adjusts the next time-step based on the largest error. The model checks the time-step length, and when it reaches 10$^{17}$ seconds in less than $\sim$100~time-steps, we consider the model to be converged.

To run coupled climate-photochemistry experiments, we pass the converged profile from VPL Climate (temperature and gas mixing profiles) to the chemistry code, which is run to convergence. The modified gas mixing ratios from the chemistry code are then passed back to {VPL Climate} to compute a new equilibrium temperature structure and condensible gas mixing profiles. These profiles are then passed back to the chemistry model and this process is repeated until global convergence is achieved. 

\subsection{Atmospheric Escape Model}
 
Plausible upper limits of water and oxygen inventories of post-runaway atmospheres are modeled using the atmospheric escape framework introduced by \citet{Luger:2015a} and \citet{Luger:2015b}. Briefly, we compute the extreme ultraviolet (XUV)-driven, energy-limited escape rate of hydrogen modified by the hydrodynamic drag of oxygen as in \citet{Hunten:1973}, assuming photolysis is fast enough that it is not the rate-limiting factor. As in \citet{Luger:2015b}, we assume surface sinks are inefficient at removing photolytically-produced oxygen, yielding a strict upper bound on the amount of oxygen retained in the atmosphere. Unlike \citet{Luger:2015b}, we model the decrease in the efficiency of the flow due to the increase in the mixing ratio of oxygen, which causes the flow to transition to diffusion-limited escape \citep[i.e.][]{Schaefer:2016}. We also relax the assumption of constant escape efficiency $\epsilon_{xuv}$, instead modeling it as a function of the incident XUV flux as in \citet{Bolmont:2017}. For the TRAPPIST-1 planets, $\epsilon_{xuv}$ varies between about 0.01 (at high XUV flux) and 0.10 (at moderate XUV flux).

\subsection{Model Inputs} \label{sec:inputs}

\begin{table*}
\centering
\caption{Stellar and planetary system parameters as model inputs for TRAPPIST-1. }
\label{table:planets}
{\iftwocol
    \small\selectfont
\else
    \footnotesize\selectfont
\fi
  \begin{tabular}{llllllll}
    \hline \hline
    \textbf{Parameter}        & \textbf{Modeled} & \multicolumn{6}{l}{\textbf{Measured$^\text{a}$}}                                                  \\
Star                      &                  & \multicolumn{6}{l}{TRAPPIST-1---2MASS J23062928-0502285}                                \\
Magnitudes$^\text{b}$    &                  & \multicolumn{6}{l}{$V$ = 18.80$\pm$0.08, $R$ = 16.47$\pm$0.07, $I $= 14.0$\pm$0.1, $J$ = 11.35$\pm$0.02, $K$ = 10.30$\pm$0.02}          \\
Mass ($M_\odot$)          &                  & \multicolumn{6}{l}{0.0802 $\pm$ 0.0073}                                                   \\
Radius ($R_\odot$)        & 0.117            & \multicolumn{6}{l}{0.117 $\pm$ 0.0036}                                                    \\
Luminosity ($L_\odot$)    & 0.000524         & \multicolumn{6}{l}{0.000524 $\pm$ 0.000034}                                               \\
Effective Temperature (K) & 2500             & \multicolumn{6}{l}{2,559 $\pm$ 50}                                                        \\
Metallicity {[}Fe/H{]}    & 0.               & \multicolumn{6}{l}{+ 0.04 $\pm$ 0.08}                                                     \\
Log $g$                   & 5.               & \multicolumn{6}{l}{4.8}                                                           \\ \hline
\textbf{Planets$^\text{a}$}          & \textbf{b}       & \textbf{c}   & \textbf{d}   & \textbf{e}   & \textbf{f}   & \textbf{g}   & \textbf{h$^\text{c}$}   \\
Period (days)             & 1.510870         & 2.42182      & 4.0496       & 6.0996       & 9.2067       & 12.353       & 18.767       \\
Semi-major axis (AU)      & 0.01111          & 0.01521      & 0.02144      & 0.02817      & 0.0371       & 0.0451       & 0.0595 \\
Irradiation ($S_\odot$)   & 4.245            & 2.265        & 1.140        & 0.6603       & 0.381        & 0.258        & 0.148        \\
Radius$^\text{d}$ ($R_\oplus$)        & 1.121$^{+0.032}_{-0.031}$  & 1.095$^{+0.031}_{-0.030}$        & 0.784$\pm0.023$        & 0.910$^{+0.027}_{-0.026}$        & 1.046$^{+0.030}_{-0.029}$        & 1.148$^{+0.033}_{-0.032}$        & 0.773$^{+0.027}_{-0.026}$        \\
Mass$^\text{d}$ ($M_\oplus$)          & 1.107$^{+0.143}_{-0.154}$    & 1.156$^{+0.131}_{-0.142}$ & 0.297$^{+0.035}_{-0.039}$ & 0.772$^{+0.075}_{-0.079}$          & 0.934$^{+0.078}_{-0.080}$          & 1.148$^{+0.095}_{-0.098}$   & 0.331$^{+0.049}_{-0.056}$          \\
Density$^\text{d}$ ($\rho_\oplus$)          & 0.726$^{+0.091}_{-0.092}$    & 0.883$^{+0.078}_{-0.083}$ & 0.616$^{+0.062}_{-0.067}$         & 1.024$^{+0.070}_{-0.076}$          & 0.816$^{+0.036}_{-0.038}$          & 0.759$^{+0.033}_{-0.034}$   & 0.719$^{+0.102}_{-0.117}$          \\
Gravity$^\text{d}$ (m/s$^2$)         & 7.94              & 9.46        & 4.74         & 9.15         & 8.37         & 8.55        & 5.43         \\ 
 Impact parameter $b$ ($R_*$) & 0.126$^{+0.092}_{-0.078}$ & 0.161$^{+0.076}_{-0.084}$ & 0.17$\pm0.11$ & 0.12$^{+0.11}_{-0.09}$ & 0.382$\pm0.035$ & 0.42$\pm0.031$ & 0.45$^{+0.06}_{-0.08}$ \\ 
\hline
  \end{tabular}
  \flushleft
  \textbf{Note:} We use nominal values for our modeling. Error bars (1$\sigma$) are shown for reference. Standard errors on the period are smaller than the precision reported here. Error on semi-major axis is approximately 3\% of the quoted values. Irradiation is computed from the reported semi-major axis, assuming the luminosity quoted here. Gravity is given without error bars, and is for the nominal values of mass and radius. \\
  $^\text{a}$ Data from \citet{Gillon:2017} unless otherwise noted.   \\
  $^\text{b}$ Data from \citet{Gillon:2016}. \\
  $^\text{c}$ Data from \citet{Luger:2017b}. \\
  $^\text{d}$ Data from \citet{Grimm:2018}.
  
}

\end{table*}

Our climate, photochemical, and atmospheric escape models require a number of inputs that tailor the model to the planetary atmosphere studied. Inputs include planetary properties (e.g. orbit, radius, etc.), atmospheric gas mixing ratios for each constituent in each layer, gas absorption properties (i.e. cross-section data and line lists), thermodynamic data for condensible gases, particle optical properties for aerosols, wavelength-dependent surface albedo data, and stellar spectral energy distribution (SED), including an estimate of total XUV flux over time.

\subsubsection{Planetary Properties and Model Atmospheres}

Here we adopt the current best-fit orbital periods, radii, and masses of the TRAPPIST-1 planets \citep[see Table \ref{table:planets} and ][]{Gillon:2016,Gillon:2017,Luger:2017b,Grimm:2018}. We assume the planets are in fixed, circular orbits. For all seven known planets, we simulate post-runaway-greenhouse, \ce{O2}-dominated atmospheres with and without continuous outgassing, and Venus-like \ce{CO2}-dominated atmospheres. These atmospheres are modeled with surface pressures of 10 and 100 bars for \ce{O2}-dominated cases and 10 and 92.1~bars (Venus' surface pressure) for Venus-like cases. We also model a potentially habitable aqua planet for TRAPPIST-1 e for comparison, both with the uninhabitable cases that we present here and with other studies of TRAPPIST-1 planetary climates \citep{Wolf:2017,Turbet:2018} and observables \citep{Morley:2017}. The particulars of each case are presented in detail in \S\ref{sec:results}.

The nominal planetary atmospheres are one-dimensional and plane-parallel, consisting of 64 pressure levels, half of which are linear intervals of pressure from the surface to 10\% of surface pressure, followed by logarithmic spacing in pressure to 0.01~Pa at the top of the atmosphere. 

\subsubsection{Gas Absorption Data} \label{sec:gas_abs}

Gas absorption cross-sections are calculated from three sources: vibrational-rotational transitions at visible and infrared wavelengths, continuum absorption from electronic transitions and pre-dissociation bands at UV wavelengths, and dimer- and pressure-induced absorption bands at visible to infrared wavelengths. Absorption lines associated with vibrational-rotational transitions are calculated using the line-by-line model, LBLABC (see Appendix \ref{app:vplc}) with the HITRAN2012, HITEMP2010, or Ames line databases \citep{HITRAN:2012,HITEMP:2010,Huang:2017}. These line lists assume Earth-like isotopic abundances. We include foreign broadening by the dominant gases in the atmosphere (\ce{CO2}, \ce{N2}, and/or \ce{O2}). 

Collisional-induced absorption data is used for \ce{CO2}-\ce{CO2} \citep{Moore:1971,Kasting:1984,Gruszka:1997,Baranov:2004,Wordsworth:2010,Lee:2016}, \ce{N2}-\ce{N2} (\citealp{Schwieterman:2015b} based on \citealp{Lafferty:1996}), and \ce{O2}-\ce{O2} \citep{Greenblatt:1990,Hermans:1999,Mate:1999}.

UV--visible cross section data is incorporated from a variety of primary sources available from the MPI-Mainz UV/VIS Spectral Atlas of Gaseous Molecules of Atmospheric Interest\footnote{\url{http://satellite.mpic.de/spectral_atlas}} \citep{Keller:2013}. Updated sources are listed in Appendix~\ref{app:atmos}.

\subsubsection{Aerosol Optical Properties} \label{sec:input_aer}

The single-scattering optical properties of clouds and aerosols are defined in pre-computed tables for a range of specific particle types. Each particle type is defined by its composition, associated wavelength-dependent refractive indices, and its particle size and shape distributions. Given these properties, a single scattering model is used to compute the wavelength-dependent extinction and scattering efficiencies ($Q_\text{ext}$ and $Q_\text{abs}$) and the phase function moments (for Mie scattering) or the particle asymmetry parameter ($g$ for some non-spherical particles) used in our radiative transfer calculations.

For the cloudy aqua planet, we specify Earth-like cirrus (water-ice) and stratocumulus (liquid water) clouds, each of cumulative optical depth $\tau=5$. For cirrus clouds, the optical properties from B. Baum's Cirrus Optical Property Library \citep{Baum:2005}\footnote{\url{http://www.ssec.wisc.edu/\~baum/Cirrus/
Solar\_Spectral\_Models.html}} are used. The particles consist of a distribution of 45\% solid columns, 35\% plates, and 15\% 3D bullet rosettes, spanning 2--9500~\um{} with a cross-section weighted mean diameter of 100~\um{}. Stratocumulus cloud optical properties are based on refractive indices of water from \citet{Hale:1973} and calculated using Mie scattering, with a two-parameter gamma distribution ($a=5.3$, $b=1.1$) and mean particle radius 4.07~\um{}.

For Venus-like sulfuric acid aerosols, we use refractive indices for a range of specific acid concentrations between 25 and 100\% from \citet{Palmer:1975}. The sulfuric acid concentration is computed assuming vapor pressure equilibrium of the \ce{H2O} and \ce{H2SO4} gases with the condensed solution. The effective optical properties are then calculated by spline interpolation of the tables. Our photochemical model calculates the monodisperse aerosol radii at every atmospheric layer given the coagulation, sedimentation, and diffusion timescales \citep[see e.g.][]{Pavlov:2001}, in phase equilibrium. For climate and spectral modeling, we convert the monodisperse distributions into log-normal distributions with the modal radii equal to the monodisperse particle radii and geometric standard deviation equal to 0.25, similar to the particle distribution findings for individual Venus aerosol modes as described in \citet{Crisp:1986}. The differential optical depth profile used in the radiative transfer calculations for each planet is calculated from the log-normal distribution, assuming that the total mass in each layer is the same as in the monodisperse distribution. The optical depth $d\tau$ is calculated by:
\begin{equation}
d\tau = dz \int_{R_\text{min}}^{R_\text{max}} \pi r^2 Q_\text{ext}(\lambda,r) n(r) dr,
\end{equation}
where $Q_\text{ext}$ is the extinction efficiency at a reference wavelength and $n$ is the number density distribution. 

\subsubsection{Thermodynamic Data}

The thermodynamic data required for condensibles in VPL~Climate (e.g.~\ce{H2O}) consist of saturation vapor temperature, pressure, and enthalpy of formation. For water, above 273~K we use data sourced from NIST\footnote{\url{http://webbook.nist.gov/chemistry/fluid/}}. Below 273~K, we use the saturation vapor temperature and pressure from \citet{Wagner:1994}. The enthalpy of condensation is from a fit to data provided in \citet{Rogers:1989}.
For sulfuric acid (\ce{H2SO4}), we use a fit calculated by \citet{Gao:2015} (see references therein).

\subsubsection{Surface Spectral Albedo}

\begin{figure}[h]  
\centering
\iftwocol
    \includegraphics[width = \columnwidth]{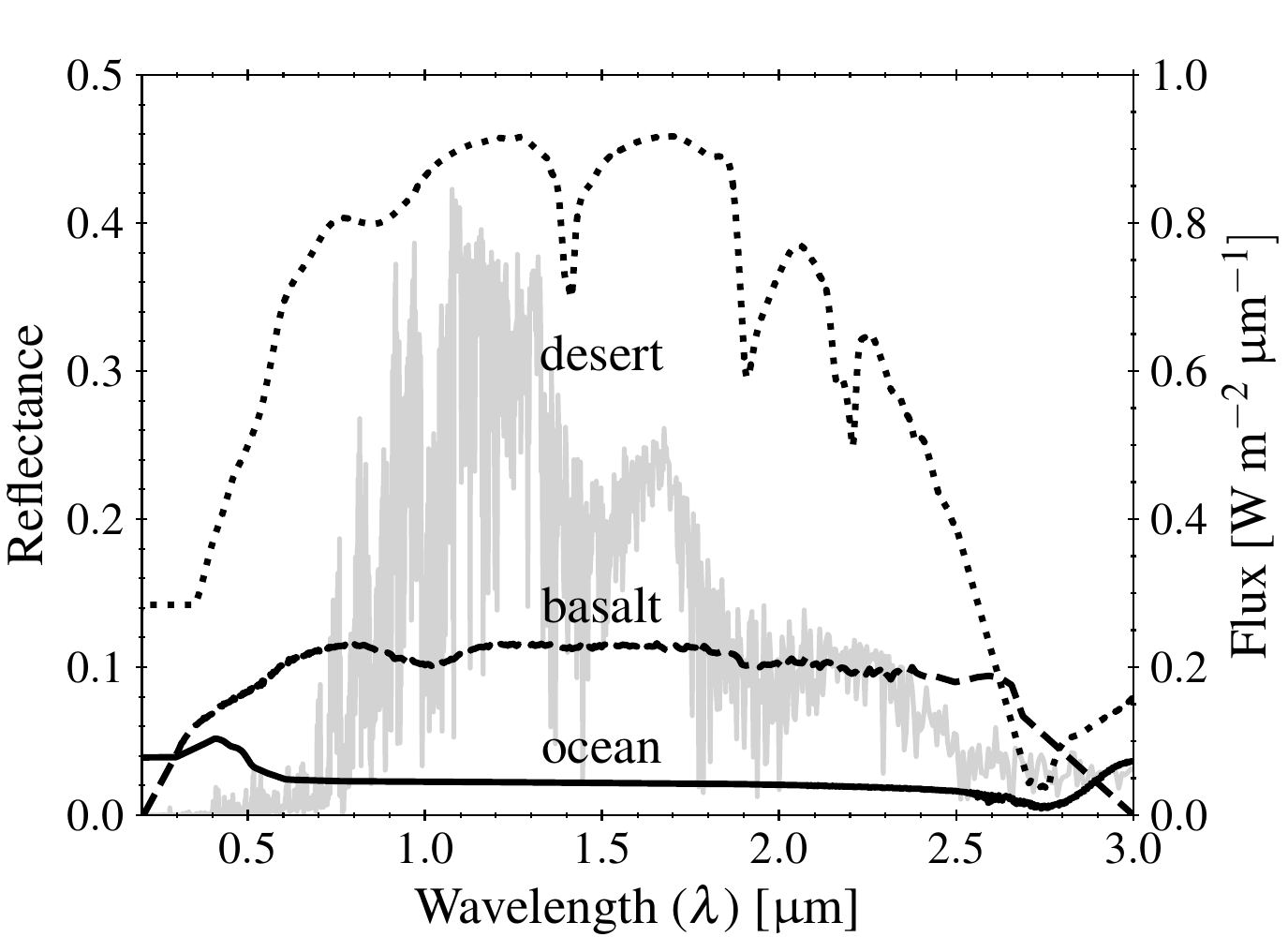}
\else 
    \includegraphics[width = 0.5\textwidth]{albedo}
\fi
\caption{Wavelength-dependent surface reflectance for modeled planetary environments with stellar irradiation plotted for comparison. The desert surface is primarily kaolinite and was used for the desiccated \ce{O2}-dominated climates. The basaltic surface was used for Venus-analog climates. The ocean surface was used for the habitable aqua planet. These data are from the U.S.G.S. spectral library \citep{Clark:2007}$^{\ref{foot:usgs}}$, except for the ocean longward of 2~\um{}, which is from the ASTER spectral library \citep{ASTER2.0}$^{\ref{foot:ASTER}}$.
\label{fig:albedo}} 
\end{figure}

We use different surface compositions based on the type of climate we are simulating.  Basalt was used for the volcanic, Venus-like planets and a desert surface (primarily kaolinite) was used for the post-ocean-loss \ce{O2}-dominated atmospheres. For the aqua planet, we used open ocean. Surface reflectance data is from the U.S.G.S. spectral library \citep{Clark:2007}\footnote{\label{foot:usgs}\url{http://speclab.cr.usgs.gov/spectral.lib06}}, except for the ocean longward of 2~\um{}, which is from the ASTER spectral library \citep{ASTER2.0}\footnote{\label{foot:ASTER}\url{https://speclib.jpl.nasa.gov/}} (see Figure~\ref{fig:albedo}).
The integrated, energy flux-weighted surface albedos for these surfaces for our TRAPPIST-1 stellar spectrum are: 0.09 (basalt), 0.37 (desert), 0.02 (ocean). For comparison, the same albedo of snow used in the VPL Spectral Earth Model \citep{Robinson:2011} is 0.16.

\subsubsection{Stellar Spectra}

 \begin{figure*}
  \centering
  \includegraphics[width = \textwidth]{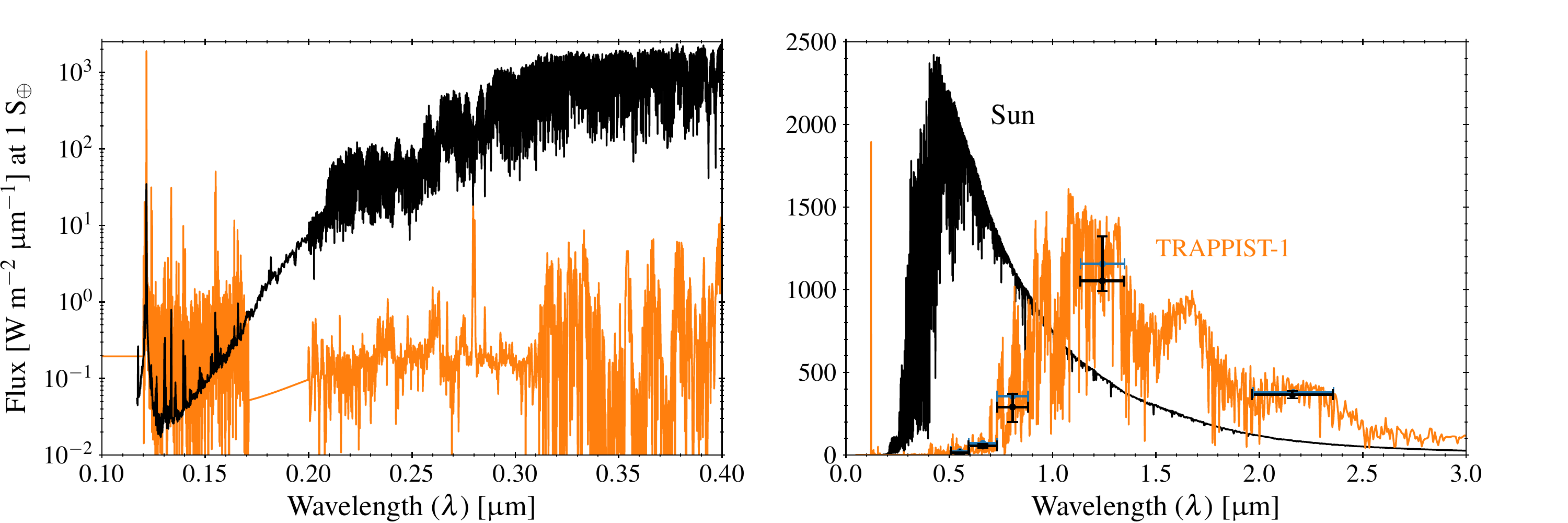}
  \caption{The input stellar spectral energy distribution (shown in orange) for TRAPPIST-1, an M8V star \citep{Liebert:2006}, using the PHOENIX v2.0 spectral database \citep{France:2013}, normalized to $1S_\oplus$. The UV is from Proxima Centauri \citep{Meadows:2018}, scaled to the measured Lyman-$\alpha$ flux \citep{Bourrier:2017}. The Solar spectrum is shown in black for comparison. Photometric bands (\textit{VRIJK}) are shown as points with error bars (black are from \citealt{Gillon:2016}, blue are band-integrated PHOENIX model fluxes), where the horizontal bars are the full-width half-maximum of the filter bands, and the observed flux errors are 3$\sigma$. Late-type M dwarf stars like TRAPPIST-1 have different effects on planetary photochemistry and climate than G-type stars, due to the strong UV flux and energy peak in the NIR. These spectra are available online using the VPL Spectral Explorer$^{\ref{footnote:specex}}$, or upon request. \label{fig:sed}} 
\end{figure*}

Characterizing the climate and computing spectra require the stellar spectrum of TRAPPIST-1 (2MASS J23062928-0502285). The available data on this star in the UV to IR is limited to photometry and to measurements of Lyman-$\alpha$ from the Hubble Space Telescope \citep{Bourrier:2017}. TRAPPIST-1 has an effective temperature of 2559 $\pm$ 50~K, [Fe/H] = 0.04 $\pm$ 0.08, $M_* = 0.0802 \pm 0.0073~M_\odot$, and $R_* = 0.117 \pm 0.0036~R_\odot$ \citep{Gillon:2017}. To model the panchromatic UV--NIR stellar spectrum necessary for line-by-line climate-photochemical modeling, we use the 2500~K, [Fe/H] = 0.0, log $g$ = 5.0 spectral model of the PHOENIX v2.0 spectral database, which spans from 0.25 to 5.5~\um{}, and has been shown to faithfully reproduce the visible--NIR spectra of M dwarfs \citep{France:2013}. We compared this PHOENIX model to the photometric observations \citep{Gillon:2016}, and found that the model SED, binned to available photometric bands, is within 3$\sigma$.

The intensity and wavelength dependence of the stellar UV spectrum is a critical factor affecting atmospheric chemistry, and subsequent atmospheric composition and climate. For example, \citet{Grenfell:2007} noted  changes in the atmospheres of planets around F-G-K stars as a function of orbital distance, including the impact of outgassed species on climate, while \citet{Grenfell:2014} noted that UV changes within 200--350~nm has the largest effect on ozone levels for planets around late-type M~dwarfs. In the absence of complete UV observations of TRAPPIST-1, we fit a UV spectrum compiled for Proxima Centauri \citep{Meadows:2018}. We scaled Proxima's UV spectrum to the Lyman-$\alpha$ flux ratio of TRAPPIST-1 to Proxima Centauri, a factor of 1/6 at 1 AU \citep{Bourrier:2017}.
The UV flux was stitched to the PHOENIX SED at around 0.3~\um{}, where photospheric emission began to dominate (i.e. where the flux from the PHOENIX model began to rise above the UV). This method provides only a possible UV flux environment for the planets of TRAPPIST-1 and does not represent the star's actual activity (except the reconstructed Lyman-$\alpha$). Given the variable activity of M~dwarfs, UV observations of TRAPPIST-1 are needed to properly characterize its planets.
Our compiled spectrum is shown in Figure \ref{fig:sed}, and is available for download online using the VPL Spectral Explorer\footnote{\label{footnote:specex}\url{http://depts.washington.edu/naivpl/content/vpl-spectral-explorer}}.

\subsubsection{XUV Flux}

We model the evolution of stellar XUV luminosity as in \citet{Luger:2015b}. We use the power law of \citet{Ribas:2005} with exponent $\beta=-1.23$, a saturation fraction $L_\text{xuv}/L_\text{bol} = 2.5\times10^{-4}$ (based on the measurements of \citealt{Wheatley:2017}), and a saturation time of 1~Gyr. We use the \citet{Baraffe:1998} stellar evolution tracks to model the pre-main sequence phase of TRAPPIST-1 and the habitable zone boundaries of \citet{Kopparapu:2013} to decide whether the planets are experiencing hydrogen escape, as in \citet{Luger:2015b}. Like \citet{Schaefer:2016}, we assume energy-limited escape transitions to diffusion-limited escape due to the accumulation of oxygen atoms.

\section{Results} \label{sec:results}

\begin{table*}[]
\centering
\caption{Modeled Planetary States and Their Environmental Parameters}

\label{table:experiments}
{\iftwocol
    \small\selectfont
\else
    \scriptsize\selectfont
\fi
\begin{tabular}{lllll}
\hline
\textbf{Planetary State} & \textbf{Surface} & \textbf{clouds}      & \textbf{gases}                                           & \textbf{surface pressure} \\ \hline
Aqua planet, clear sky   & ocean            & None                 & 80\% \ce{N2}, 20\% \ce{O2}, trace \ce{H2O}, \ce{CO2}, \ce{SO2}, \ce{H2S}, \ce{OCS}, NO, \ce{CH4} & 1 bar                     \\
Aqua planet, cloudy      & ocean            & 50\% water, 50\% ice & 80\% \ce{N2}, 20\% \ce{O2}, trace \ce{H2O}, \ce{CO2}, \ce{SO2}, \ce{H2S}, OCS, NO, \ce{CH4} & 1 bar                     \\
\ce{O2}, desiccated           & desert           & None                 & 95\% \ce{O2}, 4.5--4.95\% \ce{N2}, 0.5~bar \ce{CO2}                             & 10, 100~bar                    \\
\ce{O2}, outgassing           & desert           & None                 & 95\% \ce{O2}, 4.5--4.95\% \ce{N2}, 0.5~bar \ce{CO2}, trace \ce{H2O}, \ce{SO2}, OCS, \ce{H2S}   & 10, 100~bar                    \\
Venus-like, clear sky    & basalt           & None                 & 96.5\% \ce{CO2}, 3.5\% \ce{N2}, trace \ce{H2O}, \ce{SO2}, OCS, \ce{H2S}, NO, HCl     & 10, 92 bar                    \\
Venus-like, cloudy       & basalt           & \ce{H2SO4}                & 96.5\% \ce{CO2}, 3.5\% \ce{N2}, trace \ce{H2O}, \ce{SO2}, OCS, \ce{H2S}, NO, HCl  & 10, 92 bar                    \\ \hline
\end{tabular}}
\end{table*}


We first describe the potential atmospheric evolution, including water loss and oxygen production, for the seven TRAPPIST-1 planets due to the pre-main-sequence evolution of their host star.  These results inform our chosen \ce{O2} levels for the subsequent environmental modeling. We present calculated climates and compositions for \ce{O2}- and \ce{CO2}-dominated atmospheres, and their spectra and observational discriminants. To span plausible pressure and compositional ranges for these evolved atmospheres, model 10 and 100~bar
\ce{O2}-dominated atmospheres, with and without trace constituents due to surface outgassing or volatile delivery; and 10 and 92.1~bar
\ce{CO2}-dominated Venus-like atmospheres, with and without Venus-like haze/cloud aerosols. As a potentially habitable comparison case, we also calculate climate and trace composition of a 1~bar atmosphere for an ocean-covered TRAPPIST-1~e ``aqua planet''. 
We show simulated pressure-temperature structures and mixing ratio profiles. A summary of our experiments is shown in Table~\ref{table:experiments}.

A number of assumptions underlie our modeling in this work. Our atmospheres are assumed to be in hydrostatic equilibrium, though hydrodynamic processes could be occurring on planets b, c, and d, inward of the inner edge of the habitable zone. We assume the modeled atmospheres are stable against collapse on the night side, due to sufficiently high atmospheric mass (10--100~bar) to reduce day-night temperature contrasts \citep{Turbet:2018} or ocean circulation for the 1~bar aqua planets \citep{Hu-Yang:2014}. We assume these planets began in a warm state due to the super-luminous pre-main-sequence phase, and that their secondary atmospheres were formed during this time. We ignore ion and particle effects, though these non-thermal processes may be important drivers of atmospheric loss \citep{Airapetian:2017,Garcia:2017} and atmospheric chemistry \citep{Segura:2010,Airapetian:2017,Tilley:2018}. The recent densities inferred from transit timing variation (TTV) measurements by \citet{Grimm:2018} currently indicates high volatile content planets, perhaps with the exception of planet e, which suggests that the most desiccated atmospheres are less probable, and supports ocean-bearing surfaces or large outgassing fluxes. However, the error bars on the planet densities remain quite large, and do not rule out more refractory compositions, supporting the possibility of desiccated atmospheres.

\subsection{Atmospheric Escape and Oxygen Build-up} \label{sec:results-atmesc}

\begin{figure*}
  \centering
  \includegraphics[width = \textwidth]{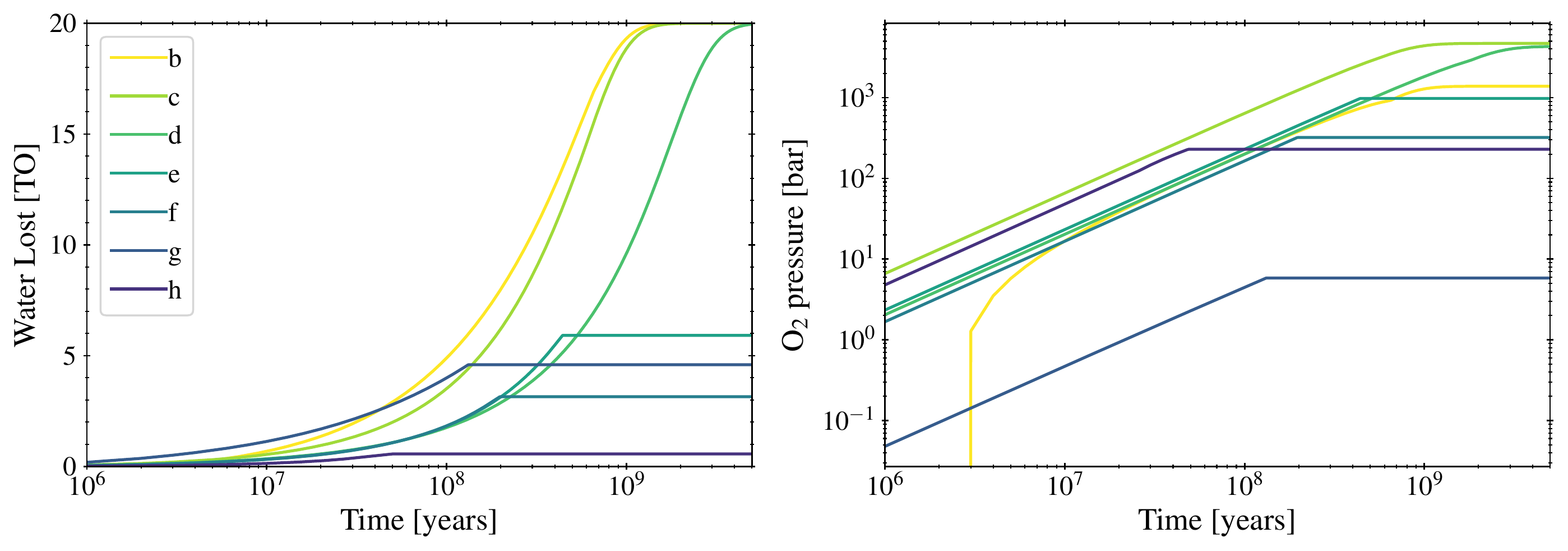} 
  \caption{Computed pre-main sequence evolution of water loss and oxygen production for the TRAPPIST-1 planets assuming inefficient surface sinks. These results represent the maximum build-up of oxygen in the atmosphere. The maximum ocean inventory assumed was 20 Earth oceans, all of which may be lost for planets b, c, and d. Planet h may lose less than one ocean of water. \label{fig:evol}} 
\end{figure*}

To determine plausible oxygen inventories for our \ce{O2}-dominated atmospheres, we computed the loss of hydrogen and oxygen during the pre-main sequence phase of TRAPPIST-1 (see Figure \ref{fig:evol}). The maximum initial water inventory considered was 20 Earth oceans (by mass), the approximate maximum ocean loss for planets b, c, and d under our assumptions. Planets e, f, and g may each lose between 3--6 Earth oceans of water, while h may lose less than one ocean.  This ocean loss results in the generation of dense \ce{O2} atmospheres with maximum pressures (assuming no sinks) from 22--5000~bar. 

The amount of water each planet may have lost and the subsequent surface pressure of atmospheric oxygen vary due to instellation, atmospheric scale height, and planetary surface area. The degree of \ce{H2O} loss is a strong function of instellation and time spent interior to the HZ inner edge. For example, as the host star dims, within 50--400~Myrs after formation, planets h, g, f, and e successively enter the habitable zone and cease to lose significant amounts of \ce{H2O}, while planets b, c and d continue to lose \ce{H2O} (Figure~\ref{fig:evol}). Although the orbits (and levels of instellation received) for these planets are well-spaced in resonance chains \citep{Luger:2017b}, they exhibit variations in \ce{H2O} loss not related to their instellation levels. Under diffusion-limited escape, we assume that \ce{H2O} loss is proportional to the inverse of scale height \citep{Hunten:1973}, which is in turn proportional to temperature and inversely proportional to gravity and mean molecular mass. The small variations in \ce{H2O} loss among b--d and e--h are in part due to the different scale heights for each planet. 

In comparison to ocean loss, the resultant surface pressure of equivalent atmospheric oxygen generated by water photolysis does not show a similar increasing trend with instellation; the sequence we calculated for the TRAPPIST-1 system (from highest to lowest \ce{O2} surface pressure) is c, d, b, e, g, f, h. Compared to planet b, c loses similar amounts of \ce{H2O} and is similar in size, but because c has a much larger mass and stronger surface gravity, it builds up a higher surface pressure of equivalent atmospheric \ce{O2}. That is, c can accrue a similar mass of remaining \ce{O2}, but the stronger surface gravity results in a larger surface pressure (since surface pressure is the product of surface gravity and the integrated atmospheric mass per unit area). The sequence of e, g, f, and h are also affected by each planet's respective gravitational acceleration, as opposed to an effect of radius; while f and g are similar in radius, their nominal surface gravities are very different, which results in a higher surface pressure of oxygen for g.

\subsection{Planetary Environmental States}

\begin{table*}
  \begin{center}
  \caption{Surface Temperatures [K] \label{table:results}}
  \includegraphics[width = 14cm]{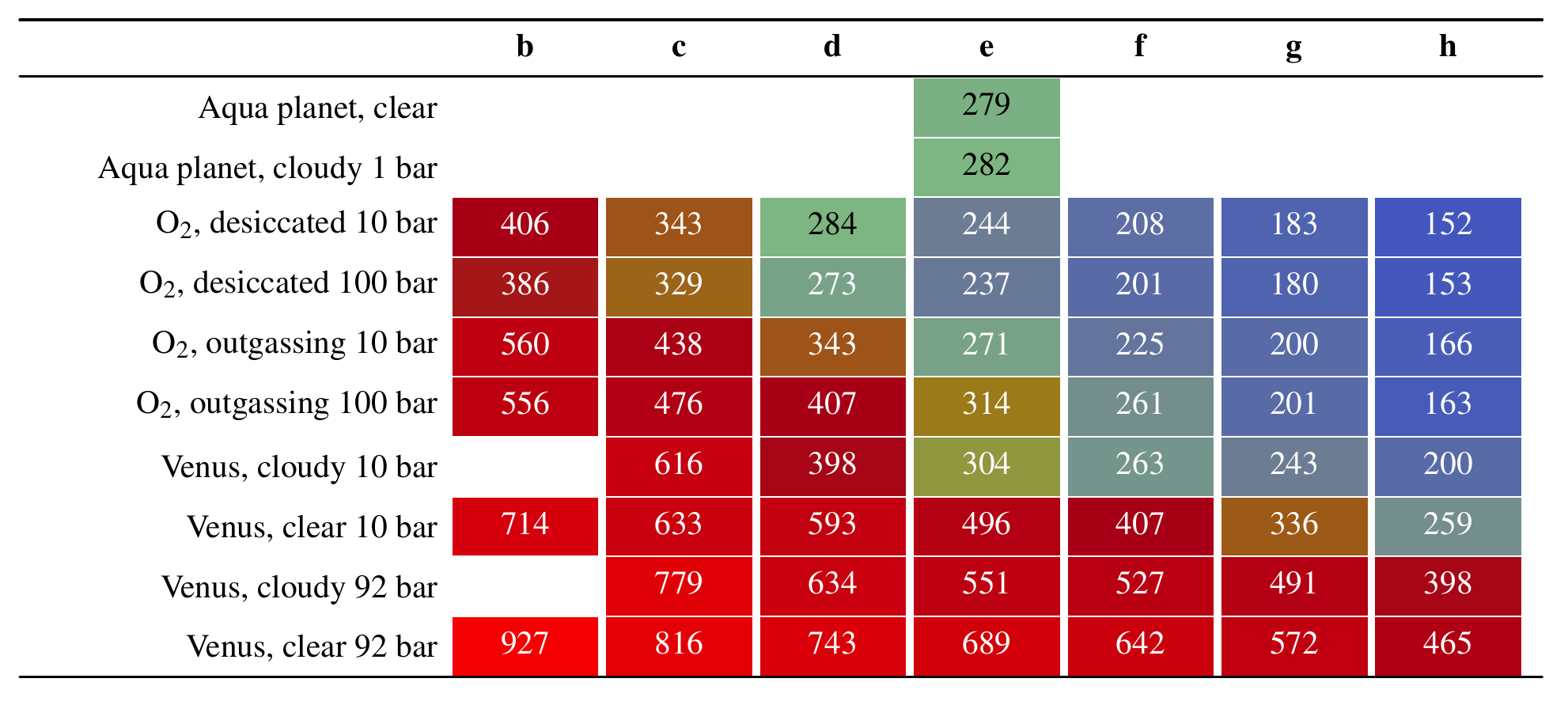}
  \end{center}
  
  {\small\selectfont Surface temperatures [K] for every modeled planet and environment combination. Colors represent surface climate: blues are significantly below the freezing point of water, reds are above the runaway greenhouse temperature, and greens are within potentially habitable surface temperatures. The aqua planet environment was only modeled for TRAPPIST-1~e. Venus trace gases have lower boundary values of 30~ppm \ce{H2O}, 28~ppm \ce{SO2}, 1~ppm OCS, and photochemical abundances of CO and \ce{H2S}.
  Temperate surfaces do not imply habitability due to the inhospitable character of the environmental states, including very low water abundance. Planets in the habitable zone may not necessarily have habitable temperatures.
  } 
\end{table*}

\begin{table*}[]
\centering
\caption{Ozone Column Densities [cm$^{-2}$]}
\label{table:ozone}
{\small\selectfont
\begin{tabular*}{\textwidth}{l@{\extracolsep{\fill}}ccccccc}
\hline
\textbf{Planets}                        & \textbf{b} & \textbf{c} & \textbf{d} & \textbf{e} & \textbf{f} & \textbf{g} & \textbf{h} \\ \hline
Aqua planet, clear, 1 bar &         &         &         & $3.1 \times 10^{18}$        &         &         &         \\
Aqua planet, cloudy, 1 bar  &         &         &         & $2.9 \times 10^{18}$        &         &         &         \\
\ce{O2}, desiccated, 10 bar  & $2.5 \times 10^{18}$        & $6.3 \times 10^{18}$        & $1.1 \times 10^{19}$        & $4.7 \times 10^{19}$        & $2.2 \times 10^{22}$        & $6.0 \times 10^{22}$        & $1.3 \times 10^{23}$        \\
\ce{O2}, desiccated, 100 bar  & $7.2 \times 10^{17}$        & $2.4 \times 10^{18}$        & $6.4 \times 10^{18}$       & $8.8 \times 10^{18}$        & $6.1 \times 10^{22}$        & $2.7 \times 10^{23}$        & $1.4 \times 10^{24}$        \\
\ce{O2}, outgassing, 10 bar   & $1.6 \times 10^{18}$        & $5.2 \times 10^{18}$        & $9.9 \times 10^{18}$        & $2.7 \times 10^{19}$        & $8.5 \times 10^{20}$        & $2.4 \times 10^{21}$        & $4.0 \times 10^{21}$        \\
\ce{O2}, outgassing, 100 bar  & $1.7 \times 10^{19}$        & $2.1 \times 10^{20}$        & $2.7 \times 10^{21}$       & $2.4 \times 10^{22}$        & $7.2 \times 10^{23}$        & $6.6 \times 10^{24}$        & $1.8 \times 10^{25}$        \\ 
Venus-like, clear, 10 bar  & $2.3 \times 10^{15}$        & $2.1 \times 10^{15}$        & $5.4 \times 10^{15}$        & $1.0 \times 10^{15}$        & $6.9 \times 10^{15}$        & $8.6 \times 10^{15}$        & $1.5 \times 10^{16}$        \\
Venus-like, cloudy, 10 bar  &  ~   & $1.0 \times 10^{15}$        & $1.1 \times 10^{16}$        & $4.6 \times 10^{15}$        & $7.6 \times 10^{15}$        & $9.1 \times 10^{15}$        & $1.5 \times 10^{16}$        \\
Venus-like, clear, 92 bar  &  $1.8 \times 10^{15}$    & $1.9 \times 10^{15}$        & $5.1 \times 10^{15}$        & $1.3 \times 10^{15}$        & $6.7 \times 10^{15}$        & $9.1 \times 10^{15}$        & $1.6 \times 10^{16}$    \\
Venus-like, cloudy, 92 bar  &  ~   & $1.0 \times 10^{15}$        & $9.5 \times 10^{14}$        & $4.2 \times 10^{15}$        & $7.8 \times 10^{15}$        & $9.8 \times 10^{15}$        & $1.6 \times 10^{16}$   \\
\hline
\end{tabular*}
  \flushleft
  Compare to global-average Earth ozone column depth of $8 \times 10^{18}$~cm$^{-2}$ (\url{https://ozonewatch.gsfc.nasa.gov/}).
  }
\end{table*}

In the following subsections, we describe our input assumptions and resultant environmental states (atmospheric temperature profiles as a function of pressure, and vertical profiles of principal atmospheric constituent abundances) for each atmosphere. Globally-averaged surface temperatures for all modeled climates are given in Table~\ref{table:results}, and \ce{O3} column densities are given in Table~\ref{table:ozone}.

\subsubsection{\ce{O2}-Dominated Atmospheres---Desiccated}

\begin{figure*} [!]
  \centering 
  \includegraphics[width = \textwidth]{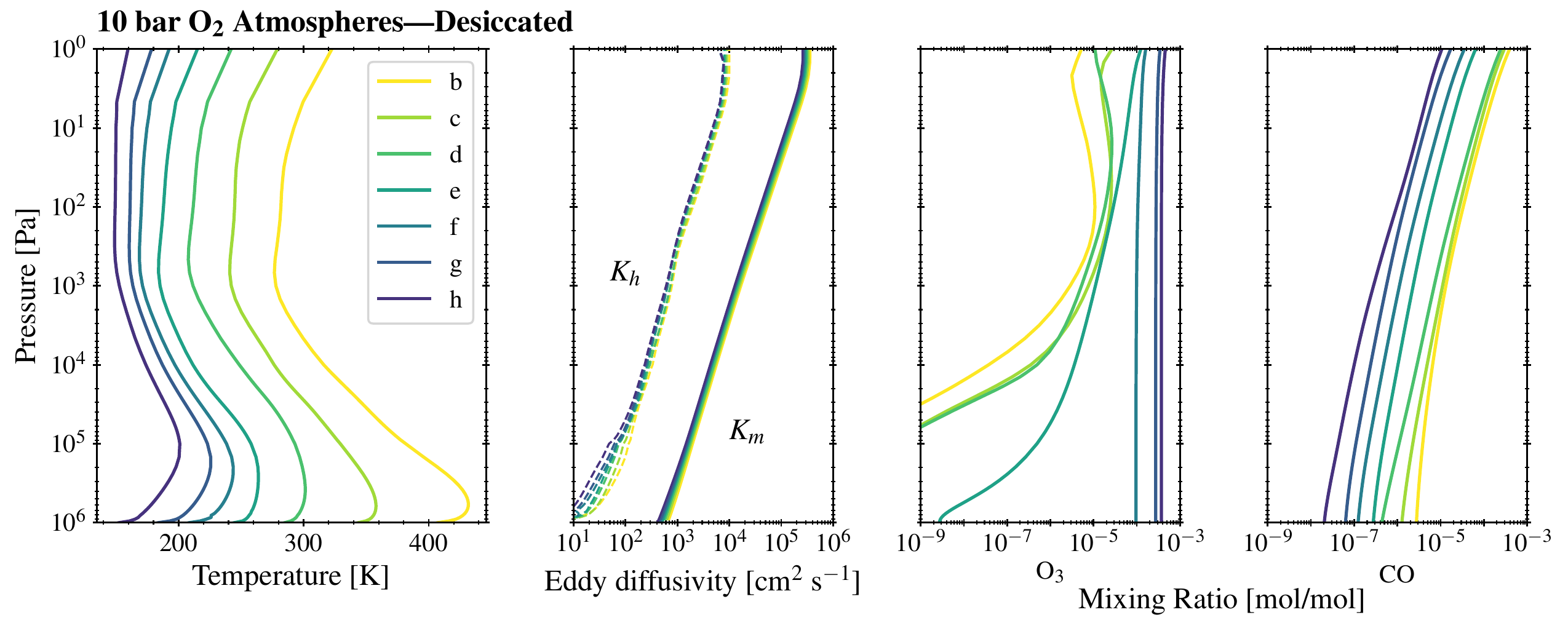}
  \includegraphics[width = \textwidth]{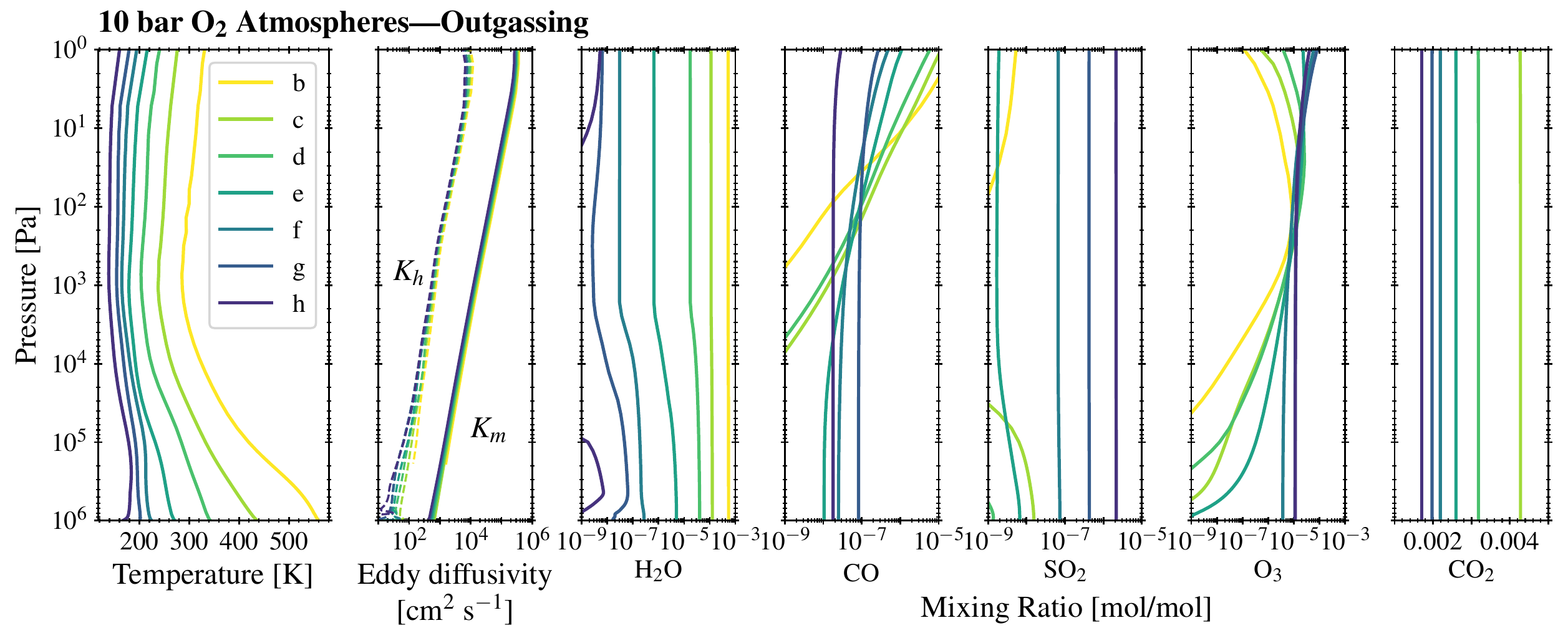}
  \caption{Converged structures for the 10-bar desiccated (\textit{upper panel}) and outgassing (\textit{lower panel}) \ce{O2}-dominated TRAPPIST-1 atmospheres. The 10~bar atmospheres are similar in structure to the 100~bar atmospheres above 10~bars ($10^6$ Pa). The desiccated atmospheric structures exhibit a strong temperature inversion in the lower atmosphere due to absorption by \ce{O3} and \ce{O2}-\ce{O2}. 
The outgassing atmospheric structures and associated surface temperatures exhibit a continuum from Venus-like hotter atmospheres to colder temperature profiles similar to the desiccated cases.
   \label{fig:PT-O2}}
\end{figure*}

In Figure \ref{fig:PT-O2}, we present environmental states for completely desiccated, oxygen-dominated atmospheres for the seven TRAPPIST-1 planets, assuming 10 or 100~bars surface pressure.  These surface pressure values were chosen as representative of possible states, because although TRAPPIST-1's pre-main-sequence evolution may produce up to thousands of bars of \ce{O2} in the most extreme case (Figure~\ref{fig:evol}), much of this \ce{O2} is likely to be re-incorporated into the land surface, ocean, lithosphere, or mantle \citep[e.g.][]{Hamano:2013,Schaefer:2016,Wordsworth:2018}, or depending on each planet's magnetic field, lost to non-hydrodynamic, top-of-atmosphere processes over time \citep[e.g.][]{Khodachenko:2007,Lammer:2007,Lammer:2011,Ribas:2016}, such as the ``electric wind'' on Venus \citep{Collinson:2016}. While an abiotic \ce{O2}-dominated planetary atmosphere has yet to be discovered, and Venus does not retain significant quantities of atmospheric oxygen, it may be possible to test the existence of this hypothesized ocean-loss-generated atmosphere with JWST.

We initialized these atmospheres with 95\%~\ce{O2}, 0.05~bar \ce{CO2} (consistent with \citealp{Meadows:2018}), and the remainder \ce{N2}, each specified as a fixed surface mixing ratio, with no outgassing. Since this case assumes a planet that was severely volatile-depleted in both its atmosphere and mantle, these atmospheres are considered completely desiccated (i.e. they contain no water, nor any other hydrogen-bearing species). The remaining composition is dominated by photochemically-produced trace gases.

The composition of photochemically-produced trace gases, principally \ce{O3} and CO, increase monotonically as a function of irradiation from the parent star (i.e. as a function of semi-major axis). 
The equilibrium concentration of ozone is due to the competing effects of vertical transport and the Chapman cycle, whereby photolysis of \ce{O2} produces atomic oxygen that can combine with \ce{O2} to form \ce{O3}, which is directly lost by photolysis if atomic oxygen combines to form \ce{O2}. For the hotter planets (b--d), \ce{O3} photolysis is efficient down to the surface from absorption in the broad Chappuis band. However, the colder planets receive less UV flux as a result of distance from the star, so the photolysis rates are slower. In the coldest atmospheres (f, g and h), eddy transport can overwhelm photolysis and reaction rates for \ce{O3}, resulting in well-mixed vertical profiles and saturated \ce{O3} absorption bands.
Furthermore, photolysis of \ce{O2} drops faster with semi-major axis because its cross-section is limited to the FUV and easily saturates compared to \ce{O3}, which is includes a strong NUV band and the weak Chappuis band in visible wavelengths, which are not easily saturated.

The monotonic increase in CO with semi-major axis in these atmospheres is due to \ce{CO2} photolysis and lack of recombination. In the terrestrial Solar System atmospheres, the recombination to \ce{CO2} is efficient due to photolytic water products (e.g. \ce{OH^-}). Since this environment is devoid of hydrogen, CO can only recombine with atomic oxygen under a slow, density-dependent three-body reaction \citep{Gao:2015}.

These modeled environments exhibit surface temperatures (152--406~K, from h to b) that are similar to their equilibrium values (173--400~K; \citealt{Gillon:2017,Luger:2017b}) largely due to low greenhouse gas abundances. 
The 100~bar cases are cooler at the surface than the 10~bar cases, which indicates an anti-greenhouse effect is occurring. 
The atmospheric structures for these completely desiccated planets each exhibit a stratospheric temperature inversion, like Earth, due to the radiative effects of the photochemically-produced \ce{O3} and \ce{O2}-\ce{O2} collision-induced absorption (CIA), which is significant in these dense \ce{O2} atmospheres.
A sensitivity test removing each constituent to determine the radiative effect suggests that \ce{O3} and \ce{O2}-\ce{O2} contribute equally to this heating structure, but via different mechanisms: \ce{O2}-\ce{O2} directly absorbs M~dwarf NIR irradiation, while \ce{O3} traps outgoing thermal radiation.  

The pressure level at which the stratosphere peaks is different than Earth due to the \ce{O2}-\ce{O2} CIA.
Earth's stratosphere is caused by \ce{O3} NUV absorption and peaks around $\sim$0.001~bar, while the pressure level at which the peak in these \ce{O2} atmospheres occurs is much closer to the surface, around 1--10~bar, due to the density dependence of CIA. This stratospheric absorption induces heating in the atmosphere, exceeding heating at the planetary surface, causing a surface temperature inversion similar to Earth and Mars at night \citep[e.g.][]{Andre:1982,Hinson:2004}, and inhibiting a troposphere from forming. The causes for these conditions on Earth and Mars at night are due instead to rapid cooling of the surface, rather than warming of the atmosphere. 

\subsubsection{\ce{O2}-Dominated Atmospheres---Outgassing} \label{sec:O2_vapor_results}

 Here we consider the case where a planet has lost its oceans, but continues to outgas from a volatile-rich mantle, as is likely the case on Venus \citep[e.g.][]{Bullock:2001}. For other planetary systems, including TRAPPIST-1, this could also be due to delivery by impactors \citep{Chyba:1990,Kral:2018}. 
 For all seven planets we specify Earth-like outgassing fluxes ($F$), which are relatively well-known:  $1.68\times10^{11}$~cm$^{2}$~s$^{-1}$ \ce{H2O} ($\sim10F_{\ce{CO2}}$; \citealp{Burton:2000}), $1.0\times10^{8}$~cm$^{2}$~s$^{-1}$ \ce{H2S}, $1.35\times10^{9}$~cm$^{2}$~s$^{-1}$  \ce{SO2} \citep{Carn:2017}, $6.88\times10^{9}$~cm$^{2}$~s$^{-1}$ \ce{CO2} ($\sim3.5F_{\ce{SO2}}$; \citealp{Williams:1992}), and $3.7\times10^{5}$~cm$^{2}$~s$^{-1}$ \ce{OCS} ($\sim10^{-4} F_{\ce{CO2}}$; \citealp {Belviso:1986}). These fluxes represent volcanic outgassing, so are distributed within one scale height of the surface ($\sim$6--20~km). We specify minimal non-zero dry deposition rates of $10^{-8}$~cm~s$^{-1}$ for all outgassed species to prevent a ``runaway'' state.
 
As shown in Figure~\ref{fig:PT-O2}, identical outgassing fluxes result in different levels of trace gases across the seven planets. Some species vary monotonically with irradiation (\ce{H2O}, \ce{O3}, \ce{CO2}), while others do not (CO, \ce{SO2}). These trends are modulated by surface gravity (planets b, d, and h have low surface gravity) and atmospheric temperature.
There are several interacting chemical networks:  carbon, ozone, and sulfur. First considering the carbon network, the primary carbon species are \ce{CO2} and its photolytic product, CO. The CO profiles demonstrate the competing effects of photolysis and vertical transport. The hotter planets have a steep CO gradient, driven by strong photolysis, while the colder planets have well-mixed CO profiles, where vertical transport overcomes photolysis rates and the peak abundance drops.  The \ce{CO2} levels are not wholly a function of photolysis; for a given surface pressure, they have different atmospheric column masses (as discussed in \S\ref{sec:results-atmesc} on atmospheric escape). Even though the hotter planets have larger \textit{mixing ratios} of \ce{CO2}, because of hotter temperatures their atmospheres are less dense, and have smaller \textit{number densities}. Photolysis of \ce{CO2} from identical outgassing fluxes results in lower \ce{CO2} inventories (i.e. number densities) in the more irradiated atmospheres. 
These \ce{CO2} results differ from the fully desiccated environments due the Earth-like active surface fluxes we assumed here.
 
The ozone profiles are very similar to the desiccated \ce{O2}-dominated planets. The production of \ce{O3} is the same, but the outgassing planets have more sinks, primarily water vapor. As a result, the outgassing cases overall have lower column densities of \ce{O3}. Since the outer planets can be very cold, they cannot maintain much water vapor (Figure~\ref{fig:PT-O2}), so they too can accumulate substantial quantities of \ce{O3}, particularly g and h.
 
  \begin{figure*}  [!]
  \centering 
  \includegraphics[width = \textwidth]{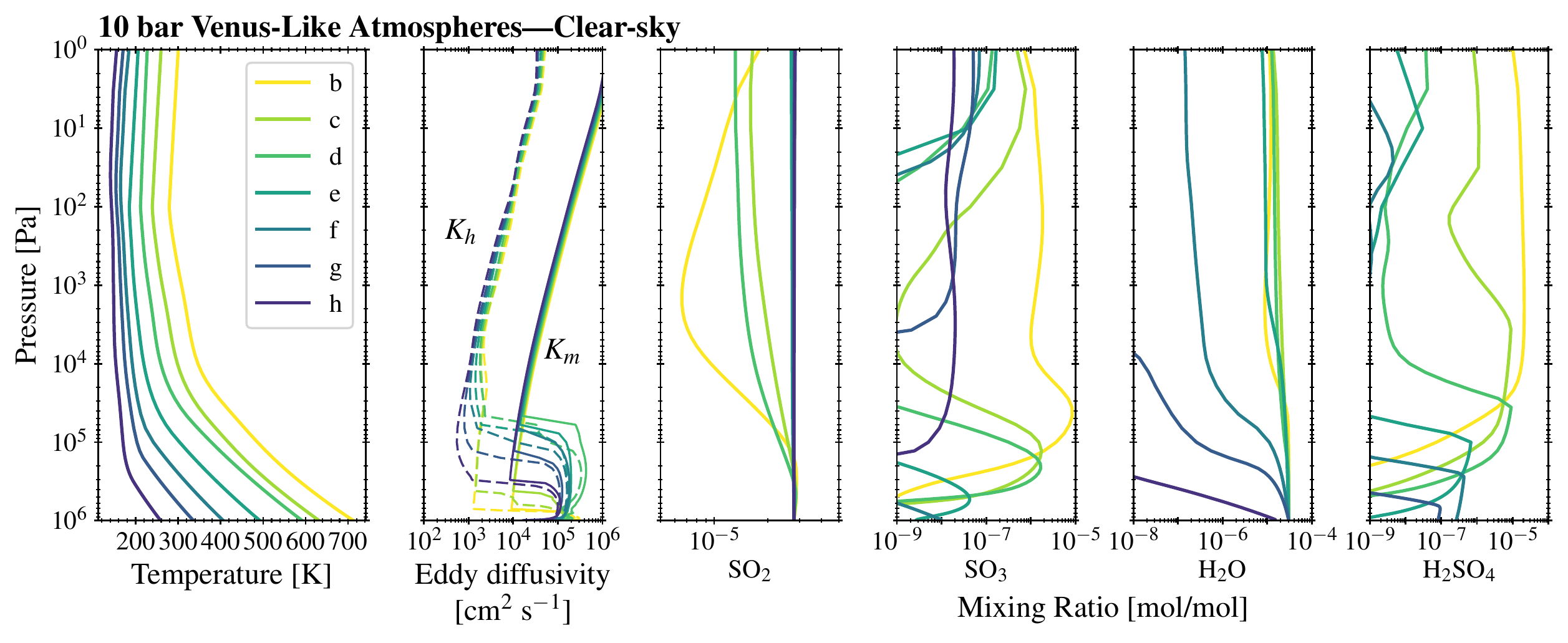}
  \includegraphics[width = \textwidth]{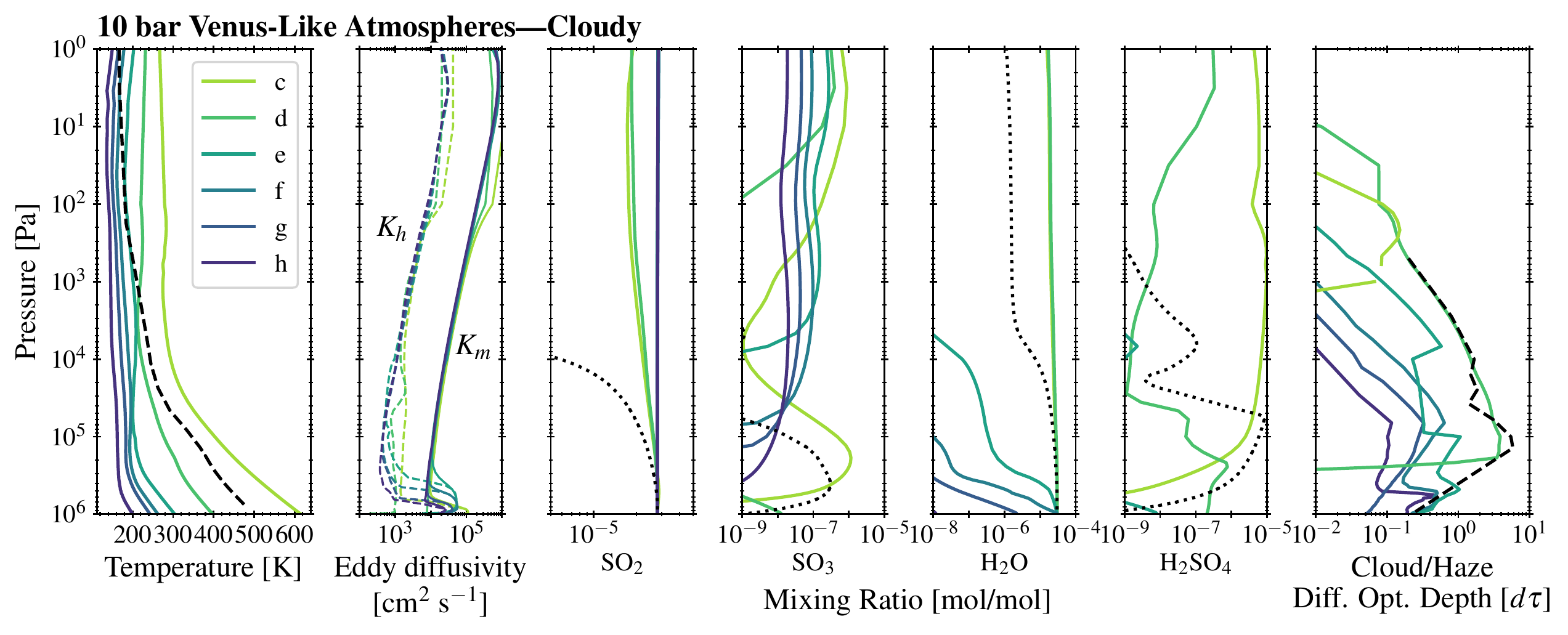}
  \caption{Converged structures of the 10-bar 
  clear-sky (\textit{upper panel}) and cloudy (\textit{lower panel}) Venus-like TRAPPIST-1 atmospheres. Note: b is not included in the cloudy case, as aerosols were not formed. The Venus International Reference Atmosphere (VIRA) temperature structure is shown in the cloudy panel (black dashed line). The 92~bar atmospheres are not shown here, but show similar structure to the 10~bar cases. Here we show \ce{H2O}, \ce{SO2}, \ce{SO3}, and \ce{H2SO4} gases only, as the most relevant to climate and cloud formation. The bulk gases \ce{CO2} and \ce{N2} are 96.5\% \ce{CO2} and 3.5\% \ce{N2} for the 10~bar and 92~bar atmospheres, respectively.  With a (clear-sky) Venus-like atmosphere, TRAPPIST-1~b may exhibit surface temperatures in excess of Venus (due to higher instellation) and is too hot to form sulfuric acid aerosols in our model, even though it was most effective at forming high-altitude \ce{H2SO4} vapor. Sulfur dioxide survives more readily in these atmospheres due to the lower MUV--NUV flux. 
   \label{fig:PT-CO2}}
\end{figure*}

The abundance and vertical distribution of species in the sulfur network (\ce{SO2}, \ce{SO3}, and \ce{H2SO4}) is a balance between water vapor availability, temperature, and photolysis rates. The colder atmospheres are water-limited due to condensation and lower photolysis rates, so atmospheric \ce{SO2} inventories are highest. The formation of \ce{SO3} is from combination of \ce{SO2} with free oxygen atoms. Sulfuric acid is a primary sink for \ce{SO3} and is formed by \citep{Krasnopolsky:2012}:
\begin{equation} \label{eq:h2so4}
     \ce{SO3 + H2O + H2O -> H2SO4 + H2O},
\end{equation}
and can either thermally decompose or condense and rain out. The inner planets receive sufficient irradiation to permit efficient conversion of \ce{SO2} to \ce{SO3} and \ce{H2SO4}.
 
Volatiles in these atmospheres have a large effect on planetary climate and bridge the results of the \ce{O2} desiccated and Venus-like atmospheres. The coldest planets (g and h) maintain temperature profiles similar to the desiccated cases, including near-surface temperature inversions, because they are too cold to maintain quantities of water vapor sufficient to heat the lower atmospheres. Conversely, the remaining planets generate sufficient water abundances to maintain a positive lapse rate. To prevent a full runaway greenhouse, we have limited the stratospheric water relative humidity to 0.1\% for b and 1.0\% for c, while the remaining planets are limited to 10\%. Planets b, c, and d show the tendency to continue into a runaway greenhouse state, which is supported in our results by their abundances of water vapor well-mixed throughout the atmospheric column. Reducing the outgassing of \ce{H2O} in our model would more closely approximate Venus-like \ce{H2O} abundance. Only TRAPPIST-1~e has a moderate surface temperature (271--314~K) in this outgassing case, which supports the possibility for it to outgas an ocean and recover habitability.

\subsubsection{Venus-Like Atmospheres}

\begin{figure} 
  \centering
  \iftwocol
    \includegraphics[width = \columnwidth]{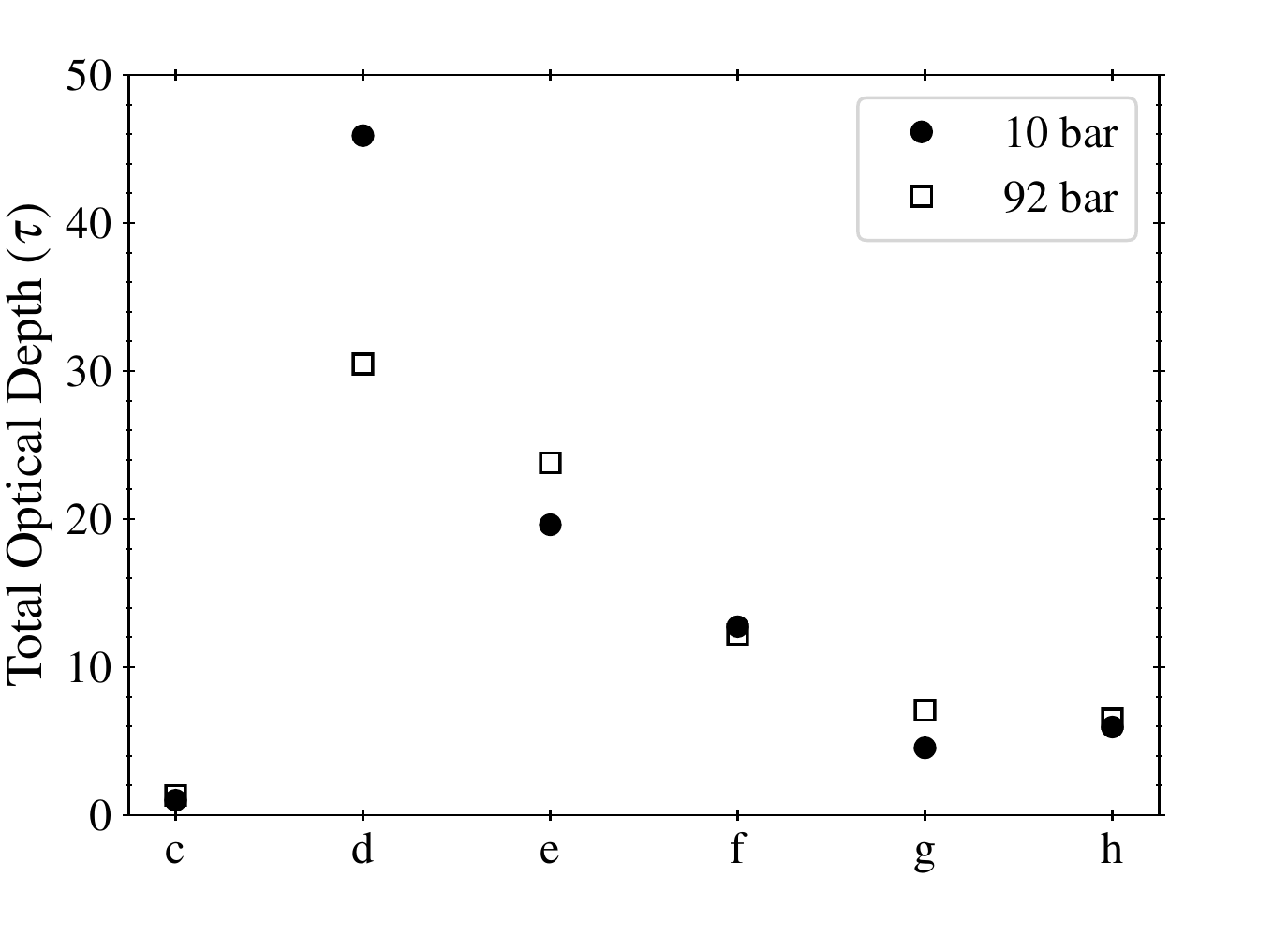}  
    \else
    \includegraphics[width = 0.5\textwidth]{optdepth}  
    \fi
  \caption{Sulfuric acid cloud total optical depths for each cloudy Venus atmosphere. TRAPPIST-1 b did not condense sulfuric acid in our models. TRAPPIST-1~c maintained only thin clouds, because sulfuric acid absorption warms the cloud layer and the local warming can cause the clouds to evaporate. Optical depths peak for planet d, which has vigorous sulfuric acid formation and has lower surface gravity, so can more easily loft aerosols. The planets generally decline in optical depth with distance from the star due to lower photolysis rates and for the 10~bar atmospheres, the lower abundance of water vapor (in the 10~bar cases). Note the 92~bar atmosphere photochemistry and cloud formation were truncated at 10~bar. Planets f, g, and h had cloud formation to 10~bar, and so could also form thicker clouds deeper toward the surface.
   \label{fig:optdepth}}
\end{figure}

Here we model Venus-like environments, which may be common around M~dwarfs \citep{Kane:2014}. For this environmental state, we assume that the complete loss of oceans during the super-luminous pre-main-sequence phase removed a potentially strong sink for soluble gases (e.g. \ce{CO2} and \ce{SO2}).  Subsequent loss or sequestration of oxygen over time \citep{Schaefer:2016}, and initial and subsequent outgassing of volatiles, may have allowed a Venus-like high-\ce{CO2} atmosphere to develop. Even though planet h may not have lost an entire Earth ocean (see Figure~\ref{fig:evol}), it could have initially outgassed significant amounts of \ce{CO2}. We modeled Venus-like atmospheres for all seven planets (Figure~\ref{fig:PT-CO2}) using the 10-bar boundary conditions used to validate observations of Venus, following \citet{Krasnopolsky:2012} (see Appendix~\ref{app:validation}). Venus atmospheric chemistry is typically modeled in sections: by thermochemistry, which dominates the lower atmosphere below the cloud deck \citep[e.g.][]{Krasnopolsky:2013}, and by photochemistry, which is responsible for the directly observable altitudes at and above the cloud deck \citep[e.g.][]{Krasnopolsky:2012,Zhang:2012}. We use photochemistry to model these atmospheres but do not include thermochemical reactions, which could be a useful improvement in future work for the planets with the highest surface temperatures.

Our Venus atmospheres are initiated as either 10 or 92~bars of total pressure with Venus-like levels of \ce{CO2} (96.5\%) and \ce{N2} (3.5\%). For the 92~bar cases, we assumed a constant mixing ratio for every species from 10~bars to the surface. We initiated the photochemical model with trace gas lower boundary conditions required to reproduce observations of Venus following the modeling of \citet{Krasnopolsky:2012}, consisting of fixed mixing ratios at 10~bar of 30~ppm \ce{H2O}, 4.5~ppm \ce{H2}, 5.5~ppm NO, 28~ppm \ce{SO2}, 1~ppm OCS, and 400~ppm HCl. We calculate the photochemical-kinetic equilibrium profiles of these and other trace gases subject to these boundary conditions. These trace gases absorb at wavelengths that are complementary to \ce{CO2} and can have a substantial effect on climate  \citep{Bullock:2013,Lee:2016}.

For purposes of climate-photochemical modeling for observational discriminants, we separately model both clear-sky and cloudy Venus-like atmospheres, except for planet b, which does not condense sulfuric acid aerosols in our photochemistry model.
Both clear-sky and cloudy atmospheres display a similar, Venus-like profile (dashed black line in Figure~\ref{fig:PT-CO2}), despite being irradiated by a very different SED than the Sun. The temperature profiles are characterized by long adiabats from the surface each to a $\sim0.05$--1.0~bar tropopause. The surface temperatures are the hottest of the environments we simulated, ranging from 259--465~K (200--398~K cloudy) for planet h (outside the outer edge of the HZ), to 714--927~K for b, which receives approximately twice Venus' stellar flux.

As in the \ce{O2} outgassing cases, the Venus-like atmospheric photochemistry is dominated by the formation of sulfuric acid from the combination of water with \ce{SO3}, a photolytically-derived product of \ce{SO2}.  This sulfur chemistry is not driven as strongly as Venus' due to lower MUV--NUV flux from TRAPPIST-1, so the rates of both \ce{SO2} and \ce{SO3} photolysis are much lower than for Venus. However, \ce{SO2} is still susceptible to photolytic destruction, which results in the trend seen in Figure~\ref{fig:PT-CO2}, that \ce{SO2} is more abundant further from the star. The abundance of \ce{SO3} is a balance between its formation rate via \ce{SO2} photolysis and the rate at which it combines with water to form \ce{H2SO4}.  In our simulations, water vapor decreases as a function of semi-major axis in both clear-sky and cloudy cases due to this condensation into \ce{H2SO4} aerosols, as well as into water and water-ice in the cooler planets (e, f, g, and h; although we do not model the water clouds that would form as a result of this process).

The prevalence of \ce{H2SO4} vapor depends on its chemical formation, where and whether it condenses, and its destruction via thermal decomposition in the lower atmosphere.  In our simulations, \ce{H2SO4} formed in abundances similar to Venus (up to 17~ppm vs 8.5~ppm for Venus, \citealt{Krasnopolsky:2015}) for the hotter TRAPPIST-1 planets and drops in abundance with semi-major axis due to the reduced availability of \ce{H2O} at lower temperatures.  The highest abundances of \ce{H2SO4} and \ce{SO3} are seen in the stratospheres of the hotter planets (b and c), with peak vapor concentrations decreasing with semi-major axis and at lower altitudes. Similarly, the peak optical depths of the clouds drop with semi-major axis due to less effective formation processes. Since \ce{H2SO4} can persist closer to the surface in the colder atmospheres, the semi-major-axis-dependent total optical depth peaks for planet d, due to the competing effects of formation/condensation and thermal decomposition/rainout (see Figure~\ref{fig:optdepth}. Notably, for planet b, the modeled atmosphere was too hot for condensation. Planet d exhibited the highest aerosol optical depth due to its low surface gravity, which allows aerosols to be more easily lofted in the atmosphere.

The sulfuric acid aerosols in our cloudy Venus atmospheres substantially reduced the surface temperatures. The net top of atmosphere stellar irradiance was reduced by $\sim$50--60$\%$, resulting in temperature ranges of 200--616~K (10~bar) and 398--779~K (92~bar), h to c. As a caveat, it is possible that sulfuric acid production in Venus' atmosphere may vary by orders of magnitude \citep{Gao:2014}, and it is unknown how an alien M~dwarf spectrum may affect the more uncertain aspects of Venus-like atmospheric chemistry, particularly of the unknown UV absorber, which we exclude.

  \subsubsection{Aqua Planet}
  
  \begin{figure*}  
  \centering 

  \includegraphics[width = \textwidth]{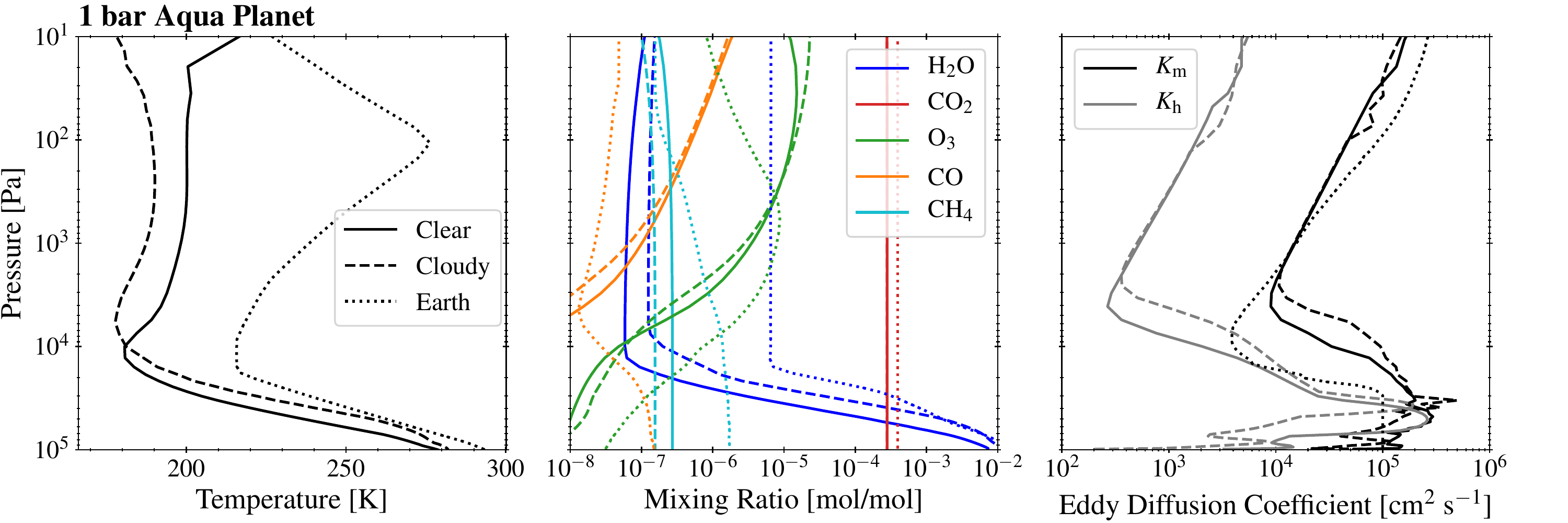}
  \caption{Climates, gas mixing ratios, and eddy diffusion profiles for TRAPPIST-1~e, clear-sky and cloudy aqua planets. Included are data from Earth: the temperature and mixing ratio profiles are from case 62 of the Intercomparison of Radiation codes in Climate Models (ICRCCM), and the eddy diffusion rate is that retrieved by \citet{Massie:1981}.  The temperatures of our modeled atmospheres are somewhat colder than modern Earth, but still able to maintain surface water. The mixing ratios of trace constituents differ from modern Earth, except \ce{CO2}.
   \label{fig:PT-H2O}}
\end{figure*}

For TRAPPIST-1~e only, we modeled a potentially habitable, ocean-covered planet. Our initial work and that of recent 3D~GCM studies \citep{Wolf:2017,Turbet:2018} suggest that the other habitable zone planets (f and g) require \ce{CO2}-dominated atmospheres for stability, and cannot be Earth-like. This temperate, Earth-like  atmosphere could exist if e had a sufficient volatile inventory to survive the stellar pre-main-sequence, had subsequent outgassing or water delivery, or if these planets migrated after formation \citep{Luger:2015a,Meadows:2018}. 
We simulated this environment with an atmosphere of 80\%~\ce{N2} and 20\%~\ce{O2}, similar to Earth. We assume the \ce{O2} could be either biogenic or have resulted from early ocean loss. In addition to the fluxes assumed for the outgassing \ce{O2} cases, we specify an Earth-like geological source of \ce{CH4} ($6.8~\times~10^8$~cm$^2$~s$^{-1}$; \citealp{Guzman:2013}). Water vapor is calculated self-consistently using our mixing-length convection and condensation routine, described in \S\ref{sec:condensation}, assuming 100\% surface relative humidity.

We model both cloudy and clear-sky atmospheres for the aqua planet. In the cloudy case, we specify cirrus and altostratus clouds, each of cumulative optical depth $\tau=5$. These are based on the VPL 3D Spectral Earth Model, which used Earth observing data with our SMART radiative transfer model to faithfully reproduce Earth's spectrum \citep{Robinson:2011}. 

We use a wavelength-dependent open ocean surface albedo for this planet. We found that typical snow-covered ice surfaces with albedo up to 0.73 in the visible had little impact on surface temperature, since the SED of TRAPPIST-1 and other late-type M~dwarfs emit most of their radiation longward of 1~\um{}, and both ice and water are efficient absorbers in the NIR \citep{Shields:2013}. 

These aqua planet cases are able to maintain temperate surfaces (279~K clear, 282~K cloudy). The clouds have a slight greenhouse effect, as the energy lost by scattering incoming stellar light is more than balanced by a colder emission temperature.  The cloudy environment exhibits a much cooler tropopause and weak stratosphere, but similar tropospheric profile, since the troposphere is driven by moist convective processes in both cases.

With Earth-like outgassing, the aqua planets exhibit pre-industrial Earth-like abundances of \ce{CO2}, but different profiles of other trace gasses. The stratospheric water abundance is reduced by a cold tropopause, more so for the cloudy case. The ozone profile is different than Earth's, with a broader, higher altitude (lower pressure) peak abundance, although the column density is only slightly lower (Table~\ref{table:ozone}). Ozone failed to produce a significant stratosphere, because the stellar spectrum provides very little flux at NUV--visible wavelengths absorbed by \ce{O3}. For this modeled environment, we limited CO to photochemical production, so the tropospheric levels are significantly less than Earth's CO, which is primarily from biologically-related sources \citep{Seinfeld:2006}. Similarly, the geological fluxes result in a correspondingly lower tropospheric methane abundance but similar abundance in the upper stratosphere.

\subsection{Spectra and Observational Discriminants} 

\begin{figure*} 
  \centering
  \includegraphics[width = \textwidth]{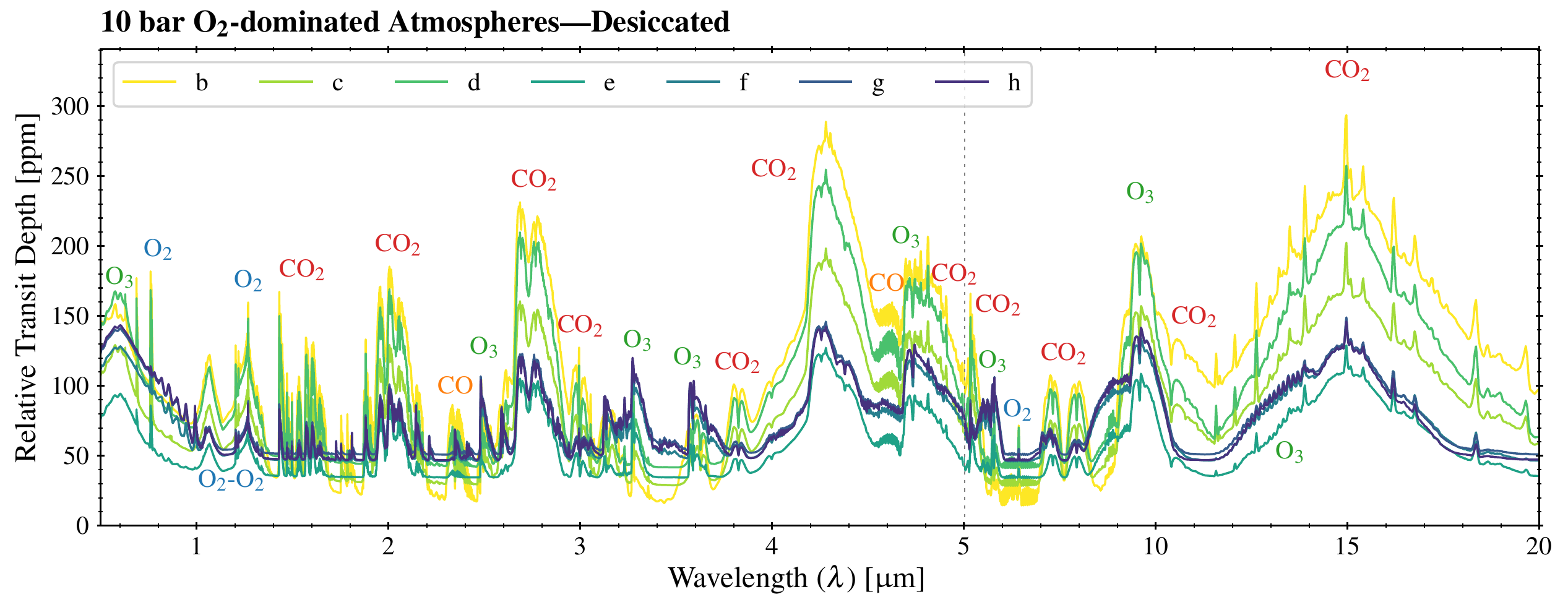}
  \includegraphics[width = \textwidth]{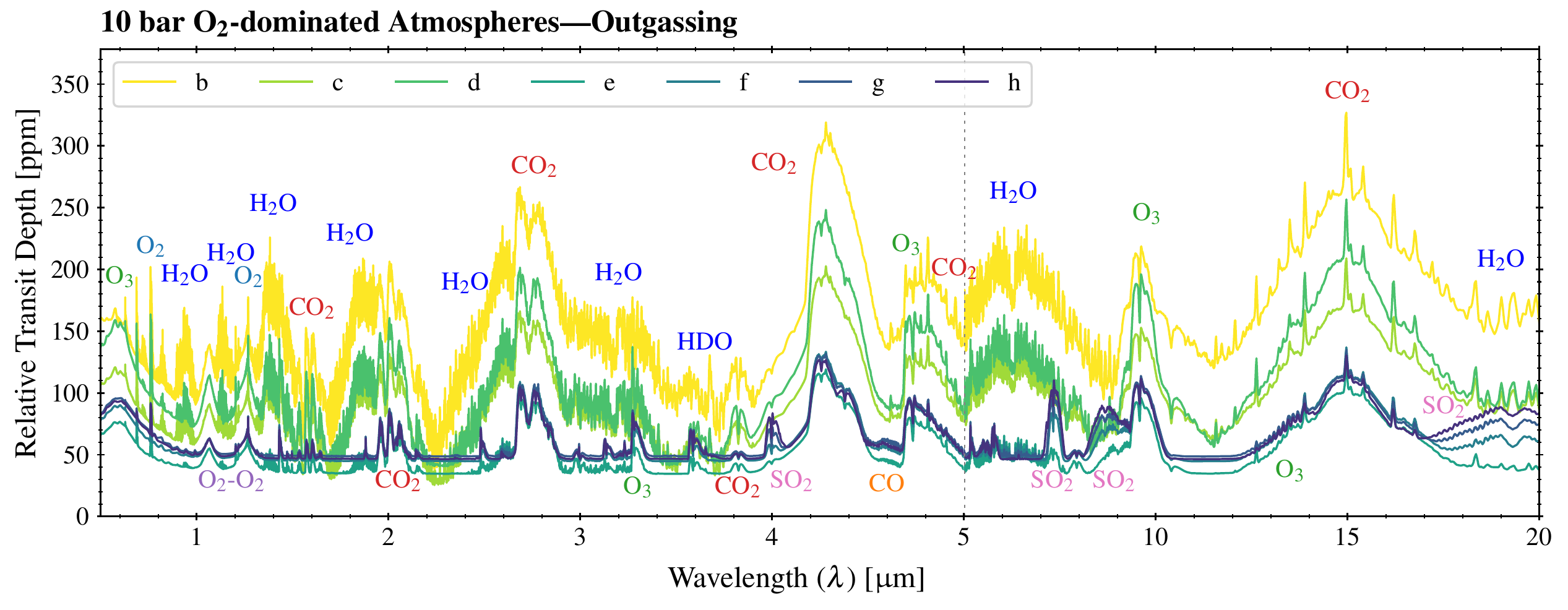}
  \caption{Simulated transit transmission spectra for the \ce{O2} desiccated (\textit{upper panel}) and \ce{O2} outgassing (\textit{lower panel}) atmospheres. The y-axes are the relative transit depths, and show the modeled atmospheric signal. We show only the 10~bar atmospheres, as the 100~bar atmospheres are qualitatively similar. The desiccated atmospheres are dominated by \ce{CO2} and \ce{O2}-\ce{O2}. The presence of weaker \ce{O3} bands and CO are indicative of the desiccated environment. The outgassing atmospheres have additional features from \ce{H2O} and \ce{SO2}. The hotter atmospheres maintain substantial stratospheric \ce{H2O} and the outer planets, too cold for maintaining \ce{H2O}, build up \ce{SO2}, which would otherwise condense with \ce{H2O} to form sulfuric acid. 
  \label{fig:trn-evol}}
 
\end{figure*}

\begin{figure*} 
  \centering
  \includegraphics[width = \textwidth]{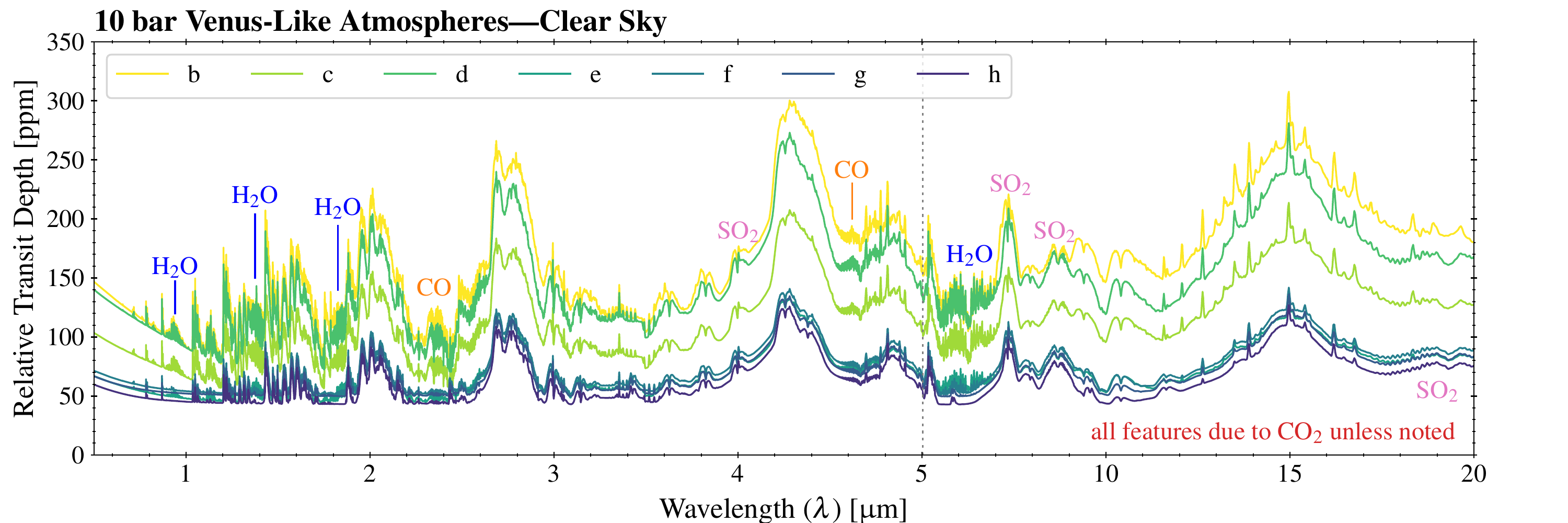}
  \includegraphics[width = \textwidth]{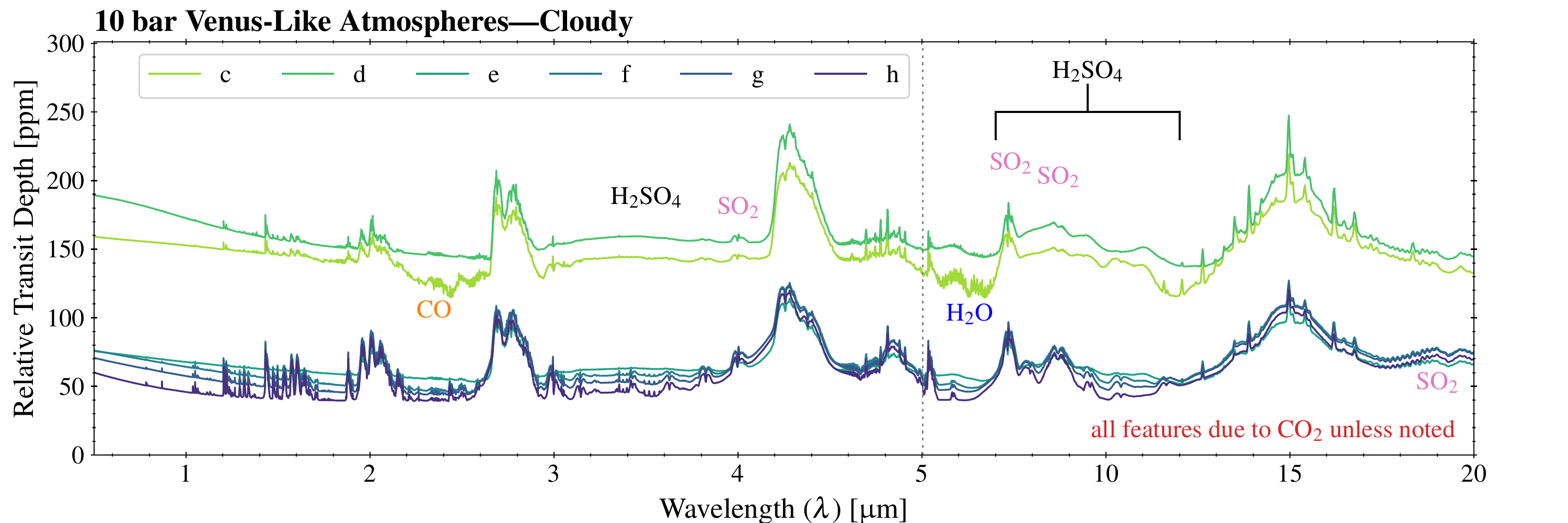}
  \caption{Simulated transit transmission spectra for the Venus-like clear-sky (\textit{upper panel}) and cloudy (\textit{lower panel}) atmospheres. The y-axes are the relative transit depths, and show the modeled atmospheric signal. We show only 10~bar atmospheres, as the 92~bar atmospheres are quantitatively similar. All unlabeled features are \ce{CO2}, which dominate these spectra. The generally flat, higher transit depths of the cloudy Venus-like spectra are due to the sulfuric acid aerosols. TRAPPIST-1~b is not included, because it did not condense \ce{H2SO4} in our model. The colder cloudy atmospheres have lower cloud decks (see Figure~\ref{fig:PT-CO2}), revealing deeper relative transit depths.
  \label{fig:trn-venus}}
 
\end{figure*}

Here we show simulated transit transmission and emission spectra for our evolved worlds to support observation planning with JWST for this planetary system. These are noiseless spectra generated using SMART, sampled at 1~cm$^{-1}$ and convolved with a 1~cm$^{-1}$ half-width at half-max slit function, which can be used as the model input for instrument and observation simulators that calculate noise sources and instrument sensitivity.  These spectra can also be compared with known or anticipated instrument noise floors to determine zeroth order signal detectability. In a subsequent paper \citep{Lustig:2018}, these spectra will be used to quantify the detectability of our simulated atmospheres with JWST and assess optimum observing techniques for identifying key planetary characteristics.

We have also produced direct imaging reflectance spectra.
Reflectance spectra of late-type M~dwarfs may first be observed by large ground-based instruments, or later by HabEx/LUVOIR.
The spectra we simulated, including those not presented here, are available online using the VPL Spectral Explorer\footnote{\url{http://depts.washington.edu/naivpl/content/vpl-spectral-explorer}}, or upon request.

\subsubsection{Transit Transmission Spectra}  


\begin{figure*}[ht]
  \centering
  \includegraphics[width = \textwidth]{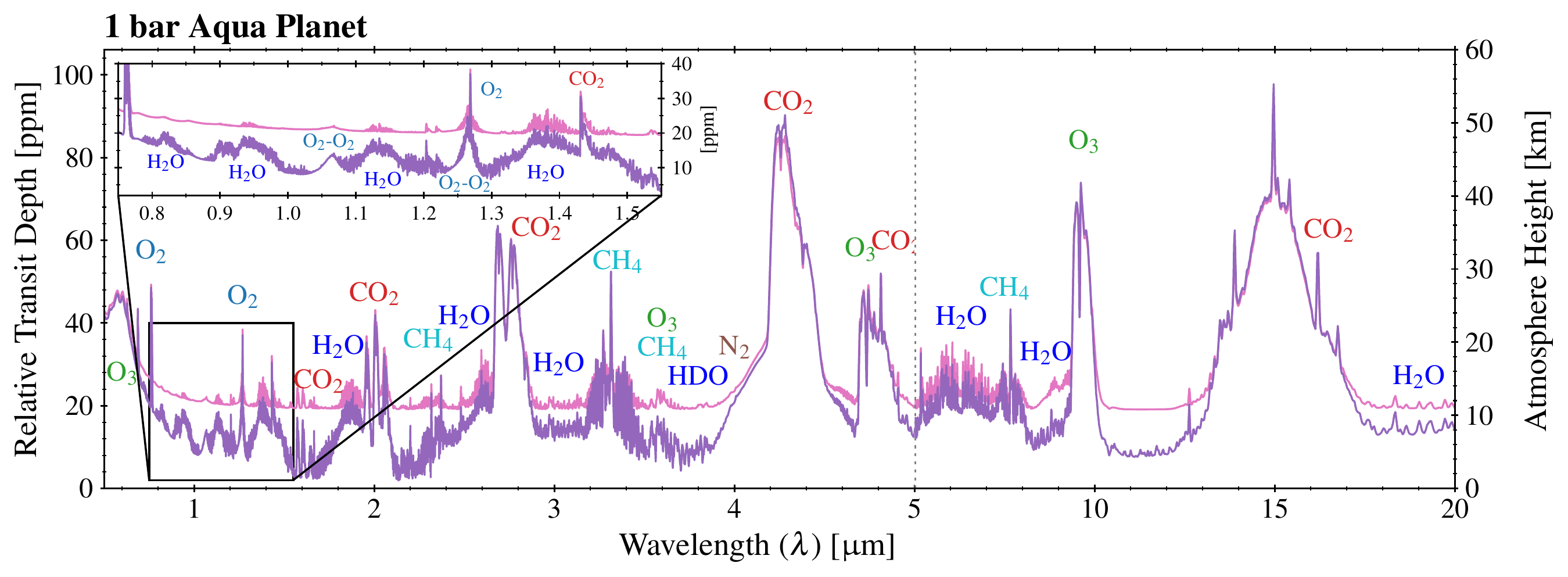}
  \caption{Transit transmission spectrum for clear sky and cloudy TRAPPIST-1~e aqua planets. Most of this simulated planet's important and abundant gases, including \ce{H2O}, \ce{O2}, \ce{CH4}, and \ce{O3}, have features in this spectral range.
  These cloudy and clear spectra demonstrate that a potentially habitable planetary atmosphere, some of its trace outgassed constituents, and photochemical byproducts are accessible using transmission spectroscopy, which could be attempted with JWST.   \label{fig:trn-H2O}}
\end{figure*}

In Figures \ref{fig:trn-evol}--\ref{fig:trn-H2O}, we show transit transmission spectra of our modeled TRAPPIST-1 planetary environments, covering nearly the entire wavelength range of potential JWST exoplanet observations (0.5--20~\um{}). These spectra are presented as ``relative transit depth,'' i.e.~the signal originating from the atmosphere as compared to the solid planetary body. Note in this paper we do not account for stellar limb darkening, so ``transit depth'' here refers to $(R_p/R_*)^2$, where $R_p$ is the radius of the solid body and $R_*$ the radius of the star. To assess the signal originating from the atmosphere, we can expand the transit depth:
\begin{equation} \label{eq:dF_F}
    \begin{split}
   \frac{dF}{F} = \left( \frac{R_p + R_a}{R_*} \right)^2 = \newline \left( \frac{R_p}{R_*} \right)^2 +\frac{2 R_p R_a}{R_*^2} + \left( \frac{R_a}{R_*} \right)^2,
   \end{split}
\end{equation}
and define the relative transit depth of the atmosphere, valid for atmospheres with small scale height:
\begin{equation} \label{eq:dF_F_approx}
    \frac{dF_a}{F} \approx \frac{2 R_p R_a}{R_*^2},
\end{equation}
where $R_a$ is the vertical extent of the atmosphere. Although the radius of the solid body and altitude of the atmosphere for exoplanets cannot be separately measured, this equation allows us to relate predicted variations in transit depth of the modeled spectra for different-sized bodies to the physical extent of the modeled atmospheres. 

The seven planets exhibit a variety of features in transmission, the strengths of which depend on each planet's characteristics. TRAPPIST-1~b displays the strongest features, owing to its higher temperature and moderately low surface gravity, both of which increase the atmospheric scale height, and therefore its transmission signal. Planet b shows molecular signals exceeding 200~ppm, which is well above the putative 20, 30, and 50~ppm noise floors for JWST NIRISS SOSS, NIRCam grism, and MIRI LRS, respectively \citep{Greene:2016}. Planet d exhibits molecular signals nearly as strong as b, in large part because d has the lowest surface gravity of these planets.
Due to geometry and the effects of refraction \citep{Misra:2014,Meadows:2018}, the surface and near-surface atmospheres of the 10--100~bar atmospheres cannot be probed with JWST transmission spectroscopy. 
All of the modeled atmospheres we present include \ce{CO2}, and despite a broad range of abundances, they all exhibit strong \ce{CO2} absorption at 2.0, 2.8, 4.3, and 15~\um{}, though the hotter and/or higher abundance \ce{CO2} atmospheres have additional \ce{CO2} features, particularly short-ward of 2~\um{}. In the following subsections, we present the spectra for each environment. 

\paragraph{\ce{O2}-Dominated Atmospheres, Desiccated}

Transit transmission spectra for the desiccated, high-\ce{O2} atmospheres are presented in Figure~\ref{fig:trn-evol}.  As in \citet{Schwieterman:2016} and \citet{Meadows:2018}, the characteristic spectral features of a high-\ce{O2} atmosphere are strong \ce{O3} and \ce{O2}-\ce{O2} CIA bands together. The CIA bands are strongest at 1.06 and 1.27~\um{}. A mixture of \ce{O2}, \ce{O3}, and \ce{O2}-\ce{O2} are present at 0.5--1.3~\um{}, and a mixture of \ce{CO2}, \ce{CO}, and \ce{O3} from 1.5--6~\um{}. The strongest \ce{O3} bands are the Chappuis band at $\sim$0.6~\um{} and the 9.6~\um{} feature, with weaker bands at 2.5, 3.3, 3.6, 4.75, and 5.6~\um{}. These weaker bands are generally only available here because of high levels of \ce{O3} combined with relatively low levels of \ce{CO2}. Most of this wavelength range could be accessible to JWST NIRISS and NIRSpec instruments. Features of \ce{CO2} and \ce{O3} are also present from 6--20~\um{}, which could be probed by the JWST MRS instrument. In addition to the complete lack of any water features, the presence of CO features (2.35~\um{} and 4.6~\um{}) is characteristic of this severely desiccated atmosphere  \citep[e.g.][]{Gao:2015}, though these will be very challenging to observe with JWST.

\paragraph{\ce{O2}-Dominated Atmospheres, Outgassing}

The transmission spectra of the outgassing atmospheres are shown in Figure~\ref{fig:trn-evol}. The detectability of outgassed trace species varies by planet, particularly \ce{H2O} and \ce{SO2}. Water features at 1.4, 1.9, 2.6, 3.0--3.3, and 6.3~\um{} rival the strength of \ce{CO2} bands for the hottest planets (b, c, and d), due to stratospheric water vapor in excess of the moist greenhouse limit, indicative of planets experiencing water loss. The colder planets have insufficient \ce{H2O} to generate \ce{H2SO4}, so maintain higher levels of \ce{SO2}, which shows small features of less than 60~ppm at 4.0, 7.3, 8.7, and 19~\um{}. Much larger \ce{SO2} fluxes and/or lower water vapor abundances would be required to generate \ce{SO2} features above the putative noise floor of JWST \citep[20--50~ppm][]{Greene:2016}.

\paragraph{Venus-Like Atmospheres}

Clear-sky and cloudy Venus-like transmission spectra are shown in Figure~\ref{fig:trn-venus}. With a clear sky, Venus-like atmospheres exhibit deep absorption features, primarily from \ce{CO2}. Water vapor is present in the clear-sky spectra between 0.9--2.0~\um{} and at 6.3~\um{}. Sulfuric acid aerosols truncate the minimum altitude probed by the transmission measurement, severely reducing the strength of the absorption features. These Venus-like planets may have absorption features in transmission approaching $\sim90$~ppm.

\paragraph{Aqua Planet}

Clear-sky and cloudy aqua planet transmission spectra for TRAPPIST-1~e are shown in Figure~\ref{fig:trn-H2O}. Due to the large angular size of TRAPPIST-1 as seen from the planet, the clear-sky atmosphere can be probed nearly to the surface, while the cloud deck truncates the spectrum at 11~km. A number of molecular absorption bands are present, including \ce{O3} ($\sim$0.6, 4.7, 9.6~\um{}), \ce{O2} (0.76, 1.27~\um{}), \ce{CH4} weakly (3.3 and 7.7~\um{}), and \ce{H2O} (6.3~\um{}). Even with an abundance of only $\sim$290~ppm, \ce{CO2} is the most observable molecule in these atmospheres (1.6, 2.0, 2.7, 4.3, 15~\um{}). The bulk atmospheric constituent, nitrogen, exhibits a 4.1~\um{} CIA band, which could provide a sensitive probe of its partial pressure \citep{Schwieterman:2015b}.

\begin{figure*}
  \centering
  \includegraphics[width = 0.49\textwidth]{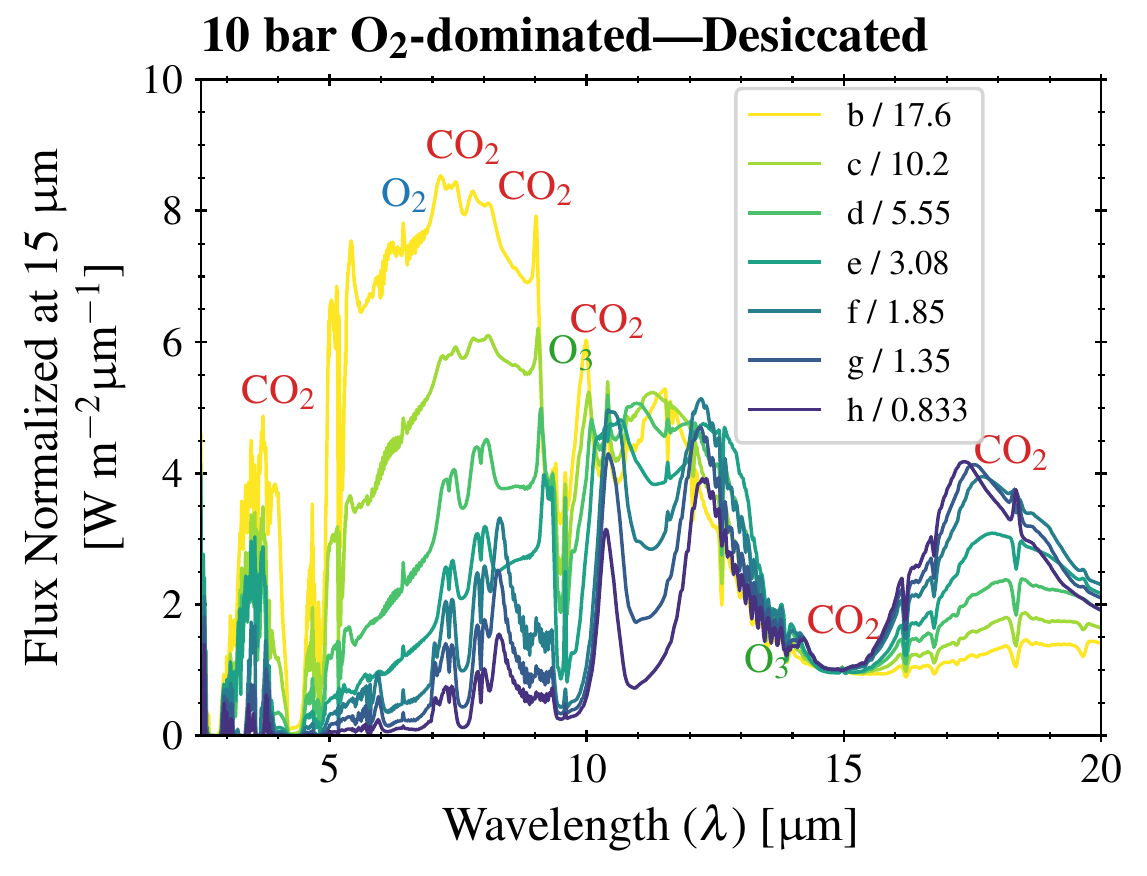}
  \includegraphics[width = 0.49\textwidth]{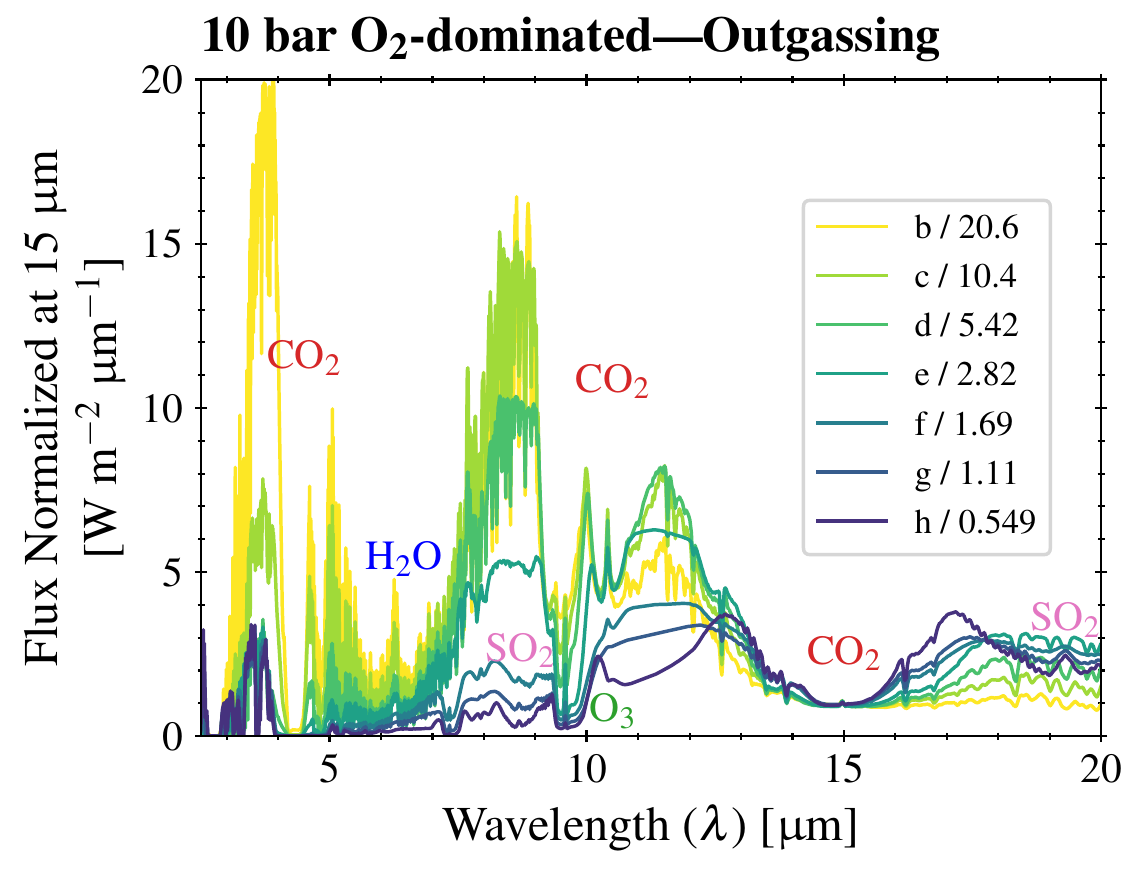}
  \includegraphics[width = 0.49\textwidth]{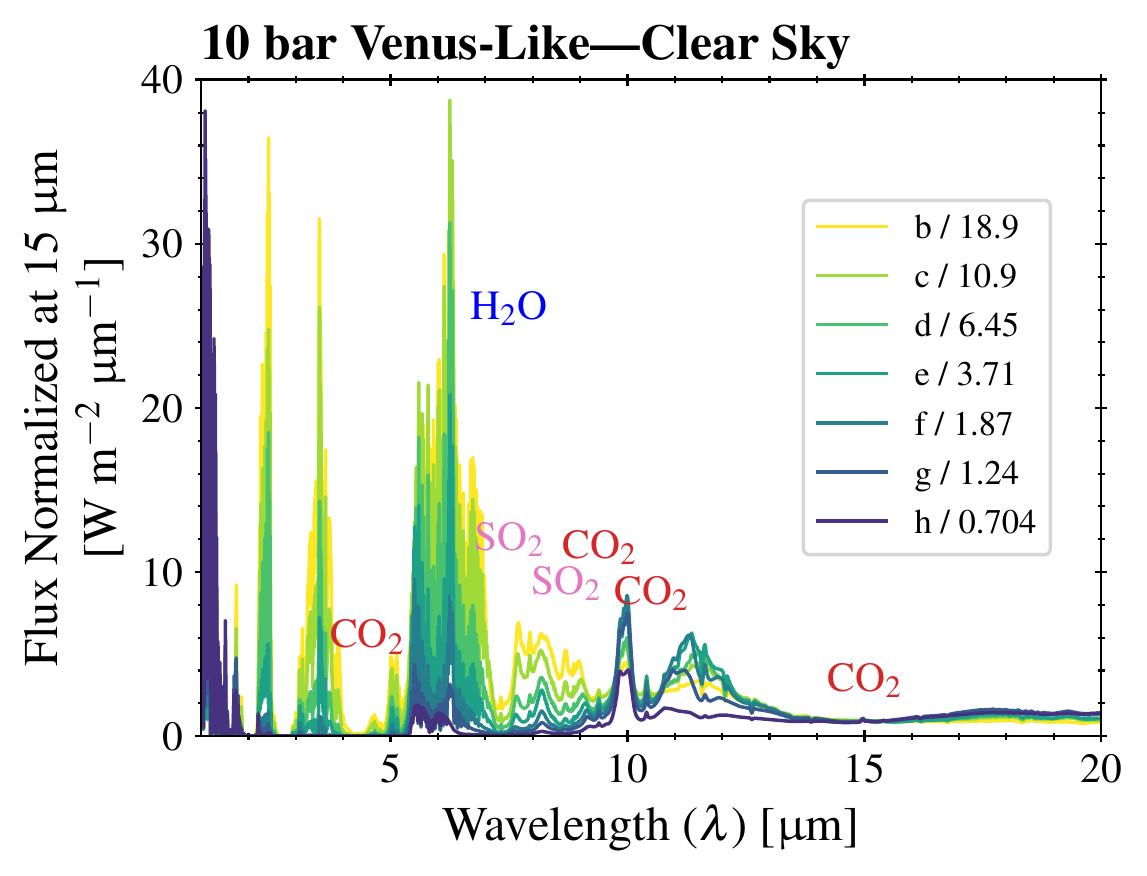} 
  \includegraphics[width = 0.49\textwidth]{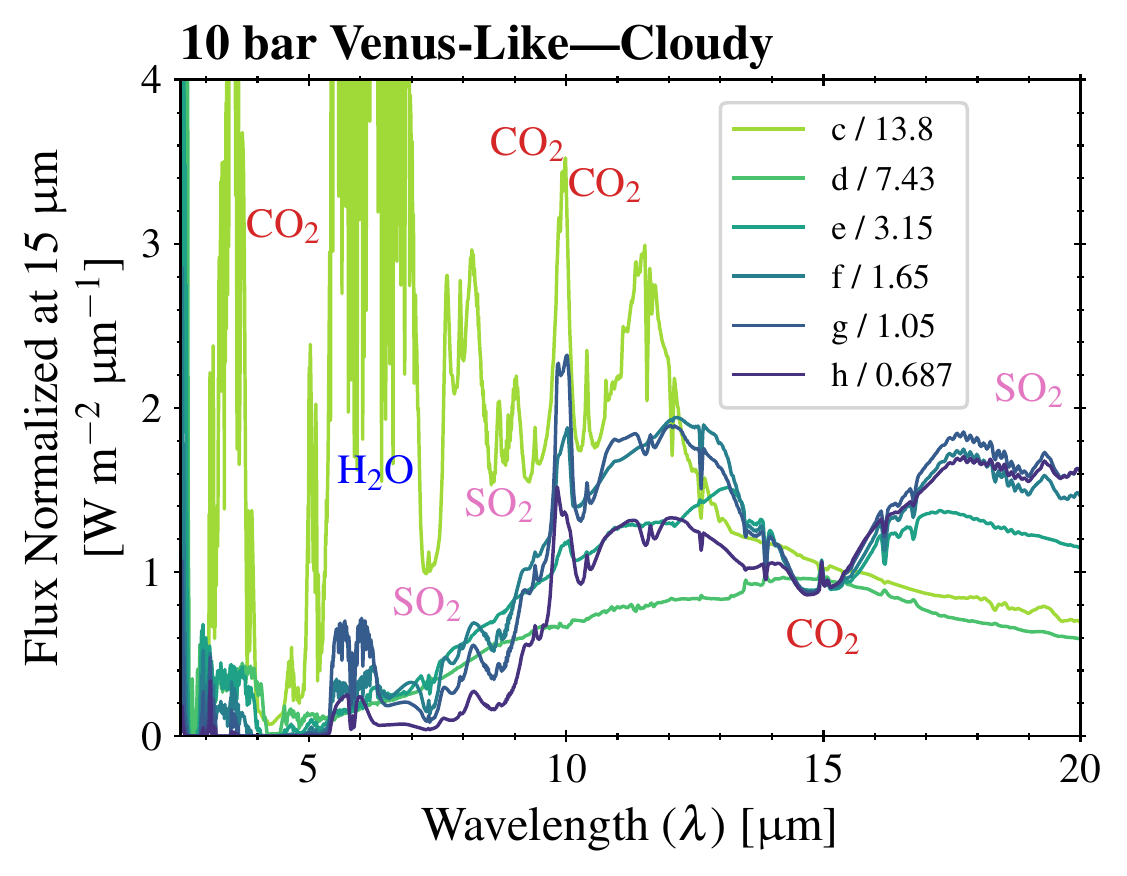}
  \caption{Normalized emission (outgoing flux) spectra for the desiccated atmospheres. 
  The legends provide the factors by which the spectra were reduced to normalize the flux to 15~\um{} for each. The emission spectra are dominated by \ce{CO2} features in both absorption and emission, but also include \ce{H2O}, \ce{O3}, and \ce{SO2}.
  \label{fig:emit-evol}}
\end{figure*}

\subsubsection{Emission Spectra}

\begin{figure}
  \centering
  \iftwocol
      \includegraphics[width = \columnwidth]{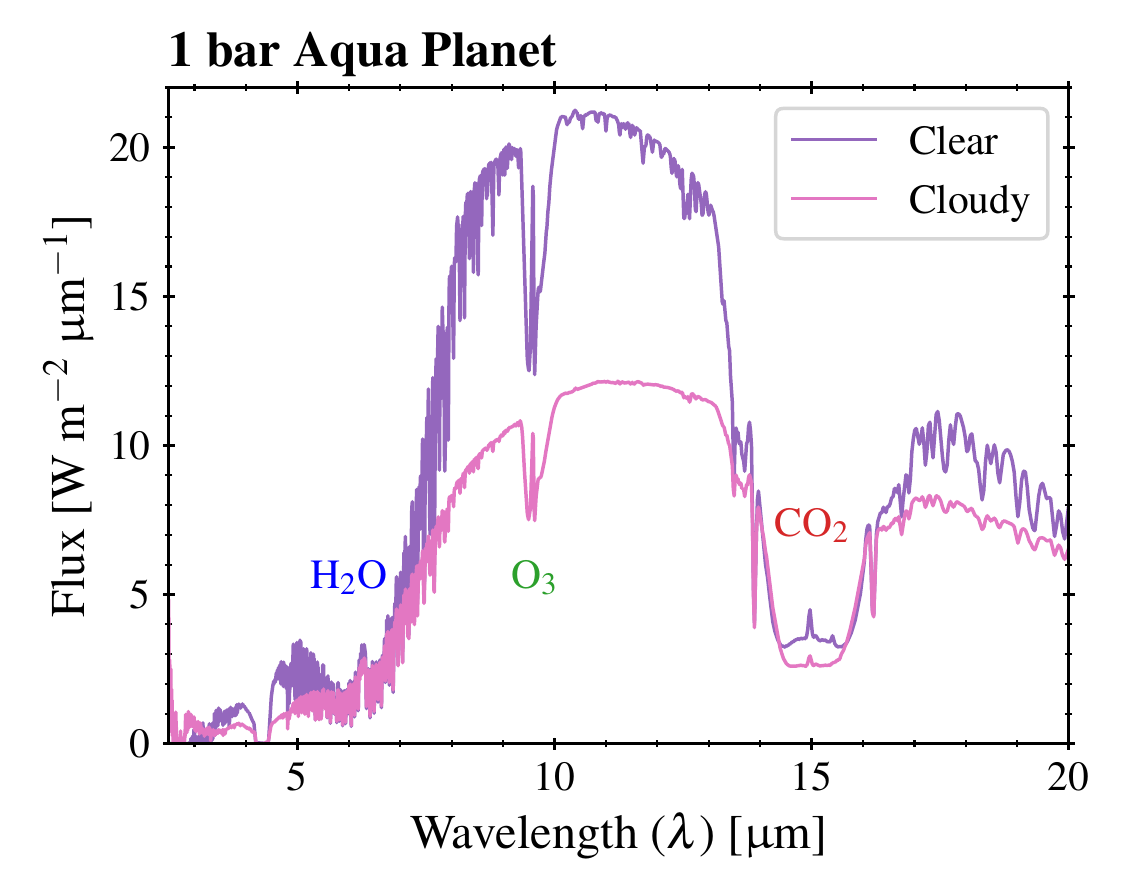}
\else
    \includegraphics[width = 0.5\textwidth]{h2o_1bar_emit_overlap_small}  
\fi
  \caption{Outgoing flux spectra for clear sky and cloudy, 1~bar TRAPPIST-1 e aqua planets. These closely resemble Earth's emission spectrum \citep[c.f.][]{Robinson:2011}, characterized by the 6.3~\um{} water band and weak water features throughout, and the 15~\um{} \ce{CO2} band. The weak water features mostly disappear in the cloudy case, and the overall emission spectrum is reduced in magnitude by half. \label{fig:emit-H2O}}
\end{figure}

We present thermal emission spectra as normalized flux (Figures~\ref{fig:emit-evol} and \ref{fig:emit-H2O}).
Below $\sim5$~\um{}, stellar flux reflected from the surface or scattered in the atmosphere contributes significantly to the spectra.

\paragraph{\ce{O2}-Dominated Atmospheres, Desiccated}

The desiccated, \ce{O2}-dominated atmospheres (Figure~\ref{fig:emit-evol})
are distinctive in emission spectra, due to lower temperatures, lack of water bands, and their unusual temperature structures. The stratospheric temperature peak causes some bands to be seen in emission rather than absorption, depending on how cool the particular planet is. In the colder atmospheres, the \ce{O3} band at 9.6~\um{} exhibits varying hot wing emission at 10.4~\um{} from the stratospheric temperature peak. In the hotter atmospheres with lower \ce{O3} abundance, part of the 9.4~\um{} \ce{CO2} ``hot band'' also emerges in emission at 9.0~\um{}. \ce{CO2} exhibits varying changes between absorption and emission at $\sim$10--18~\um{}. As a potential insight into isotope fractionation for \ce{^18O/^16O}, the fundamental asymmetric stretch $v_1 = 1 \rightarrow 0$ transition of the \ce{^16O^12C^18O} isotopologue at 7.3 and 7.9~\um{} \citep{HITRAN:2012} appear in emission. The \ce{O2} band at 6.4~\um{} exhibits strong emission, but it is very narrow, and present only because of the complete lack of water vapor. Carbon monoxide is not observable. 

\paragraph{\ce{O2}-Dominated Atmospheres, Outgassing}

The emission spectra of the \ce{O2}-dominated, outgassing atmospheres are given in Figure~\ref{fig:emit-evol}. The radiative-convective processes and resultant emission spectra are largely dominated by water vapor in the atmosphere. The warmer planets, with larger quantities of water vapor throughout the atmospheric column, exhibit massive \ce{H2O} absorption in the 6.3~\um{} band. Conversely, the colder, drier planets exhibit \ce{SO2} absorption at 7.3, 8.8, and 19~\um{}. These outgassing atmospheres have much less ozone than the desiccated ones, due to destruction catalyzed by the presence of water vapor, and so their \ce{O3} features are much weaker, particularly for the warmer planets.

\paragraph{Venus-like Atmospheres}

The Venus-like atmospheres (Figure~\ref{fig:emit-evol})
emit very little flux, except through narrow windows between strong absorption bands, similar to Venus. This suppression of emitted flux is due to the combined Venus greenhouse of primarily \ce{CO2}, \ce{H2O}, and \ce{SO2}. The emission windows in the hotter atmospheres are at 2.4, 3.5, and 5.5--7.0~\um{}. The cooler, clear-sky planets emit some flux around 10~\um, between \ce{CO2} hot bands. Water vapor absorbs weakly in the 5.5--7.0~\um{} window and \ce{SO2} absorbs from 7--9~\um{} in the clear-sky case, but is obscured in the cloudy case. The remaining emission is reduced by \ce{CO2} absorption. The cloudy spectra are exhibit more shallow absorption features and are more similar to a blackbody spectrum because the thermal emission is primarily from the cloud deck, though the colder planets also emit above the cloud deck from the 15~\um{} \ce{CO2} band.

\paragraph{Aqua Planets}

The potentially habitable aqua planet atmospheres (Figure~\ref{fig:emit-H2O}) 
have emission spectra very similar to Earth \citep[c.f.][]{Robinson:2011}. The near-surface absorption from the water vapor continuum is apparent in the clear-sky case, particularly between 10--13~\um{}, and the cloudy case emits from its global cloud deck at a colder temperature. Both cases exhibit absorption from \ce{H2O} (6.3~\um{}), \ce{O3} (9.6~\um{}), and \ce{CO2} (15~\um{}).


\section{Discussion} \label{sec:discussion}

\begin{figure*} 
  \centering
  \iftwocol
    \includegraphics[width = \textwidth]{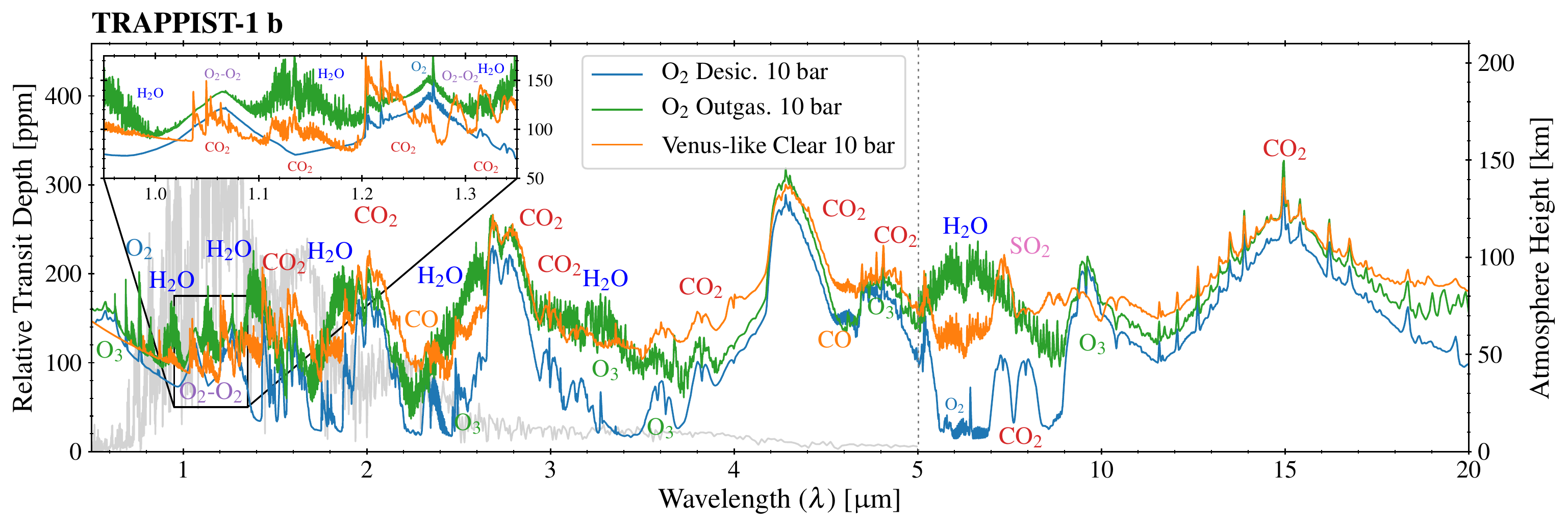}
    \includegraphics[width = \textwidth]{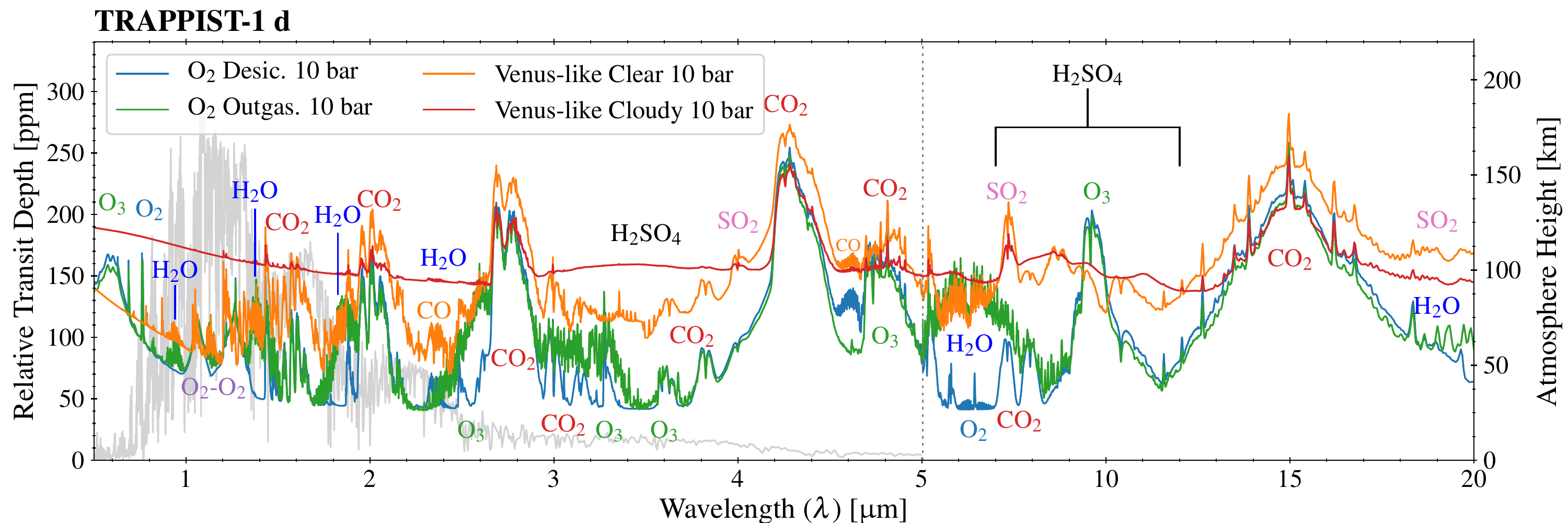}
    \includegraphics[width = \textwidth]{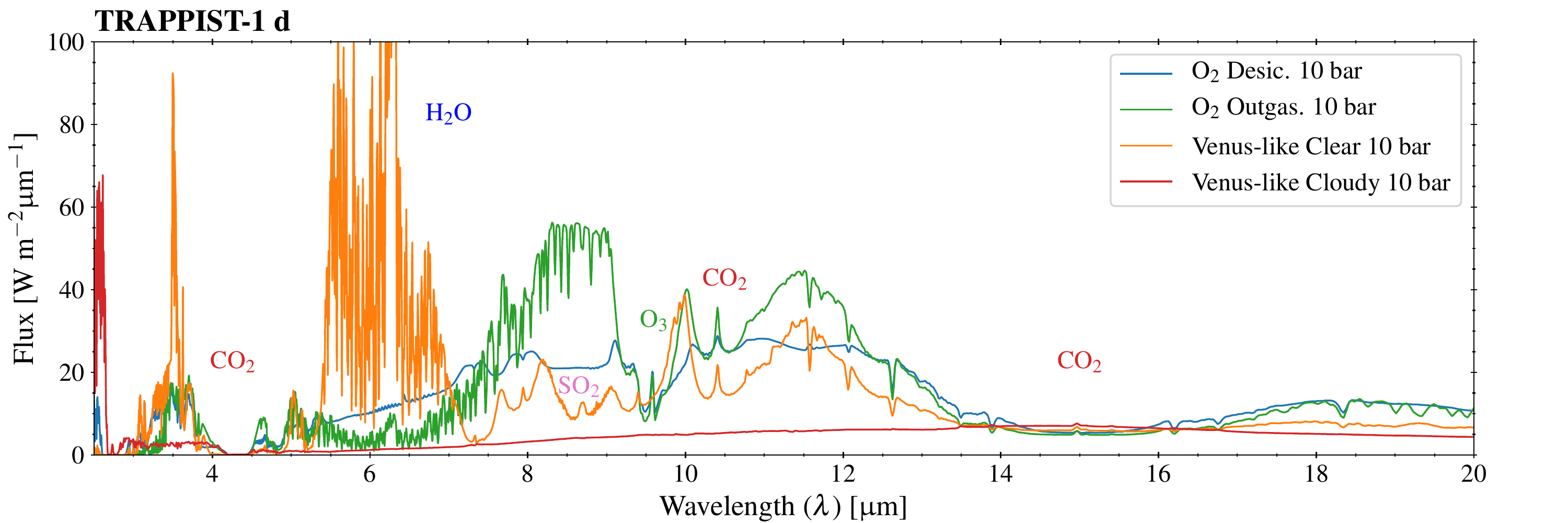}
  \else   
    \includegraphics[width = 0.66\textwidth]{trappist-1_b_trans_venus_inset}
    \includegraphics[width = 0.66\textwidth]{trappist-1_d_trans_venus}
    \includegraphics[width = 0.66\textwidth]{trappist-1_d_emit}
  \fi
  
   \caption{TRAPPIST-1~b (\textit{top panel}) and d (\textit{center panel}) transit transmission spectra, and TRAPPIST-1~d thermal emission spectra (\textit{bottom panel}) of the 10~bar atmospheres, for all environments simulated. The higher mass environments exhibit similar spectral features, but different strengths and temperatures. We include the stellar spectrum (grey) in the transmission plots to illustrate the spectral regions with the most available photons for backlighting the atmosphere in transmission. As shown in the previous figures, \ce{CO2} dominates all of these environments. Absorption by \ce{O2}-\ce{O2}, \ce{O3}, \ce{SO2}, CO, \ce{H2O}, and weaker \ce{CO2} bands can distinguish these environmental states in both transmission and emission. Note the inset for b that shows the overlapping \ce{CO2} and \ce{O2}-\ce{O2} bands that could be confused at low resolution. The \ce{H2O} feature of the clear sky Venus-like atmosphere in emission peaks at $\sim$420~W~m$^2$~\um{}$^{-1}$. TRAPPIST-1~b is scheduled to be observed by JWST and has the strongest features, due to lower gravity and hotter temperatures. Planet d also has strong signal in transmission, even with high-altitude sulfuric acid aerosols. TRAPPIST-1~b, c, and d exhibit strong features in emission due to high temperatures---d is shown here in emission to demonstrate the large difference due to aerosols. \label{fig:t-1_b} }

\end{figure*}

\begin{figure*}
  \centering
  \includegraphics[width = \textwidth]{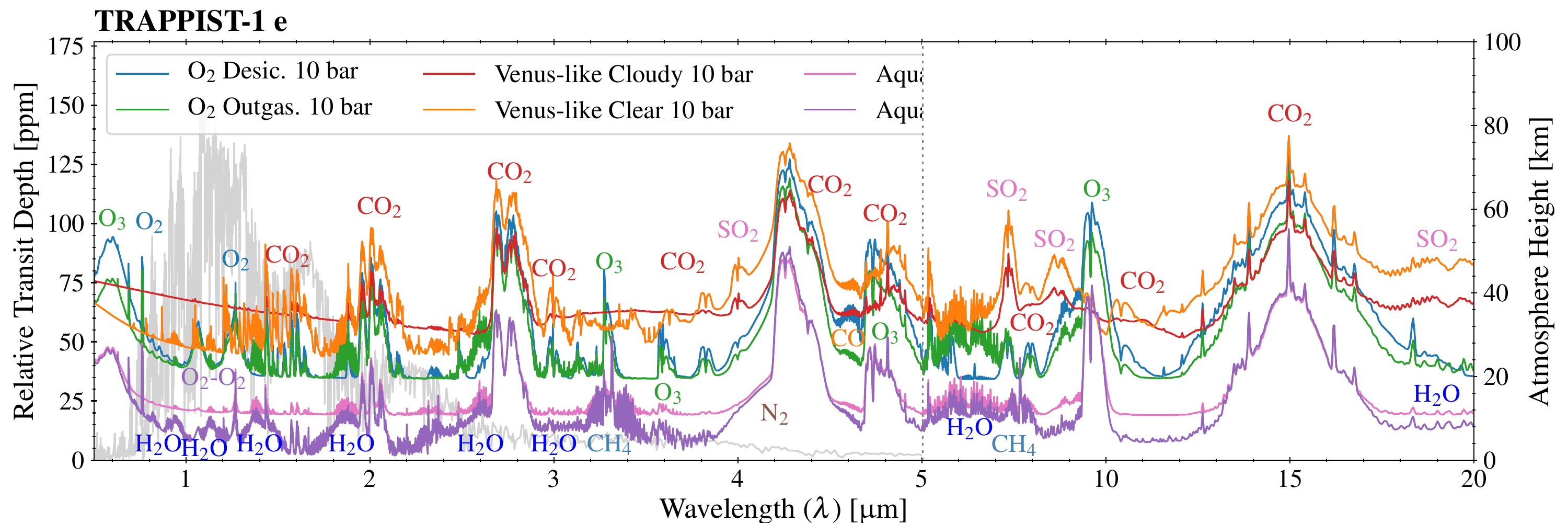}
  \includegraphics[width = \textwidth]{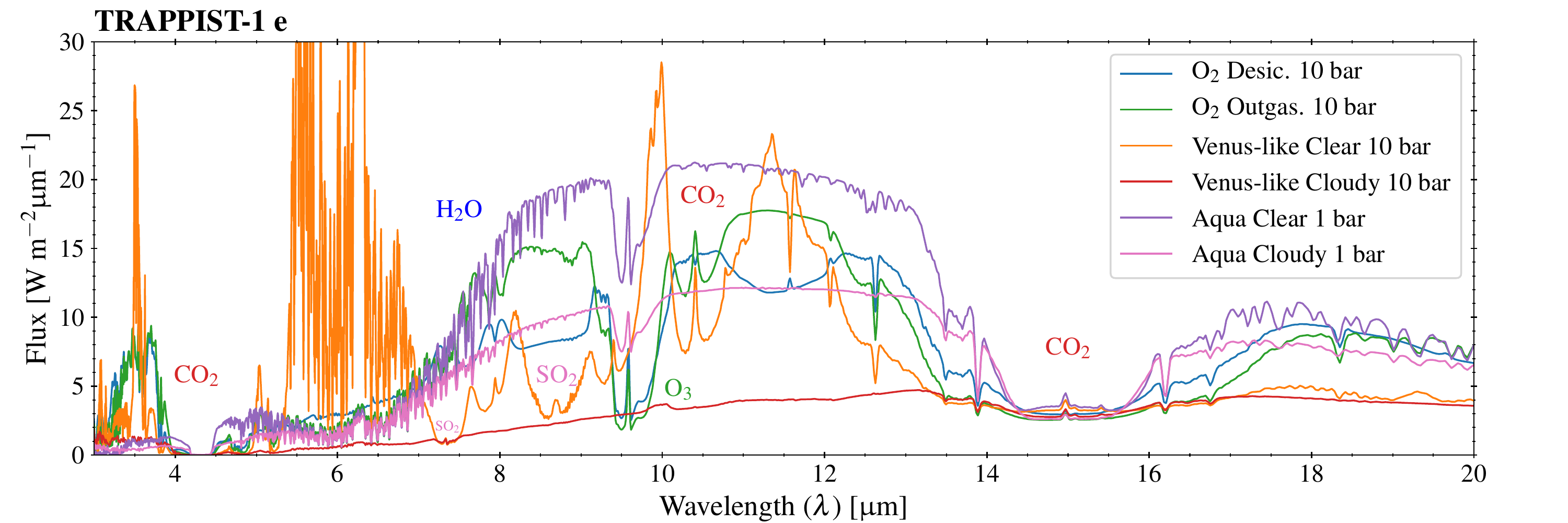}
  \caption{TRAPPIST-1 e transit transmission (\textit{top panel}) and thermal emission (\textit{bottom panel}) spectra of the 10~bar atmospheres for the four simulated evolved environments, and a self-consistent pre-industrial Earth. The higher mass environments exhibit similar spectral features, but different strengths and temperatures. We include the stellar spectrum (grey) in the transmission plot to illustrate the spectral regions with the most available photons for backlighting the atmosphere in transmission. As shown in the previous figures, \ce{CO2} dominates all of these environments. Absorption by \ce{O2}-\ce{O2}, \ce{O3}, \ce{H2O}, and weaker \ce{CO2} bands can distinguish these environmental states in both transmission and emission. The \ce{H2O} feature of the clear sky Venus-like atmosphere in emission peaks at $\sim$60~W~m$^2$~\um{}$^{-1}$. \label{fig:t-1_e}} 
\end{figure*}

We have used a new, versatile, coupled climate-photo-chemical model to generate environmental states for the TRAPPIST-1 planets for different assumed evolved atmospheres, including desiccated and water-rich.  These atmospheric states are climatically and photochemically self-consistent with the spectral energy distribution from the parent star, and show a diversity of characteristics as a function of bulk composition, orbital distance, and outgassing rates. These characteristics are very different from those expected for \ce{H2}-dominated atmospheres.  We have shown that the non-Earth-like atmospheric compositions that may result from M~dwarf stellar evolution strongly impact planetary climate and surface temperature, potentially reducing habitability within the HZ. Hot temperatures modeled beyond the outer edge require additional study to determine if the habitable zone could be extended with additional trace gases. These different environments produce spectral features that can be observed by upcoming observatories, including JWST, to discriminate among the environments. Here we discuss important implications of this work for M~dwarf terrestrial planets in the areas of atmospheric evolution and escape, climate and photochemistry, observational discriminants, and prospects for future observations of these environments.

\subsection{Atmospheric Escape, Ocean Loss, and Oxygen Accumulation}

We conducted a study of \ce{H2O} loss and \ce{O2} accumulation for the TRAPPIST-1 system using an update to the model of \citet{Luger:2015b}, and found that ocean loss affects the entire HZ.  Even planets that are beyond the outer edge during the main sequence phase (e.g.~TRAPPIST-1~h) have the potential to lose a significant ocean fraction and generate \ce{O2}.
These results are broadly consistent with a study on ultra-cool dwarf stars conducted by \citet{Bolmont:2017}. Note that \citeauthor{Bolmont:2017} based their TRAPPIST-1 work on earlier system parameters \citep{Gillon:2016}, which included only planets b, c, and a range of semi-major axes for d. Where we can compare, we predict up to an order of magnitude larger amounts of H loss and \ce{O2} accumulation, likely due to our much older stellar age (5~Gyr, \citealt{Burgasser:2017}, vs 850~Myr, \citealt{Bolmont:2017}), and assumptions of the planetary radii and masses, which we based on additional observations \citep{Grimm:2018}. 

Our model and calculation of the escape efficiency $\epsilon_\text{xuv}$ are consistent with the value inferred for low-mass planets in the \textit{Kepler} population, considering their sizes and densities as a function of irradiation \citep{Owen:2013}. Another metric, the ``cosmic shoreline,'' supports our use of thermally-driven hydrodynamic escape. The ``cosmic shoreline'' is an empirically-derived relationship between planetary irradiation and escape velocity that divides planets with atmospheres from airless rocks, and appears to apply broadly to the Solar System and the population of exoplanets with known densities \citep{Zahnle:2017}.

A thorough discussion of caveats for assessing H escape and \ce{O2} accumulation was given by \citet{Bolmont:2017}. We do not include the effects of \ce{O2} loss processes other than thermal escape, such as surface sinks \citep{Schaefer:2016,Wordsworth:2018} and non-thermal atmospheric loss \citep{Collinson:2016,Airapetian:2017}, which are important and warrant further detailed study. At the location of Proxima Centauri b for an Earth twin, the calculated non-thermal escape rate may not currently exceed Earth-like outgassing replenishment rates, which may allow a secondary atmosphere to accumulate \citep{Garcia:2017}. However, non-thermal escape could have contributed to net \ce{O2} loss during the pre-main-sequence phase. Therefore, the \ce{O2} generated by our evolution models represents maximum possible inventories, using nominal parameters for stellar and planetary properties. We reduced the \ce{O2} abundance for our climate/photochemistry and spectral models in consideration of these loss mechanisms, which we did not include in our models.  Given the importance of oxygen as a biosignature, the generation of abiotic \ce{O2}-rich, post-ocean-loss atmospheres should be considered when interpreting future observations of M~dwarf terrestrial planets.

\subsection{Climate and Photochemistry in the TRAPPIST-1 Planetary System}

These diverse environments produce a wide range of habitable and uninhabitable surface temperatures (Table~\ref{table:results}), suggesting that the region around a star where Venus-like planets may be common evolutionary outcomes \citep{Kane:2014} may extend into and beyond the classical habitable zone for late-type M~dwarfs.  Even though some of these modeled environments do not support surface liquid water, if dense \ce{CO2} atmospheres do form on terrestrial planets after an early runaway greenhouse phase, our results  suggest that surface temperatures in the habitable range or hotter may occur beyond the classical maximum greenhouse limit of the HZ outer edge (where warming from additional \ce{CO2} is offset by higher Rayleigh scattering in a temperate \ce{H2O}-\ce{CO2} greenhouse; \citealt{Kasting:1993}). Even TRAPPIST-1~h, beyond the HZ outer edge ($S_\text{h}=0.148S_\oplus$), could have a surface temperature as hot as 465~K if early atmospheric evolution resulted in a clear-sky Venus-like atmosphere, and 398~K with the expected sulfuric acid clouds. These high temperatures may be stable over long time periods, because known sinks for \ce{CO2}---such as deposition to an ocean and carbonate-silicate weathering \citep{Walker:1981}---will not be available, although a surface carbonate buffer may form and keep the atmosphere at an equilibrium \ce{CO2} value that is less than 92 bars \citep{Hashimoto:1997}. Table~\ref{table:results} also shows that across the diversity of atmospheres considered, the instellation at the modeled TRAPPIST-1~e's distance from the star was most likely to maintain a habitable surface temperature. This suggests that the habitable zone concept is still useful as an initial means of narrowing down the search for habitable conditions on terrestrial planets, even when planetary atmospheres are not truly Earth-like. 

For late-type M~dwarf stars in particular, our work indicates that the maximum greenhouse limit may not apply, or may apply at a distance much farther from the star than previously calculated \citep{Kopparapu:2013}. We have shown that the \ce{H2O}-\ce{CO2} greenhouse used in traditional habitability calculations can be enhanced significantly by other greenhouse gases, including \ce{SO2}.  While \ce{SO2} is a greenhouse gas in desiccated atmospheres like  Venus' \citep[e.g.][]{Lee:2016}, it is less likely to be present in the atmosphere of a habitable planet with an ocean because it is highly soluble. If an initially-desiccated planet beyond the HZ were to recover habitability via outgassing, and maintain a warm temperature, other gases such as \ce{CH4} (which may be relatively abundant in planetary atmospheres around M~dwarfs; \citealt{Segura:2005,Rugheimer:2015,Meadows:2018}) would be needed to support a strong greenhouse. For a habitable planet, this greenhouse could also be enhanced by collision-induced absorption, such as \ce{CO2}-\ce{CH4} CIA, as may have been the case on early Mars \citep[c.f.][]{Batalha:2015,Wordsworth:2017}. Greenhouse gas abundances can also be affected by energetic particle events, which can drive chemistry that generates trace amounts of greenhouse gases \citep{Airapetian:2016}. These events are likely to remove greenhouse gases, by the destruction of \ce{O3} in Earth-like atmospheres by repeated flaring \citep{Tilley:2018}, or by the destruction of \ce{CH4}, which may form hydrocarbon hazes if oxygen radicals are not present \citep{Arney:2017}. In the absence of sufficient greenhouse gases, the coldest atmospheres (particularly the 10~bar cloudy Venus-like h) may be subject to collapse, but most will be resistant, as the condensation temperature for \ce{O2} is very low and day-night temperature contrasts are lower in thicker atmospheres \citep{Turbet:2018}.

Our results indicate that the interplay between stellar irradiation, photochemistry, chemical kinetics, and condensation shaped our simulated  planetary atmospheric compositions, with water vapor, ozone, sulfur species, and aerosols displaying the greatest variation. The decreases in \ce{H2O} (largely due to temperature) and increases in \ce{O3} abundance (due to an interplay between photolysis rates, water abundance, and vertical transport) as a function of orbital distance are consistent with trends found by previous studies \citep[e.g.][]{Grenfell:2007}. The water vapor profile in the warmer atmospheres is representative of planets in a moist greenhouse state, with substantial water vapor well-mixed throughout the atmosphere, and with no cold trap.  Consequently these atmospheres may suffer severe water loss over Gyrs, unless the escape rate is balanced by outgassing.   Unless they begin with warm, Venus-like atmospheres, the outer planets are too cold to recover from freezing temperatures without a brightening star or larger inventories of greenhouse gases, which may eventually accumulate from outgassing or impactor delivery.

Ozone and CO both exhibit characteristic patterns across our modeled atmospheres. In those containing $\gtrsim0.2$~bar \ce{O2}, at $\sim$200~Pa, \ce{O3} is about 10$^{-5}$~mol/mol, in all the simulated atmospheres, which is likely due to different levels of irradiation, \ce{H2O}, and \ce{O2}. Above this layer, the \ce{O3} abundance is controlled by photolysis, while below, mixing and other temperature- and water-dependent loss mechanisms dominate. In desiccated atmospheres, CO abundance is controlled by the slow, three-body \ce{CO2} recombination reaction \citep{Gao:2015}, while in atmospheres with water vapor, the abundance of CO is controlled by the availability of H-bearing recombination catalysts, which are more prevalent near the planetary surface, removing the CO at lower altitudes. Large amounts of CO in the presence of \ce{O2}/\ce{O3} is a false positive discriminant \citep{Schwieterman:2016}, indicating that oxygen is more likely to have been generated by photolysis of \ce{CO2} \citep{Gao:2015} rather than photosynthesis.

Aerosols affect the planetary climates by reflecting incoming stellar radiation and either absorbing, emitting, or scattering thermal radiation.
Sulfuric acid aerosols can reduce the surface temperatures of hot, partly desiccated planets, which caused the cloudy Venus-like atmospheres we modeled to drop by up to nearly 200~K (Table~\ref{table:results}) compared to the clear-sky cases. Sulfuric acid production also varied as a function of irradiation. The warmer planets more easily formed \ce{H2SO4} due to higher water vapor abundances and more rapid \ce{SO2} photolysis. This production was offset at lower altitudes by thermal decomposition. TRAPPIST-1~b did not condense sulfuric acid in our model because atmospheric temperatures were too high. If any of the TRAPPIST-1 planets have Venus-like atmospheres, we expect that b will not form sulfuric acid clouds, but the other planets likely would.  Although we did not model the radiative effects of aerosols for the \ce{O2} outgassing environments, sulfuric acid condensation occurred in the photochemical model, so their climates could also be affected by these clouds. Water and water-ice clouds impacted the temperature structure of the aqua planet, but had little effect on surface temperature (Figure~\ref{fig:PT-H2O}). This was due to competing effects as stellar radiation interacted with the cloud properties: the upper tropospheric cirrus clouds effectively scatter stellar radiation, but the absorption properties of water and water-ice in the NIR also make them effective absorbers of M~dwarf peak stellar radiation, especially given the spectral energy distribution of TRAPPIST-1. Exoplanet aerosols for terrestrial planets are a rich field for future study. 

The parameter space for possible planetary characteristics is large, so the assumptions regarding atmospheric surface pressure, surface material, atmospheric bulk composition, geological or biological surface fluxes, and atmospheric transport can have a large impact on the equilibrium atmospheric state. The differences between our \ce{O2} desiccated and outgassing atmospheres demonstrate the effects of surface fluxes on the planetary environment. These factors are important when considering what atmospheres may be possible on a given exoplanet, and for comparing possible planetary states and among different models. \citet{Wolf:2017} found that only TRAPPIST-1~e requires more \ce{CO2} than Earth to sustain temperate conditions, but that work underestimated the reduction in bolometric albedo due to the late M~dwarf SED \citep{Turbet:2018}. It is possible that \citet{Turbet:2018} also overestimated the surface albedos of ocean-dominated TRAPPIST-1 planets, as they calculated a mean water-ice albedo of 0.21. We used wavelength-dependent surface albedos in our climate calculations that depend on the planetary environment (Figure~\ref{fig:albedo}). We found that both pure ocean and melting ice surfaces had very low stellar flux-weighted average albedos (0.02 and 0.03 respectively), while a typical snow surface averaged only 0.16 around TRAPPIST-1. \citet{Wolf:2017} found that 0.1~bar of \ce{CO2} with 1~bar \ce{N2} is required for a surface temperature of $\sim270$~K, while modern Earth \ce{CO2} levels of $\sim400$~ppm yield 240~K. \citet{Turbet:2018} calculated 250~K under similar conditions, while we find that with or without clouds, only pre-industrial Earth \ce{CO2} abundance of 280~ppm is required in a 1~bar \ce{N2}/\ce{O2} atmosphere on an ocean-covered planet to reach an Earth-like globally-averaged surface temperature. It is crucial to note that surface albedo is one of the primary drivers of climate state \citep{Godolt:2016}, particularly in an atmosphere transparent to its host star's radiation.

\subsection{Observational Discriminants of Evolved Terrestrial Atmospheres}

\edit1{
\begin{table*}[]
\centering
\caption{Observational Discriminants
}
\label{table:disc}
{\iftwocol
    \small\selectfont
\else
    \scriptsize\selectfont
\fi
\begin{tabularx}{\textwidth}{p{0.75in}p{3.5in}p{2.50in}}
\hline
\textbf{Planetary State} & \textbf{Molecules \& Wavelengths [\um{}]}  & \textbf{Notes}   \\ \hline
Aqua planet, clear-sky    & \ce{CO2} (1.6, 2.0, 2.8, 3.6, 3.8, 4.3, 4.8, {9.5, 10.5}, 15), \ce{H2O} (0.94, 1.4, 1.9, 2.6, 3, 3.3, 6.3), HDO (3.7), \ce{O2} (0.67, 0.76, 1.27), \ce{O2}-\ce{O2} (1.06, 1.27), \ce{O3} (0.6, 4.75, 9.6), \ce{CH4} (2.35, 3.3, 7.6), {\ce{N2}-\ce{N2} (4, 4.6)}
& Earth-like worlds may be discriminated by \ce{CH4} in conjunction with oxidized species.            
\\
Aqua planet, cloudy       
& \ce{CO2} (1.6, 2.0, 2.8, {3.6, 3.8}, 4.3, 4.8, {9.5, 10.5}, 15), \ce{H2O} (1.4, 1.9, 2.6, 3, 3.3, 6.3), {HDO (3.7)}, \ce{O2} (0.67, 0.76, 1.27), \ce{O2}-\ce{O2} (1.06, 1.27), \ce{O3} (0.6, 4.75, 9.6), {\ce{CH4} (2.35, 3.3, 7.6)}, {\ce{N2}-\ce{N2} (4, 4.6)}
& Tropospheric water/water-ice clouds are problematic for observing Earth-like planets.               
\\
\ce{O2}, desiccated          
& \ce{CO2} (1.6, 2.0, 2.8, 3.0, 3.2, {3.6, 3.8, 4.0}, 4.3, 4.8, 7.3, 7.9, 15), {\ce{O2}} (0.67, 0.76, {1.27, 6.4}), \ce{O2}-\ce{O2} (1.06, 1.27), {CO (2.35, 4.6)},  {\ce{O3}} (0.6, {2.5, 3.3, 3.6}, 4.75, {9.0}, 9.6, {13})
& A lower limit of \ce{O2} abundance can be constrained with the \ce{O2}-\ce{O2} bands at 1.06 and 1.27~\um{}. CO and \ce{O3} levels can indicate desiccation. Complete lack of water vapor.
\\
\ce{O2}, outgassing             
& \ce{CO2} (1.6, 2.0, 2.8, 3.0, 3.2, {3.6, 3.8, 4.0}, 4.3, 4.8, 15), \ce{H2O} (1.4, 1.9, 2.6, 3, 3.3, 6.3), HDO (3.7), {\ce{O2}} (0.67, 0.76, {1.27},{ 6.4}), \ce{O2}-\ce{O2} (1.06, 1.27), \ce{O3} (0.6, {3.3, 3.6}, 4.75, 9.6, 13), {\ce{SO2} (4.0, 7.3, 8.8, 19)}
& Water vapor is prominant in the hotter atmospheres.                 
\\
Venus-like, clear-sky    
& {\ce{CO2}} ({1.05, 1.3}, 1.6, 2.0, 2.8, 3.0, 3.2, 3.6, 3.8, 4.0, 4.3, 4.8, 7.3, 7.9, 9.5, 10.5, 15), \ce{H2O} (1.4, 1.8, 2.6, 6.3), CO (2.35, 4.6), \ce{SO2} (4.0, 7.3, 8.8, 19) 
& High \ce{CO2} abundance may be constrained by the weakest bands (e.g. 1.05 and 1.3~\um{}). 
\\
Venus-like, cloudy       
&  \ce{CO2} (1.5, 2.0, 2.8, 3.0, 3.2, 3.6, 3.8, 4.0, 4.3, 4.8, 7.3, 7.9, 9.5, 10.5, 15), \ce{H2O} (6.3), \ce{SO2} (4.0, 7.3, 8.8, 19), {\ce{H2SO4} (3.1--4, 7.5--10)}
& Sulfuric acid cloud absorption may be the best discriminant for outgassed conditions with no surface water.
\\ 
\hline
\end{tabularx}}
  \flushleft
    \small\selectfont
\textbf{Note:} Spectral features present in transmission or emission spectra are listed for our modeled environments.
\end{table*}
}

Here we compare spectra derived from the complete set of simulated atmospheres for TRAPPIST-1~b, c, and d, whose higher temperatures make them more easily observable, and e, which is a potentially habitable candidate, to identify observational characteristics that could be used to discriminate between these environments and their evolutionary histories. Spectral features that can be used to discriminate between the modeled environments are given in Table~\ref{table:disc}. Figure~\ref{fig:t-1_b} shows spectra for the atmospheres we modeled for TRAPPIST-1~b, c, and d. The stronger carbon dioxide bands are prominent in all of our planetary spectra, regardless of whether it is a bulk constituent or a trace gas,
and these \ce{CO2} bands are therefore a good indicator of the presence of a terrestrial atmosphere, but are not as useful for discriminating among different environments.  However, if the weaker \ce{CO2} bands are also present, this does indicate a higher, more Venus-like abundance. As seen in planet d's transmission spectra in Figure~\ref{fig:t-1_b}, cloudy Venus-like atmospheres are well-discriminated by sulfuric acid absorption longward of 3~\um{} (though these clouds could also form in \ce{O2}-dominated atmospheres with outgassing).

The massive \ce{O2} atmospheres may reveal themselves in transmission via the presence of \ce{O2}-\ce{O2} collision-induced absorption at 1.06 and 1.27~\um{} \citep{Schwieterman:2016}, a wavelength region strongly backlit by the host star.  Note that our photochemical model suggests that cloud formation in the outgassing atmosphere is likely, due to sulfuric acid condensation (at $\sim$1~ppb concentration). However, this will not affect the visibility of the \ce{O2}-\ce{O2} features since the aerosol scattering properties and geometry of the TRAPPIST-1 system permit transmission shortward of 2.5~\um{} for the innermost planets. The favorably compact geometry of the TRAPPIST-1 system allows transit transmission to probe relatively deep into the atmosphere, where \ce{O2}-\ce{O2} absorption is strongest, producing transit signals as high as 70~ppm \citep[vs $\sim$3~ppm for an early-type M~dwarf; ][]{Schwieterman:2016}.
However, as demonstrated in the inset in Figure~\ref{fig:t-1_b}, the broad \ce{O2}-\ce{O2} bands at 1.06 and 1.27~\um{} overlap with \ce{CO2} bands at 1.05, 1.13, 1.24, 1.29, and 1.32~\um{} in high \ce{CO2} atmospheres.
An atmosphere with high levels of both \ce{CO2} and \ce{O2} may be common if a planet that experienced complete ocean loss retained a large fraction of the \ce{CO2} that was dissolved in the oceans or subsequently outgassed from the interior.
These weak \ce{CO2} features could confuse retrieval and interpretation of the \ce{O2}-\ce{O2} features without sufficient resolution, signal-to-noise, or corroboration from stronger \ce{CO2} features elsewhere in the spectrum.

These \ce{O2}-dominated spectra also show strong features from  \ce{O3} and \ce{SO2}.  A strong ozone Chappuis band around $\sim$0.6~\um{} may indicate desiccation, as it traces ozone deep in the atmosphere, where water vapor would normally destroy it  \citep{Meadows:2018}.  
As clearly shown in Figure~\ref{fig:t-1_b} in all three panels, the 9.6~\um{} \ce{O3} feature is strong and broad for both \ce{O2}-rich atmospheres, but not present in the Venus-like spectra primarily due to the lack of \ce{O2} in the atmosphere. Detection of \ce{SO2} absorption in either transmission or emission would be indicative of an outgassing interior, and possibly an atmosphere and surface with low water abundance, either due to early desiccation, or temperatures below freezing. The \ce{O2} outgassing case is unique among the environments considered for the hotter planets in showing very strong water features, especially near 1.4 and 6.3~\um{}. This is because these planets have high levels of water vapor (up to $0.1\%$~mol/mol) in their stratospheres, so are in a moist greenhouse state, approaching a runaway greenhouse. These strong \ce{H2O} bands are indicative of a planet in the process of losing water vapor, rather than indicating habitable, ocean-bearing conditions. 

The bottom panel of Figure~\ref{fig:t-1_b} demonstrates the differences between the evolved environments in emission using TRAPPIST-1~c as an example of the hotter planets. The water band at 6.3~\um{} is the primary discriminant. At high temperatures and high abundance (e.g. the \ce{O2} outgassing case), \ce{H2O} absorption can dominate over absorption from other interesting molecules between 5--9~\um{}, such as \ce{O2}, \ce{SO2}, and \ce{CO2}. For the Venus-like clear-sky atmosphere, the low abundance of \ce{H2O} $\lesssim1$~ppm fails to suppress the outgoing radiation from the hot lower atmosphere,  resulting in an emission peak at these wavelengths.  The cloudy Venus case shows that clouds can adversely affect both outgoing flux and the strength of molecular features by truncating the atmosphere at higher, colder altitudes.

Figure~\ref{fig:t-1_e} compares the spectra of the \ce{O2}- and \ce{CO2}-dominated atmospheres with the clear and cloudy aqua planets for TRAPPIST-1~e.  The relative transit depths are naturally offset due to the truncating effects of clouds, which allow transmission to probe only altitudes above the cloud deck, and the obscuring effect of atmospheric opacity for atmospheres of different total atmospheric pressure (i.e.~1 vs 10~bar).  In the thermal infrared, clouds reduce the overall outgoing flux by emitting at colder temperatures. The features and their diagnostics are similar for the TRAPPIST-1~e water-poor atmospheres compared to d, with \ce{CO2} features dominating the spectra, prominent (but narrower) \ce{O3} features for the \ce{O2}-rich atmospheres, and a more prominent difference in the \ce{O3} Chappuis band strengths between the outgassing and more desiccated case.The \ce{O2} desiccated and Venus-like atmospheres for TRAPPIST-1~e display features and corresponding observational discriminants similar to the hotter planets. However, the \ce{O2} outgassing environment was not hot enough to maintain high levels of stratospheric \ce{H2O}, so exhibits a more Earth-like, modestly absorbing 6.3~\um{} feature. The overall emission spectrum of that environment is very similar to the aqua planets, but with additional \ce{CO2} bands due to the higher levels of \ce{CO2} we assumed.
 
Water is not a good discriminant among some of these atmospheres, so discriminating the habitable aqua planet from the uninhabitable environments is not straightforward. The aqua planet has water vapor features in the near-infrared in transmission at 0.94, 1.15, 1.4, 1.9, 2.6, 3.7, and most strongly at 6.3~\um{}.
Unfortunately, these water features have comparable signals among the clear-sky atmospheres (except the completely desiccated case), and among the cloudy cases, despite vastly different levels of tropospheric \ce{H2O}, due to the cold trapping of water in the aqua planet atmosphere.

The bulk compositition of these atmospheres can potentially be distinguished by weak spectral effects of their dominant gases \ce{O2}, \ce{CO2}, and \ce{N2}. The Earth-like aqua planet atmosphere exhibits absorption from the collision-induced \ce{N2}-\ce{N2} band at 4.1~\um{}, which is sensitive to the partial pressure, and could be diagnostic of atmospheric pressure if this challenging observation could be made \citep{Schwieterman:2015b}. At high abundances of \ce{CO2} (as low as 0.5~bar here), the wings of the 4.3~\um{} \ce{CO2} band overpower the \ce{N2}-\ce{N2} CIA. The presence of the weak NIR--MIR \ce{CO2} bands (including the numerous bands shortward of 2~\um{}, the \ce{^16O^12C^18O} isotopologue absorption bands at 3.6, 3.8, 4.0, 7.3, and 7.9~\um{}, and hot bands at 9.5 and 10.5~\um{}) are indicative of a warm, \ce{CO2}-rich atmosphere. As in Earth's atmosphere, \ce{O2}-\ce{O2} CIA is weakly present in the aqua planet atmospheres, but the bands are considerably stronger at higher \ce{O2} abundance, which would suggest an abiotic, post-ocean-loss source for the oxygen was more likely.

Potentially habitable environments may be constrained by indicators of desiccation, such as \ce{O3}, \ce{SO2}, and CO, as these gases react with water vapor and its photolytic byproducts. Both the modeled \ce{O2}-dominated and aqua planet atmospheres exhibit ozone absorption at 0.6 and 9.6~\um{}. However, a strong indication of desiccation is high ozone abundance, as constrained by the presence of a long wavelength tail of the Chappuis band (out to 1~\um{}), the 2.5 and 3.6~\um{} features, and additional weak bands at 9 and 13~\um{}, which arise in the \ce{O2} cases, particularly the colder and desiccated environments. Retrievals of ozone abundances in conjunction with climate and photochemical modeling could help constrain the likely abundances of tropospheric water vapor. The detection of \ce{SO2} would also suggest a desiccated atmosphere, because \ce{SO2} easily combines with \ce{H2O} to form sulfuric acid. The presence of CO, if not masked by \ce{CO2} or \ce{O3}, would indicate desiccation because \ce{H2O} photolytic byproducts such as hydroxyl are efficient catalysts at recombining \ce{CO2} from CO and free oxygen. While the absence of water vapor absorption does not necessarily indicate the lack of tropospheric water vapor or surface water, desiccation indicators from \ce{O3}, \ce{SO2}, and CO, even with weak water vapor features, would suggest a desiccated and likely uninhabitable environement.

Signs of atmospheric escape and ocean loss may be probed by observing isotopologue bands HDO and \ce{^18O^12C^16O}. Measuring the D/H ratio in \ce{H2O} and the \ce{^18O/^16O} ratio in \ce{CO2} could be used to probe isotopic fractionation, which occurs during atmospheric escape and ocean loss because lighter elements are preferentially lost over heavy elements \citep[e.g.][]{Hunten:1982}. D/H fractionation from atmospheric loss has been measured for both Venus \citep[D/H$\sim$120 vs Earth;][]{deBergh:1991} and Mars \citep[D/H$\sim$4 vs Earth;][]{Encrenaz:2018}. In particular, such D/H fractionation would indicate an environment that had the majority of its ocean reservoir depleted \citep{Hunten:1982}. Future work should assess oxygen fractionation due to massive ocean loss in an environment with some retained atmospheric oxygen.

\subsection{Planetary System Effects on Terrestrial Atmospheres and Spectra}

\begin{figure*}
    \centering
    \includegraphics[width=\textwidth]{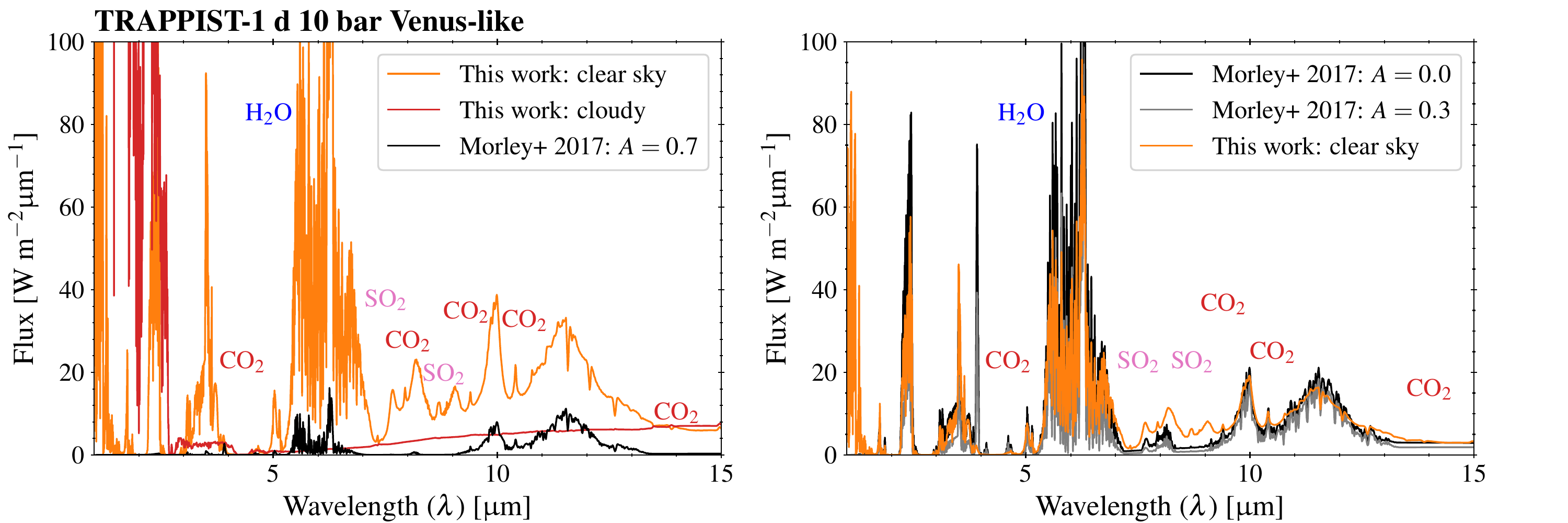}
    \includegraphics[width=\textwidth]{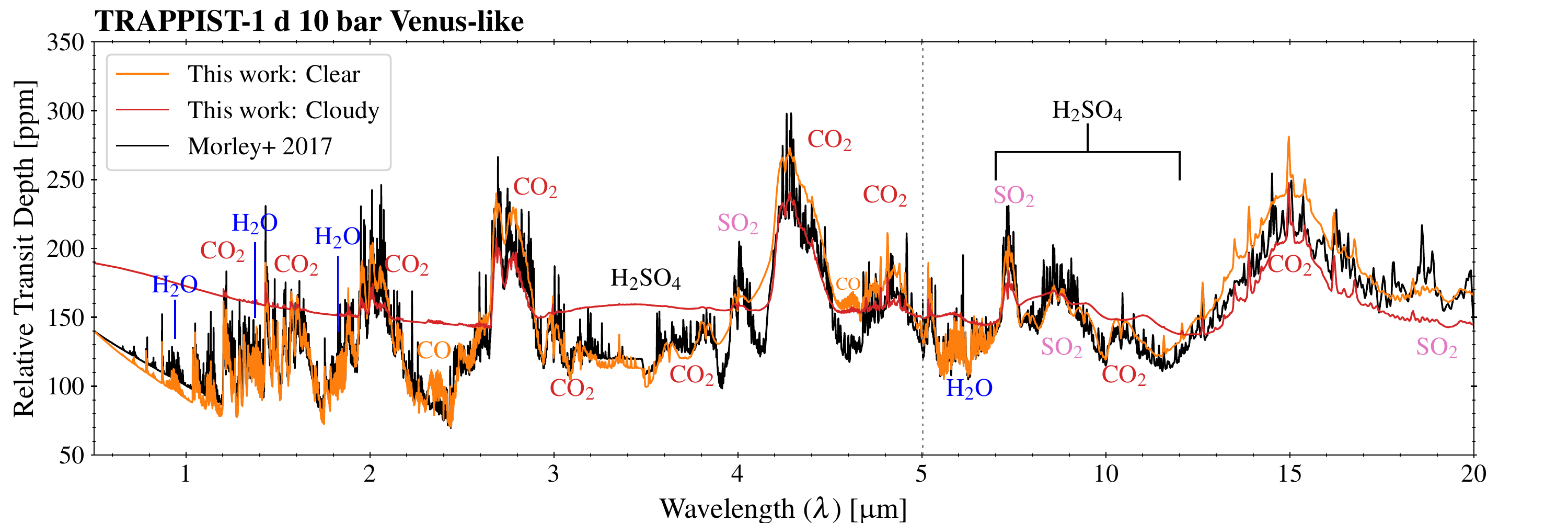}
    \caption{Comparison spectra of the 10~bar, TRAPPIST-1~d Venus analogs. To directly compare, \citet{Morley:2017} spectra were sampled at 1~cm$^{-1}$ and convolved with a 1~cm$^{-1}$ half-width at half-max slit function. \textit{Upper left panel}: our clear-sky and cloudy Venus spectra compared to the high-albedo ($A=0.7$) thermochemical equilibrium Venus of \citet{Morley:2017}. The radiative effects of sulfuric acid aerosols are much more complex than their effect on planetary albedo.
    The planet effectively emits from the cloud deck, which is not captured in models that assume a higher surface albedo to account for clouds. \textit{Upper right panel}: our clear-sky Venus is qualitatively similar to the lower albedo ($A=0.0-0.3$) of \citet{Morley:2017}.
    \textit{Lower panel}: Transmission spectra showing relative transit depth. Clouds substantially impairing the ability to observe gas absorption features. The strong \ce{SO2} features in the \citeauthor{Morley:2017} spectra are likely due to uniform vertical gas profiles and colder emission temperatures. 
    }
    \label{fig:morley_venus}
\end{figure*}

\begin{figure*}
    \centering
    \includegraphics[width=\textwidth]{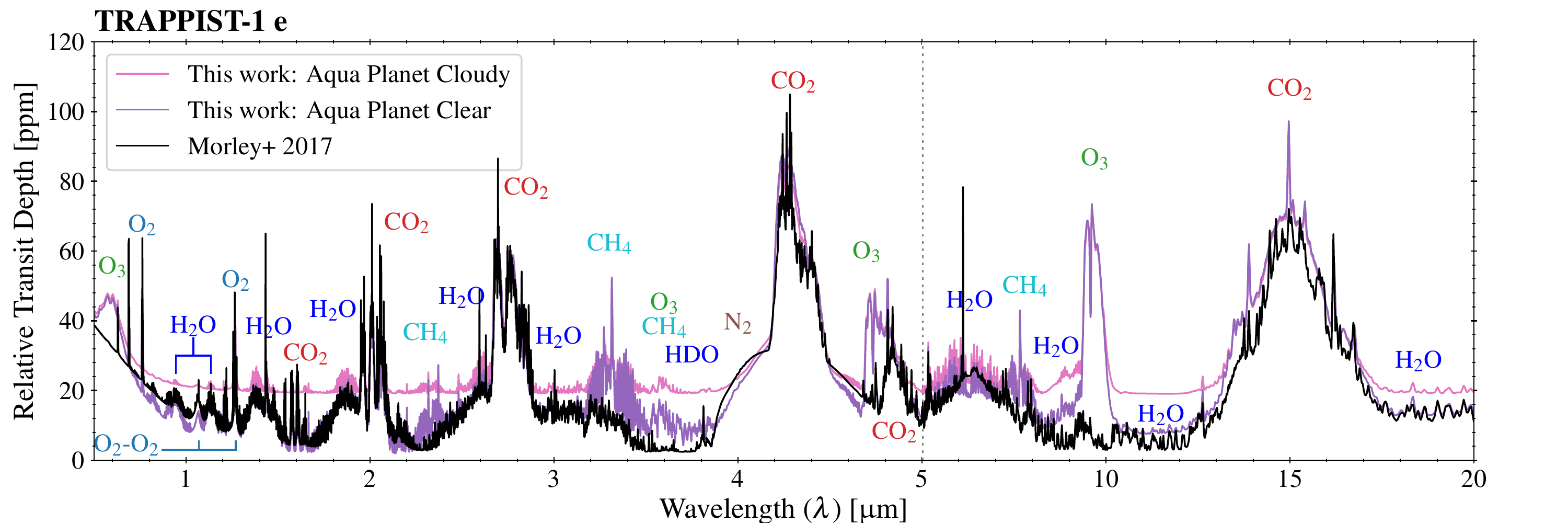}
    \caption{Transmission spectra of the 1~bar, TRAPPIST-1~e Earth analogs: our aqua planet with Earth geological outgassing compared to the thermochemical equilibrium Earth of \citet{Morley:2017} (sampled at 1~cm$^{-1}$ and convolved with a 1~cm$^{-1}$ half-width at half-max slit function, scaled to 30\%, and offset for this plot). The  \citeauthor{Morley:2017} atmospheres do not have \ce{O3}, an abundant stratospheric photochemical by-product of \ce{O2}, or \ce{CH4}, which together are the most distinctive observational features of Earth \citep{Selsis:2002,Schwieterman:2017}.
    }
    \label{fig:morley_earth}
\end{figure*}

Our results and those of others demonstrate that the atmospheres and surfaces of M~dwarf terrestrials have likely been strongly shaped by several factors, including stellar evolution, atmosphere and ocean loss, photochemistry, and geological (and biological) surface fluxes \citep{Segura:2005,Grenfell:2014,Rugheimer:2015,Meadows:2018,Meadows:2018b}. These interconnected processes will impact atmospheric composition, climate, and the resultant spectra of these environments.  To best interpret current and upcoming constraints and data on terrestrial exoplanets, the most significant of these processes need to be included in models of the planetary environments.  The stellar spectral energy distribution and activity affect atmospheric loss and photochemistry, and bulk planetary properties (e.g. composition, size) affect outgassing species and rates, which all affect the composition of a terrestrial planet atmosphere, as we have demonstrated here.  Atmospheric composition, including greenhouse gases and UV absorbers, in turn affect planetary climate, including the atmospheric temperature profile and surface temperature.  This in turn can modify vertical distribution of constituents (e.g. cold trapping water vapor below a stratosphere), which affects features visible in transmission spectra.  The thermal temperature structure (e.g. tropospheric lapse rates, thermal inversions and greenhouse warming) also affect the strength and shape of features seen in the thermal infrared. For example, in our desiccated, \ce{O2}-dominated atmospheres, the \ce{O3}- and \ce{O2}-\ce{O2}-induced stratospheres caused thermal emission in the stratosphere from \ce{CO2} (Figure \ref{fig:emit-evol}). In particular, interior outgassing and photochemistry are key processes for modeling terrestrial exoplanets, as both drive disequilibria in the atmosphere. Our results show a strong effect from outgassing (e.g. from \ce{SO2} and \ce{H2O}), which together with their photochemical byproducts and reactions with other species, shape the temperature inversions and lapse rates as well as produce radiative greenhouse (and anti-greenhouse) effects. Although our environments do not explicitly contain biogenic fluxes, surface fluxes from life processes also drive disequilibria and this is one of the principal means that will be used to search for life on exoplanets \citep{Simoncini:2013,Krissansen:2016,Meadows:2017a}.  Photochemistry is critically important in assessing planetary composition and in understanding the formation and destruction of key species, some of which may be outgassed and/or biogenic.  As we have shown, this is especially true for planets at different orbital distances, including those orbiting M~dwarfs, where the alien UV spectrum can drive chemistry that the Sun does not induce in the Earth's atmosphere. The species outgassed or generated from photochemistry may be directly observed, and their vertical distribution may affect other gas absorption features due to their effects on climate.

To illustrate the effects of modeling factors---including outgassing and photochemistry---on predicted spectra,
we directly compare a sampling of our spectra for TRAPPIST-1~d and e with \citet{Morley:2017}, who used radiative transfer and thermochemical equilibrium models to conduct a broad assessment of spectroscopic observables for the TRAPPIST-1 planets. For planet d, we compare our radiative-convective-photochemical-equilibrium 10~bar clear-sky and cloudy Venus-like atmospheres with the Venus-like atmospheres from \citeauthor{Morley:2017}, identified by their albedos ($A$ = 0.0, 0.3, and 0.7), which correspond to clear-sky to global cloud coverage (Figure~\ref{fig:morley_venus}). The spectra of clear-sky Venus-like atmospheres modeled in both approaches are similar in both transmission and emission, except for stronger \ce{SO2} features from the thermochemical model, which are reduced in our model (and Venus' true atmosphere) by \ce{H2SO4} formation. The \ce{SO2} in the thermochemical model suppresses outgoing flux at wavelengths between 7--10~\um{} when compared to both our clear and cloudy atmospheres.  Although the integrated flux of the $A=0.0$ atmosphere is very close to our clear-sky Venus, the distribution of flux is different. Our absorption features are not as deep (e.g. 7--9~\um{}), which is likely due to our atmospheres having warmer stratospheres.    
Our atmospheres also show higher emission near 6 microns as weaker water vapor absorption permits escape of radiation from the hot lower atmosphere.  This is likely due to different assumptions about the water vapor profile between the two models. 

Our spectra differ considerably from \citet{Morley:2017} for cloudy Venus-like cases. Photochemical models produce results tailored to the spectrum of the host star, whereas thermochemistry is driven purely by the atmospheric temperature profile at a given orbital position, and so is more generic.
The cloudy Venus differs significantly from the clear-sky case in both transmission and emission due to the truncation of the atmospheric column by cloud scattering and absorption at higher, colder altitudes, which reduces the outgoing thermal flux and can raise the tangent height when observed in transmission. Although the effects of clouds on surface temperature can be approximated by changing the Bond albedo in the thermochemical models, this process does not take into account the radiative effect of the clouds on the temperature profile, or on transmission and emission spectra. Sulfuric acid scattering and absorption in the cloudy Venus-like transmission spectrum, and emission from the cloud deck in direct imaging or secondary eclipse, are distinctive features of a Venus-like planet not captured in models without vertical cloud distributions.

The Earth-like atmospheres also show significant differences in composition between the two modeling approaches, primarily because we include the effects of water vapor clouds, photochemically-produced gases such as \ce{O3}, and outgassed species such as \ce{CH4}.  These characteristic features of Earth's spectrum are not seen in the spectra generated by \citet{Morley:2017} using a thermochemical equilibrium model. Our photochemically-consistent clear-sky aqua planet atmosphere also produces significantly stronger water vapor absorption bands (likely due to the inclusion of isotopologue (HDO) absorption at 3.6 and 9~\um{}), although the cloudy case is perhaps the more realistic model.

\subsection{Assessing Current and Future Observations}

\begin{figure} 
  \centering
  \iftwocol
    \includegraphics[width = \columnwidth]{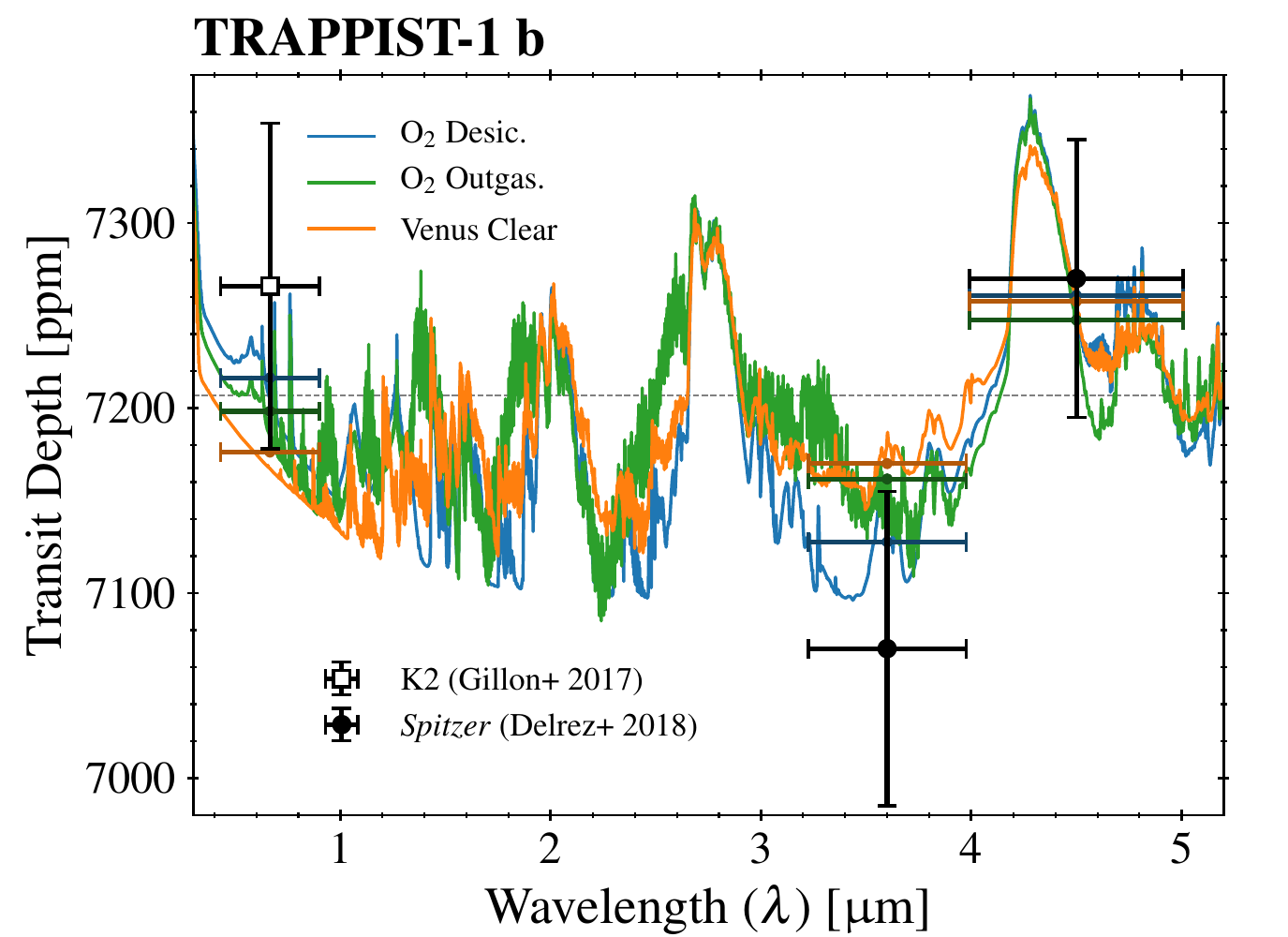}
  \else
    \includegraphics[width = 0.5\textwidth]{photometry.pdf}
  \fi
  \caption{Comparison of available photometric transit data from K2 and \textit{Spitzer} \citep{Gillon:2017,Delrez:2018} with our 10~bar model spectra for TRAPPIST-1~b. The colored points correspond to our model spectra convolved with the appropriate photometric filters. A vertical offset is applied to each model transmission spectrum so that the models optimally overlap the observations. These offsets correspond to the difference between our assumed radius for TRAPPIST-1~b and the solid body radius assuming a model atmosphere and its associated absorbing radius above the surface. The offsets are -293~km for the desiccated \ce{O2}, -308~km for \ce{O2} with outgassing, and -312~km for the clear sky Venus.
  The desiccated, \ce{O2}-dominated atmospheres fits the \textit{Spitzer} data within 1$\sigma$, while the other two atmospheres fit within 2$\sigma$.  
  \label{fig:t-1_b_obs}}  
\end{figure}

Several preliminary observational constraints on atmospheric composition have been obtained of planets in the TRAPPIST-1 system, to which we can compare our modeling results. Spectra taken with the \textit{Hubble Space Telescope} \citep{DeWit:2016,DeWit:2018} ruled out low molecular weight atmospheres, although these spectra were also consistent with flat spectra within the error bars of $\sim200-350$~ppm. For TRAPPIST-1~b, \citet{Delrez:2018} observed a signal difference of $208\pm110$~ppm between the 3.6 and 4.5~\um{} \textit{Spitzer} bands, which they suggest is due to \ce{CO2} absorption. These results, and the K2 photometry from the \citet{Gillon:2017} discovery paper, are shown with error bars in Figure~\ref{fig:t-1_b_obs}, along with the band-integrated transit depths for our modeled spectra. We have adjusted the assumed solid body radius in these spectra to fit the observations for the solid body and wavelength-dependent atmospheric absorbing radius. Our simulated spectra are consistent with this data within $2\sigma$ error, and the \ce{O2}-dominated atmospheres fit within $1\sigma$.  The  3.6~\um{} filter bandpass contains absorption by the \ce{CO2} isotopologue bands at 3.6, 3.8, and 4.0~\um{}, plus the wings of the main isotopologue 4.3~\um{} band, whereas the 4.5~\um{} band contains the 4.3 and 4.9~\um{} \ce{CO2}, 4.6~\um{} CO, and 4.75~\um{} \ce{O3} bands. Rayleigh scattering and ozone absorption may contribute to the larger transit depth measured by K2. A different distribution of UV flux than assumed in this work could enhance ozone levels, further increasing the transit depth in the K2 band and \textit{Spitzer} 4.5~\um{} band.

\begin{figure*} 
  \centering
  \includegraphics[width = \textwidth]{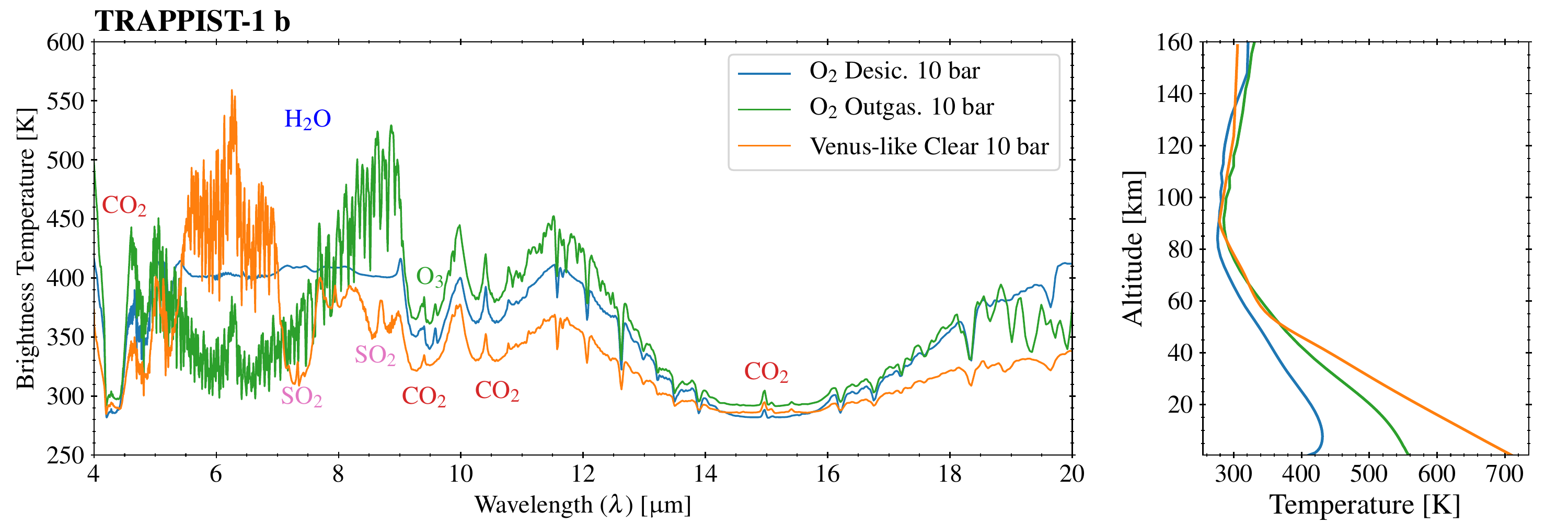}
  \caption{Brightness temperature of emergent flux for all four environments modeled for TRAPPIST-1~b (10~bar). The atmospheric structures are plotted on the right for comparison. Brightness temperature represents emergent flux plotted in units of temperature, which demonstrates the emitting layer temperature as a function of wavelength. As shown by comparing to the temperature profiles, higher brightness temperatures are emitted from lower layers, while lower temperatures represent emission from higher, cooler layers.   \label{fig:t-1_b_bright}}
\end{figure*}

\begin{figure} 
  \centering
  \iftwocol
    \includegraphics[width = \columnwidth]{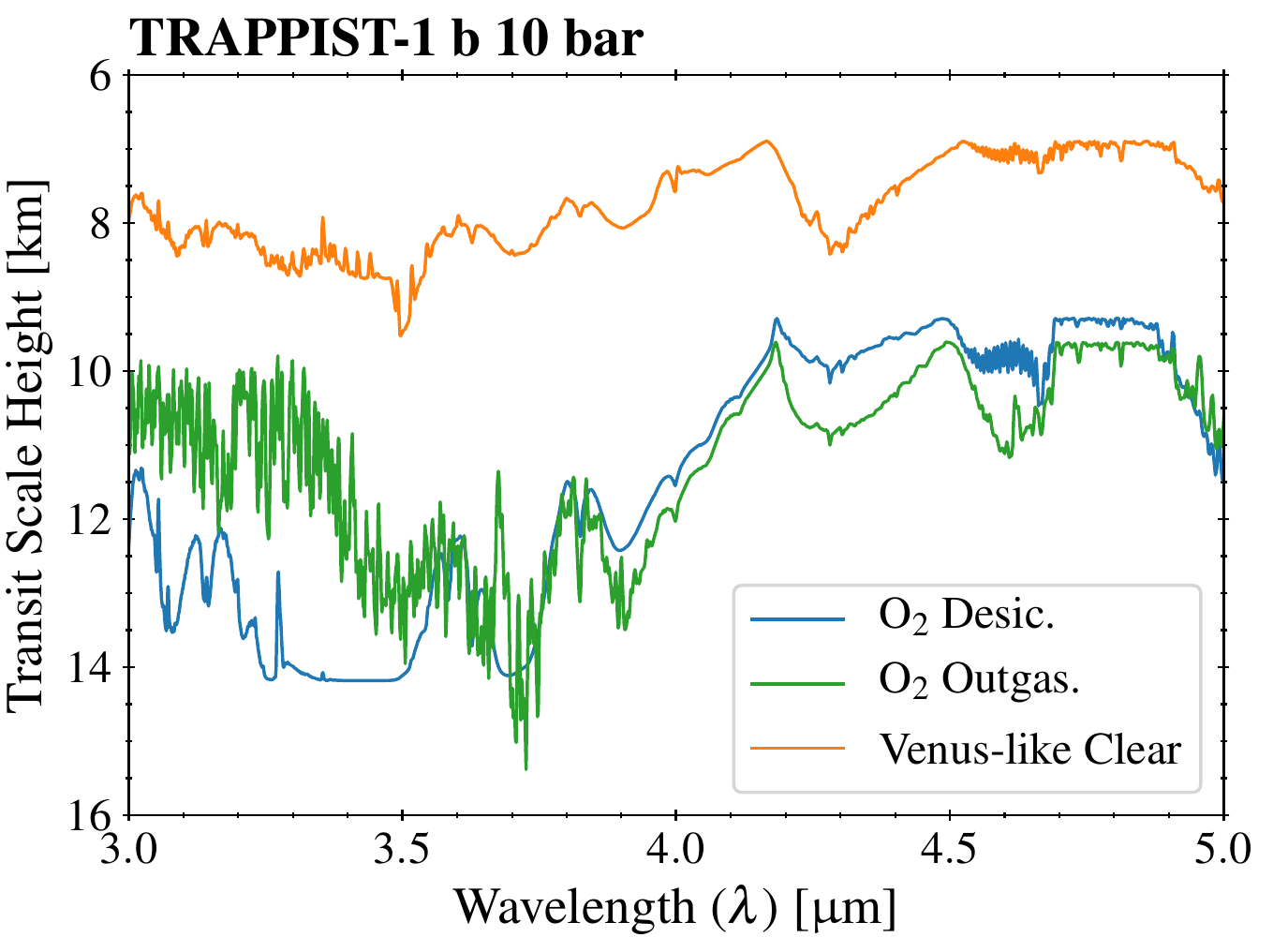}
  \else
    \includegraphics[width = 0.5\textwidth]{t-1b_scalehgts.pdf}
  \fi
  \caption{Atmospheric scale heights at the effective transit height in the wavelength range relevant for \textit{Spitzer} for modeled TRAPPIST-1~b atmospheres. The calculated scale heights are consistent with Solar System terrestrials (7--16~km). Transmission spectra sample the stratosphere, where the atmospheres are cooler, so the scale heights are smaller. The \ce{O2}-dominated atmospheres are more transparent than the Venus-like atmospheres, so their atmosphere are probed deeper into the hotter regions.   \label{fig:t-1_b_shgt}}
\end{figure}

Our simulated atmospheres are consistent with the observed data, but do not agree with the atmospheric scale height and temperature derived by \citet{Delrez:2018} for TRAPPIST-1~b.
\citeauthor{Delrez:2018} assumed the 208~ppm transit signal between the observed \textit{Spitzer} bands was due to two scale heights from the \ce{CO2} signal in an Earth-like atmosphere---with mean molecular mass 28~g~mol$^{-1}$, \ce{CO2} as the only absorbing gas, and excluding \ce{CH4} and \ce{H2O}, which have transitions in the 3.6~\um{} band.
However, their derived temperature of 1400~K and scale height of 52~km (1800~K and 100~km with water vapor included) are not consistent with the assumption of an Earth-like atmosphere.
Our self-consistent modeled environments, which include a case with a greenhouse comparable to Venus', have emission temperatures which vary with wavelength (325--575~K; Figure~\ref{fig:t-1_b_bright}), but overall stay close to the equilibrium temperature ($\sim$400~K, \citealt{Gillon:2017}, even though they have surface temperatures spanning 406--714~K.

In Figure~\ref{fig:t-1_b_shgt}, we show that the scale heights at the effective transit altitude for our modeled atmospheres vary considerably as a function of wavelength and are within values typical of Solar System terrestrials (i.e. 8--16~km for Venus and Earth; \citealt{Meadows:1996}), spanning $\sim$9--11 scale heights (3--5 scale heights when convolved with the \textit{Spitzer} bands), and are consistent with the \textit{Spitzer} observations. Scale height inferences from the size of potential features are degenerate with temperature, species, abundances, aerosols, and opacities, and so are difficult to accurately assess without properly accounting for a range of atmospheric possibilities within a retrieval framework \citep[e.g.][]{Line:2013}.
 
If the existence of an exoplanetary M~dwarf terrestrial atmosphere is confirmed, near-term observations will provide constraints on the environmental effects of M~dwarf stellar evolution on orbiting planets as a function of distance from the star and position in the habitable zone.  Our modeling work provides hypotheses for planetary states as a result of evolution and composition that could be tested by these upcoming observations, which may constrain atmospheric temperatures and species abundances using the discriminants we have presented.  In a subsequent paper, we will use the modeled planetary states and spectra with instrument noise models to predict observational requirements to distinguish the current composition and evolutionary histories of the TRAPPIST-1 planets.

\section{Conclusions} \label{sec:conclusions}

We have calculated the possible ocean loss and oxygen accumulation for the seven known TRAPPIST-1 planets, modeled potential \ce{O2}/\ce{CO2}-dominated and potentially habitable environments, and computed transit transmission and emission spectra.
These evolved terrestrial exoplanet spectra are consistent with broad constraints from recent HST and \textit{Spitzer} data.

Our evolutionary modeling suggests that the current environmental states can include the hypothesized
desiccated, post-ocean-runaway \ce{O2}-dominated planets, with at least partial ocean loss persisting out to TRAPPIST-1 h. These \ce{O2}-dominated atmospheres have unusual temperature structures, with low-altitude stratospheres and no tropospheres, which result in distinctive features in both transmission and emission, including strong collision-induced absorption from \ce{O2}. 

Alternatively, if early volatile outgassing (e.g. \ce{H2O}, \ce{SO2}, \ce{CO2}) occurred, as was the case for Earth and Venus, Venus-like atmospheres are possible, and likely stable, throughout and beyond the habitable zone, so the maximum greenhouse limit may not apply for evolved M dwarf planets. 
If Venus-like, these planets could form sulfuric acid hazes, though we find that TRAPPIST-1~b would be too hot to condense \ce{H2SO4} aerosols.

From analyzing our simulated spectra, we find that there are observational discriminants for the environments we modeled in both transit and emission, with transit signals up to 200~ppm for TRAPPIST-1~b. Detection of \ce{CO2} in all considered compositions may be used to probe for the presence of a terrestrial atmosphere. We find that the detection of water is not a good indicator of a habitable environment, as Venus-like atmospheres exhibit similar spectral features for water, so the detection of low stratospheric water abundance may be a necessary but not sufficient condition for a habitable environment.
The discriminants between these environments involve several trace gases. Careful atmospheric modeling that includes photochemistry and realistic interior outgassing is required to predict the diversity of potentially observable spectral features, to interpret future data, and to infer the underlying physical processes producing the observed features.

Nevertheless, these discriminants may be used to assess the viability of detecting evolutionary outcomes for the TRAPPIST-1 planets with upcoming observatories, particularly JWST, and this will be assessed in subsequent work.
While specifically applied here to the TRAPPIST-1 system, our results may be broadly relevant for other multi-planet M~dwarf systems.

\acknowledgements
This work was performed as part of the NASA Astrobiology Institute's Virtual Planetary Laboratory, supported by the National Aeronautics and Space Administration through the NASA Astrobiology Institute under solicitation NNH12ZDA002C and Cooperative Agreement Number NNA13AA93A. A.P.L. acknowledges support from NASA Headquarters under the NASA Earth and Space Science Fellowship Program -- Grant 80NSSC17K0468. This work was facilitated though the use of advanced computational, storage, and networking infrastructure provided by the Hyak supercomputer system at the University of Washington. This work also benefited from our participation in the NASA Nexus for Exoplanet System Science (NExSS) research coordination network. We thank the anonymous reviewer whose thoughtful comments helped us greatly improve the manuscript.

\software{SMART \citep{Meadows:1996,Crisp:1997}, LBLABC \citep{Meadows:1996}, VPL Climate \citep{Robinson:2018,Meadows:2018}, Atmos Photochem \citep{Kasting:1979,Zahnle:2006,Schwieterman:2016}, ATMESC \citep{Luger:2015a,Luger:2015b}, GNU Parallel \citep{Tange:2011}}

\bibliographystyle{aasjournal} \bibliography{linc}

\appendix
\label{appendix}

\section{VPL Climate Model Description and Updates} \label{app:vplc}

\subsection{Radiative Transfer}
 
VPL Climate is based on the line-by-line, multi-stream, multiple-scattering radiative transfer code, SMART (described in \S\ref{sec:models}). VPL Climate solves the 1D surface-atmosphere thermodynamic energy equation as an initial value problem using timestepping. To step towards equilibrium in a model atmosphere without incurring the computational penalty of a full radiative transfer calculation at every timestep, VPL Climate uses SMART to generate spectrally-resolved Jacobians describing the response of the radiative fluxes to changes in temperature and, where applicable, gas mixing ratios. The Jacobians consist of layer-by-layer, line-by-line stellar and thermal source terms and their derivatives, and derivatives of layer reflectivity, transmissivity, and absorptivity.  These quantities are used in a linear flux-adding approach to determine heating rates, as developed by \citet{Robinson:2018}. 
If, at any timestep, the atmospheric properties are outside the linear range of the Jacobians, timestepping is suspended, and updated fluxes and layer radiative properties and their Jacobians are re-computed using SMART. High-resolution fluxes from SMART are spectrally binned to 10~cm$^{-1}$ (reflected stellar) or 1~cm$^{-1}$ (emitted thermal) intervals to calculate the stellar heating and thermal cooling in each model layer at each timestep. 

Spectrally-dependent molecular absorption cross-sections for rotational-vibrational transitions of gases used by SMART are evaluated using the line-by-line model, LBLABC \citep{Meadows:1996}. LBLABC employs nested spectral grids that fully resolve the narrow cores of individual absorption lines for every atmospheric layer and include the line contributions up to 1000 cm$^{-1}$ from line center. Cross-sections for \ce{CO2} are computed using sub-Lorentzian wings with experimentally-determined $\chi$ correction factors to simulate the effects of quantum-mechanical line mixing. Lines for \ce{H2O} are computed using a line shape model with super-Lorentzian wings to parameterize the far-wing quasi-continuum absorption due to the finite duration of molecule collisions. Opacities for \ce{O2} are computed using super-Lorentzian line wings \citep{Hirono:1982}, with an exponent of 1.958. Line parameters, collisional-induced absorption, and UV--visible cross sections are obtained from a variety of sources (see \S\ref{sec:gas_abs}).

\subsection{Convective Heating}

 \begin{figure*}
  \centering
  \includegraphics[width = \textwidth]{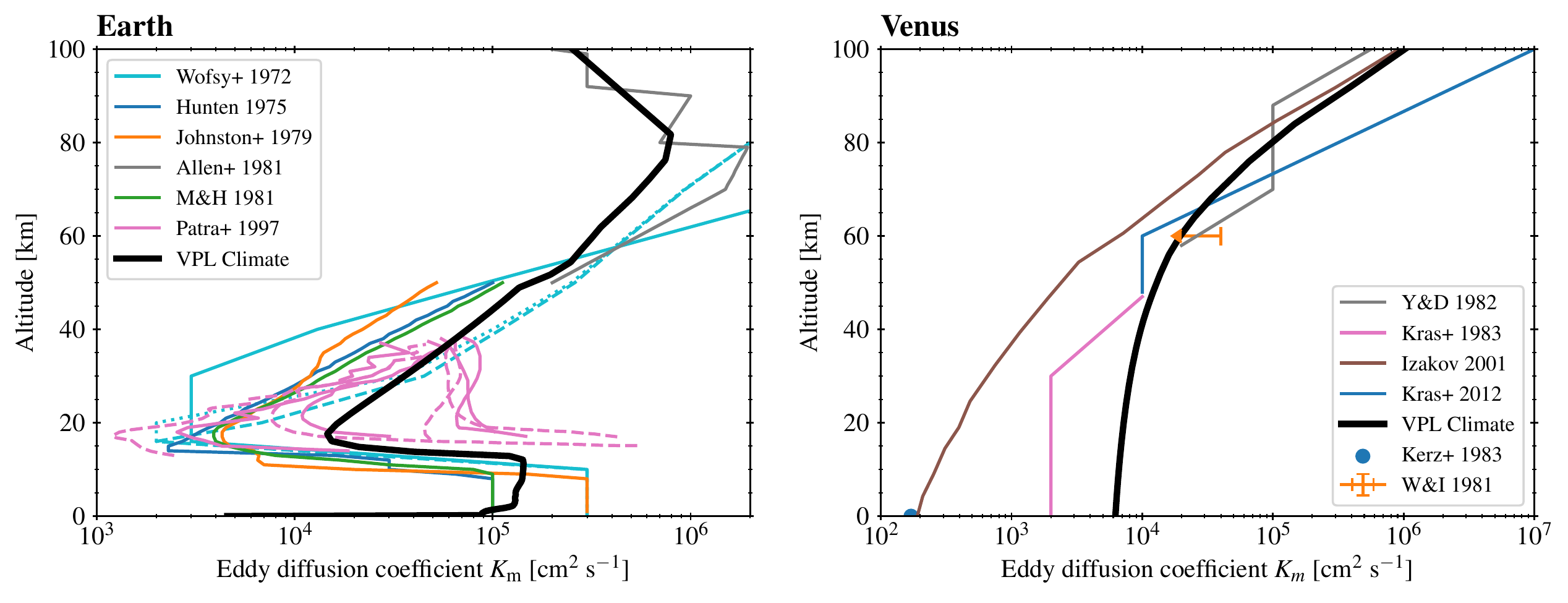}
  \caption{Eddy diffusion profiles for Earth (\textit{left panel}) and Venus (\textit{right panel}). For Earth, model fits to observations are shown from  \citet{Wofsy:1972}, \citet{Hunten:1975}, \citet{Johnston:1979}, \citet{Allen:1981}, \citet{Massie:1981}, and \citet{Patra:1997}. For Venus, we show two points derived from probes of the atmosphere \citep{Woo:1981,Kerzhanovich:1983}, profiles used with chemical kinetics modeling to fit tracer species \citep{Krasnopolsky:1983,Krasnopolsky:2012}, and theoretical modeling \citep{Izakov:2001}. Observations and studies of eddy diffusion rates vary widely.  \label{fig:eddy}}
\end{figure*}

Convective processes transport heat vertically in planetary atmospheres. Convection can result in both sensible heating (related to the energy required to change the temperature of an air parcel due to vertical mixing) and latent heating (related to the energy required to change the phase of a substance, e.g. due to condensation and evaporation). These processes can transport substantial amounts of heat from the surface to the upper troposphere, especially for planets with greenhouse gases, which can prevent the deep atmosphere (or surface) from radiating directly to space. For example, early 1D modeling results showed that convective processes reduce Earth's surface temperature relative to pure radiative equilibrium conditions by $\sim$50~K \citep{Manabe:1964}. Convective processes are also responsible for transport of trace species in the atmosphere, substantially altering atmospheric chemistry \citep[e.g.][]{Fleming:1999,Charnay:2015a}.

In VPL Climate, mixing length theory is used to calculate convective heat fluxes and heating rates,
as this approach is more physically rigorous and versatile than the  convective adjustment typically used in 1D exoplanet models.  Convective adjustment sets the troposphere temperature profile to a moist or dry adiabat defined by Earth's atmospheric conditions \citep[e.g.][]{Kopparapu:2013,Godolt:2016}. In these regimes, convective adjustment does not allow for atmospheric lapse rates that exceed the adiabat, which occurs on planets with more tenuous atmospheres, like Mars \cite[e.g.][]{Hinson:2004}. Convective adjustment also stabilizes atmospheric layers perfectly at every timestep, even if such instantaneous stability is unphysical (e.g.~requires convective motions faster than the speed of sound). Furthermore---and especially problematic for coupling to chemistry models---convective adjustment provides no direct insight into the vertical transport of mass in the atmosphere. In contrast, mixing length theory uses fundamental physical properties of the atmosphere and its constituents (e.g. air parcel temperature, density, specific heat capacities, and static stability) to estimate the vertical heat transport rate, expressed as a vertical ``eddy diffusivity,'' $K$. This quantity is then used to calculate the resulting heat and mass fluxes. Mixing length theory is therefore applicable to a wider variety of atmospheric temperature and composition regimes, and can be used to model convection for a diverse range of exoplanet atmospheres.

The convective heating rate is related to the divergence of the convective energy flux:
\begin{equation} \label{eq:q_c}
    q_c = -\frac{1}{\rho c_\text{p}} \frac{\partial F_\text{c}}{\partial z},
\end{equation}
where $z$ is altitude, $c_\text{p}$ is the specific atmospheric heat capacity, and $\rho$ is the air density.  The convective energy flux $F_\text{c}$ is given by:
\begin{equation} \label{eq:F_c}
    F_\text{c} = -\rho c_\text{p} K_\text{h} \left(\frac{\partial T}{\partial z} + \Gamma_\text{ad} \right),
\end{equation}
where $K_\text{h}$ is the eddy diffusion coefficient for heat and the adiabatic lapse rate is $\Gamma_\text{ad} = g/c_\text{p}$, where $g$ is the acceleration due to gravity.
The atmosphere is unstable to convection when $\partial T/\partial z > -\Gamma_\text{ad}$. In this case, the eddy diffusion coefficient for heat $K_\text{h}$ is given by:
\begin{equation} \label{eq:KH}
K_\text{h} = l^2 \left [ -\frac{g}{T} \left(\frac{\partial T}{\partial z} + \Gamma_\text{ad}  \right) \right ]^{1/2},
\end{equation}
where $l$ is the mixing length.
The nominal mixing length is given by \citet{Blackadar:1962}:
\begin{equation} \label{eq:zmix}
l = \frac{kz}{1 + kz/l_0}.
\end{equation}
Here, $k$ is von K\'{a}rm\'{a}n's constant and $l_0$ is the mixing length in the free atmosphere, which we express as
\begin{equation} \label{eq:l0}
l_0 = f_\text{z} H,
\end{equation}
where $H$ is the pressure scale height, given by:
\begin{equation} \label{eq:H}
H = \frac{RT}{\mu g}.
\end{equation}
 $R$ is the ideal gas constant, $\mu$ is the mean molar mass, and $f_\text{z}$ is the proportionality constant.

 \begin{table*}[]
\centering
\caption{Adjustable Convection Model Parameters}
\label{table:parameters}
{\iftwocol
    \scriptsize\selectfont
\else
    \small\selectfont
\fi
\begin{tabular}{lcccl}
\hline \hline
\textbf{Parameter} & \textbf{\ce{O2}-dominated} & \textbf{Venus-like} & \textbf{Aqua planet} & \textbf{Description} \\ 
$f_\text{z}$                   & {0.01} & {0.01} & {0.01}  & Mixing length fraction               \\
$U$  [m s$^{-1}$]                     & 0.1  & 0.1 & 10      & Surface wind speed              \\
$z_0$ [m]                     & 0.005  & 0.005 & 0.0002     & Surface roughness length              \\
$\rho_0$ [kg m$^{-3}$]               & {1}  & {1} & {1}  & Baseline density for eddy diffusion                                \\
$P_0$ [Pa]                 & {1} & {1}& {1}& Pressure for breaking of gravity waves in eddy diffusion                                  \\
$K_\text{m,0}$ [m$^2$ s$^{-1}$]                 & {0.5} & {1.0} & {0.5} & Eddy scaling coefficient for mass, gravity waves                               \\
$K_\text{m,1}$ [m$^2$ s$^{-1}$]                 & -- & -- & {20}  & Eddy scaling coefficient for mass, stability                  \\
$K_\text{m,1}$ [m$^2$ s$^{-1}$]                 & -- & {0.5} & {--}  & Constant minimal eddy diffusion                  \\
$\Lambda_0$                 & -- & -- & 0.1                & Minimum stability factor  \\ 
$f_\text{min}$                 & $10^{-4}$ & $10^{-4}$& $10^{-4}$& Minimum stability scaling factor for heat                                \\
$f_\text{max}$                 & {0.15}& {0.15}& {0.15} & Maximum stability scaling factor for heat                                    \\ \hline
\end{tabular}
\flushleft

}
\end{table*}

The preceding convection parameterization was used by \citet{Robinson:2018}, with $f_\text{z} = 1.0$ and with $K_\text{h} = 0$ under stable conditions. However, turbulent eddies still provide some vertical mass transport in stable conditions, due to shear instability and gravity wave breaking \citep[e.g.][]{Hodges:1969,Kondo:1978,Lindzen:1983,Canuto:2008}. 
Based on measurements of Earth and Venus eddy diffusion rates (see Figure~\ref{fig:eddy}), we specify the following parameterization for the eddy diffusion coefficient for mass transport:
\begin{equation}\label{eq:km}
  K_\text{m} = K_\text{m,0} \left( \rho/\rho_0 \right)^{-1/2} \left(1-e^{-P/P_0} \right) + K_\text{m,1} \Lambda + K_\text{m,2},
\end{equation}
where $\Lambda$ is a stability-related parameter that is nonzero only for the aqua planet. To determine $\Lambda$, we iterate the following from the surface upward, layer-by-layer:
\begin{equation} \label{eq:am}
    \Lambda_k = 
    \begin{cases}
        \Gamma/\Gamma_\text{ad}, & \text{if}\ \Gamma/\Gamma_\text{ad} > \Lambda_0 \\
        \Lambda_{k+1} e^{-z_k/z_{k+1}},    & \text{otherwise}
    \end{cases}
\end{equation}
where $\Gamma = -\partial T/\partial z$. In the first term of equation~(\ref{eq:km}), the density dependence represents the correlation of eddy diffusion with breaking gravity waves in terrestrial atmospheres \citep{Lindzen:1983, Izakov:2001} and the exponential decay term allows the reduction in eddy diffusion as the gravity wave amplitude decreases to zero around pressure $P_0$. We do not calculate molecular diffusion in our climate model, but this is calculated in the photochemical model. The second term in equation~(\ref{eq:km}) is stability-dependent and includes additional eddy diffusion in neutral and unstable conditions, such as in Earth's troposphere. Equation~(\ref{eq:am}) slowly reduces the contribution of stability in the transition between unstable and stable regions. The overall profile is then consistent with Earth examples in \citet{Massie:1981} and \citet{Brasseur:2006}, and with the eddy diffusion profiles assumed in exoplanet photochemical-kinetics studies by other authors \citep[e.g.][]{Segura:2007,Hu:2012}. The evolved environments we simulate have thick atmospheres and no oceans, and therefore are more like Mars and Venus, whose eddy diffusivities are well-modeled by breaking gravity waves alone \citep{Izakov:2001}, so in those cases $\Lambda = 0$. See Table \ref{table:parameters} for nominal values.

In stable conditions, the eddy diffusion coefficient for heat is reduced compared to that for momentum \citep[e.g.][]{Kondo:1978}. This reduction occurs gradually as a function of stability; \citet{Kondo:1978} provided a fit based on Richardson number ($Ri$). However, our mixing length code cannot compute $Ri$ in the free atmosphere, because it does not include an explicit description of the vertical shear in horizontal wind, so we use $\Gamma/\Gamma_\text{ad}$ as a representative stability parameter for this purpose and reduce the eddy diffusion coefficient for heat by:
\begin{equation}
    K_\text{h} = f_\text{h} K_\text{m} e^{-\alpha \Gamma/\Gamma_\text{ad} },
\end{equation}
where we set
\begin{align}
    \alpha \equiv & \frac{1}{2} \ln{\left( \frac{f_\text{min}}{f_\text{max}}\right)} \\
    f_\text{h} \equiv & f_\text{max} e^{\alpha}
\end{align}
The minimum and maximum scaling from $K_\text{m}$ to $K_\text{h}$ are denoted by $f_\text{min}$ and $f_\text{max}$, respectively, and given in Table \ref{table:parameters}.

Equations (\ref{eq:q_c})--(\ref{eq:km}) are comprised almost entirely of fundamental physical properties of the local atmospheric conditions: pressure, temperature, density, gravity, and specific heat capacity.
Since the vertical transport of heat, mass, and momentum in terrestrial exoplanet atmospheres are generally unknown, we chose to use this physically-motivated approach to convection, rather than convective adjustment to a pre-specified lapse rate that can be different from the adiabatic lapse rate (e.g. Earth is $\sim6.5$ K/km, Mars $\sim2.5$ K/km, and Venus $\sim8.0$ K/km; \citealt{Catling:2017}). 
The parameterizations we have chosen are able to reproduce eddy diffusion coefficients for mass consistent with measurements and kinetics studies of Earth and Venus (see Figure~\ref{fig:eddy}).

\subsection{Condensation and Kinematic Mixing} \label{sec:condensation}

To simulate a potentially habitable, ocean-covered (aqua) planet, we implemented an option for latent heat exchange and kinematic mixing of a condensible gas.
The layer-by-layer vertical mixing of a condensible species is calculated by solving the continuity-transport equation in 1D:
\begin{equation} \label{eq:continuity}
  \frac{\partial r}{\partial t} = -\frac{1}{\rho}\frac{\partial F_m}{\partial z} ,
\end{equation}
where the kinematic flux is given by:
\begin{equation}
    F_\text{m} = -\rho K_\text{m} \frac{\partial r}{\partial z}.
\end{equation}
Here, $K_\text{m}$ is the eddy diffusion coefficient for mass transport and $r$ is the mixing ratio of the condensible species.
As a result of either radiative cooling or an upwelling condensible species, a gas may become super-saturated (i.e. when the partial pressure at a level exceeds the saturation vapor pressure). We assume sufficient cloud condensation nuclei such that any condensate above the saturation point condenses immediately (i.e. within a single time-step). The saturated mass mixing ratio is determined by:
\begin{equation}
  r_\text{sat} = \frac{\mu}{\mu_\text{atm}} \frac{P_\text{sat}}{P}.
\end{equation}
At the globally-averaged temperatures, it is unlikely that the globally-averaged water mixing ratios will approach saturation, though there is nonetheless condensation, evaporation, and latent heat exchange. The simplest solution here is to use a different mixing ratio criterion for saturation. We compute condensation if relative humidity exceeds that maximum humidity using a modified form of equation 2 from \citet{Manabe:1967} for Earth:
\begin{equation}
    rh = rh_\text{s} \text{max} \left(\frac{p/p_\text{s}-0.02}{1-0.02}, 0.2 \right),
\end{equation}
where s refers to the surface value. The modification is our specification of a minimum value that depends on the surface humidity (\citealt{Manabe:1967} made a similar adjustment to obtain Earth's stratospheric \ce{H2O} abundance). While \citet{Manabe:1967} use $rh_\text{s}=0.77$, we find that $rh_\text{s}=0.70$ matches our globally-averaged Earth data best (see \S\ref{app:validation}).

To determine the heating rate due to condensation, we evaluate the conserved quantity, the specific moist enthalpy:
\begin{equation}
  h = (c_\text{p} r_\text{a} + c_\text{p,v} r_\text{v} + c_\text{p,c} r_\text{c}) T + L r_\text{v},
\end{equation} 
where the subscript $c$ refers to the condensed form, $v$ refers to the vapor-phase of the condensible, $a$ refers to the vapor-phase bulk atmosphere, and $L$ is the latent heat of evaporation. The expansion of the layer reduces the net latent heat, as a result of isobaric work ($PdV$), which per unit mass is equivalent to $(R/\mu) dT$. Including this term with the change in specific moist enthalpy due to a change in condensate and resultant temperature change, then solving for the change in temperature of the air parcel, the net heating rate during time-step $\Delta t$ is:
\begin{equation}
  q_\text{lh} = \frac{\Delta T}{\Delta t} = \frac{\Delta r_\text{c}}{\Delta t} \frac{L - T(c_\text{p,c} - c_\text{p,v})}{c_\text{p}(1 - r_\text{v} - r_\text{c}) + c_\text{p,v} r_\text{v} + c_\text{p,c} r_\text{c} - R/\mu},
\end{equation}
where the quantity being condensed is $\Delta r_\text{c} \equiv r_\text{c} - r_\text{sat}$, $r$ is the mass mixing ratio and $c_\text{p}$ is the specific heat.

\subsection{Surface Fluxes} \label{sec:surface} 

We compute the surface fluxes of sensible heat, latent heat, and condensible gas using similarity theory \citep[e.g.][pp. 395--409]{Pierrehumbert:2011}, given by:
\begin{align}
  F_\text{c,sh} =& \rho c_\text{p} C_\text{D} U (T_\text{s} - T_\text{sbl}) \\
  F_\text{c,lh} =& \rho L C_\text{D} U (r_\text{s} - r) \\
  F_\text{m} =& \rho C_\text{D} U (r_\text{s} - r)
\end{align}
where $U$ is the surface wind speed, $T_\text{s}$ is the temperature of the surface, $T_\text{sbl}$ is the temperature of the surface atmospheric boundary layer, $r_\text{s}$ is the surface mixing ratio boundary condition (determined from a humidity assumption or fixed mixing ratio), and $C_\text{D}$ is the coefficient of drag at the surface, which we calculate as in \citet{Pierrehumbert:2011}:
\begin{equation}
    C_\text{D} = \left( \frac{k}{\log{z/z_0}} \right)^2,
\end{equation}
where $z_0$ is the surface roughness height, which we specify as 0.005~m for the desiccated planets (corresponding to flatlands with no vegetation) and 0.0002~m for open ocean \citep{Davenport:2000,Jarraud:2008}. 
As the atmospheres are much thicker than Earth, we assume surface wind speeds of 0.1~m~s$^{-1}$, much less than Earth, but consistent with measurements of Venus' surface atmosphere \citep{Ainsworth:1975,Counselman:1979}. For the Aqua planet, we use 10~m~s$^{-1}$, which is roughly consistent with Earth's global average wind speed \citep[7~m~s$^{-1}$;][]{Meissner:2001}. 

\subsection{Aerosols}

We include the radiative effects of aerosols in cloudy cases for both the aqua planet and Venus-like atmospheres. Since this is a 1D climate model, we present end-member cases of global clear and cloudy cases, which can be used to assess the observational discriminants of these atmospheres. We have not yet implemented a cloud microphysics model in our climate model, so we specify altitude-dependent differential optical depths and aerosol optical properties from Earth or Venus, as appropriate. Aerosol parameters are described in \S\ref{sec:input_aer}.

\subsection{Convergence}

We first approach convergence using 4-stream radiative transfer for a single, global-average approximation solar zenith angle of $\sim$60 degrees. The final convergence of VPL Climate is conducted using 8-stream radiative transfer with four-point Gaussian quadrature integration of the stellar radiance, which increases the accuracy of the convergence result \citep{Cronin:2014,Hogan:2015}. In future work, we will quantitatively assess the impact of solar radiance integration in our climate model. \citet{Kitzmann:2016} determined that increasing the number of radiative transfer streams from 8 to 16 had no effect on the radiative fluxes.

\section{Photochemical Model Updates} \label{app:atmos}

Here we describe several updates made to our photochemical model to allow more robust modeling of the alien environments in this paper. We modified the wavelength grid to accommodate an input SED with wavelengths shorter than Lyman-$\alpha$ and have increased the resolution to 100~cm$^{-1}$, which now includes 750 flux bins versus the standard 118. We updated cross-sections for the molecules listed in Table~\ref{table:xsec}. Additionally, based on the recommendations in \citet{Burkholder:2015}, we updated the quantum yields of \ce{O3} with fits by \citet{Matsumi:2002} and \citet{Nishida:2004}.
These updates extended the \ce{O3} photochemistry to cooler temperatures necessary for modeling the range of planets in this work. 

We made several other changes to the photochemical model for this work. 
First, as a result of the cross-sections and grid update, we set the solar zenith angle to the daytime-mean of 60~degrees, to approximate the global diurnal average when using a two-stream radiative transfer model \citep[e.g.][]{Manabe:1964,Li:2017}.
For the \ce{O2}-dominated atmospheres, we  modified treatment of the water vapor profile so that it was no longer specified to a fixed humidity, moist adiabat, or Earth-like profile, but instead was treated like every other long-lived trace species. This is a requirement to be able to model atmospheres with water abundances significantly less than the Earth's, and is required for modeling a Venus-like atmosphere. Likewise, we also treated \ce{CO2}  as a long-lived gas, as in \citet{Kopparapu:2011}, rather than treated as an inert species \citep[c.f.][]{Meadows:2018}, though condensation limits still apply. Long-lived gases are included in the Jacobian solved at each time step, and transport between layers is calculated.  This modification allowed us to more accurately model the photolysis and longer recombination lifetimes of \ce{CO2} in desiccated atmospheres, where \ce{CO2} photolysis can outpace recombination, resulting in a steady-state equilibrium between \ce{CO2}, CO, \ce{O2}, and \ce{O3} \citep{Gao:2015}. Lastly, we adjusted the grid spacing for the atmospheric scale heights necessary to model each planet. Since we used 200 constant-thickness layers with altitude-based grid spacing in the chemistry model and VPL~Climate uses a variable grid, we interpolated the atmospheric structure values when passing data from one model to the other. Both model grids extend to 0.01~Pa.

\begin{table}[]
\begin{center}
\caption{Updated UV--Vis. Cross Sections}
\label{table:xsec}
{\iftwocol
    \small\selectfont
\else
    \scriptsize\selectfont
\fi
\begin{tabular}[t]{lll}
\hline
\textbf{Species} & \textbf{Wavelengths (nm)}                                                         & \textbf{References}                                                                                                                  \\ \hline
\ce{CO2}         & 121--169                                                                          & \citet{Lewis:1983}\\
~                & 169--200                                                                          & \citet{Shemansky:1972}\\
~                & 200--225                                                                          & Power law extrapolation  \\
\ce{H2O}         & 120--194                                                                          & \citet{Mota:2005}    \\
\ce{H2O2}         & 125--190                                                                          & \citet{Schurgers:1968}\\
~                & 190--350                                                                          & \citet{JPL:2010}\\
~                & 350--557                                                                          & Power law extrapolation  \\

\ce{H2S}         & 120--160       & \citet{Feng:1999}                           \\
~         & 160--260 & \citet{Wu:1998}                  \\
~        & 260--313 & Power law extrapolation                            \\
\ce{H2SO4}         & 120--200       & \citet{Lane:2008}                           \\
~                  & 512--745       & \citet{Mills:2005}                           \\
\ce{H2CO}        & 120--226       &  \citet{Cooper:1996}          \\ 
~       & 226--375    &  \citet{JPL:2010}      \\
\ce{HNO2}         & 120--184       & Power law extrapolation \\
~                 & 184--396       & \citet{JPL:2010}                           \\            
\ce{HNO3}         & 120--192       & \citet{Suto:1984}                           \\
~         & 192--350 & \citet{Burkholder:1993}                  \\
\ce{NO}          & 120--207                                                                          & \citet{Iida:1986}                                                                                                                    \\
\ce{NO2}         & 120--422                                                                          & \citet{Vandaele:1998}                                                                                      \\
~         & ~                                                                         & QY: \citet{JPL:2010}    \\
\ce{N2O}         & 120--160    & \citet{Zelikoff:1953}             \\
~        & 160--240 & \citet{JPL:2010}                 \\
\ce{N2O5}        & 152--200    & \citet{Osborne:2000}             \\
~        & 200--420 & \citet{JPL:2010} \\
\hline
\end{tabular}
\begin{tabular}[t]{lll}
\hline
\textbf{Species} & \textbf{Wavelengths (nm)}                                                         & \textbf{References}      \\ \hline
\ce{O2}          & 120--179       & \citet{Lu:2010}                       \\
~         & 179--203  & \citet{Yoshino:1992}        \\
~         & 205--240     & \citet{Yoshino:1988}            \\
\ce{O3}          & 120--845                                                                          & \citet{Serdyuchenko:2011}                                                                                                            \\
\ce{OCS}         & 120--210      & \citet{Limao:2015}                    \\
~         & 210--300    &   \citet{Molina:1981}                \\
\ce{SO2}         &  120--403      &  \citet{Manatt:1993}     \\
~         &  403--500    &  Power law extrapolation       \\
\ce{SO3}         &  120--140 &  Power law extrapolation  \\
~         &  140--180 & \citet{Hintze:2003} \\
~         &  180--330  &   \citet{JPL:2010}  \\
~         &  330--350 &  Power law extrapolation  \\
\ce{S3}          & 350--475                                                                          & \citet{Billmers:1991}                                                                                                                \\
\ce{S4}          & 425--575          & \citet{Billmers:1991} \\
\ce{CH4}         & 120--153     & \citet{Lee:2001}  \\
\ce{CH3O2}       & 205--295          & \citet{JPL:2010}  \\
\ce{CH3OOH}      & 210--405          & \citet{JPL:2010}  \\
\ce{HCl}         & 120--135          & \citet{Brion:2005} \\
~                & 135--220          & \citet{Cheng:2002} \\
\ce{ClO}         & 236--312          & \citet{Simon:1990} \\
\ce{OClO}        & 125--183          & \citet{Hubinger:1994} \\
~                & 290--460          & \citet{Bogumil:2003} \\
\ce{COCl2 }      & 168--305          & \citet{JPL:2010} \\
\ce{SO2Cl2}      & 190--300          & \citet{Uthman:1978} \\
 \hline
\end{tabular}}
\end{center}

Note: Most data obtained from the MPI-Mainz UV/VIS Spectral Atlas of Gaseous Molecules of Atmospheric Interest \citep[][\url{http://satellite.mpic.de/spectral_atlas}]{Keller:2013}.
\end{table}

\section{Photochemical Validation of Earth and Venus} \label{app:validation}

With the updates presented in the preceding Appendix, we present a new validation of Earth's and Venus' atmospheres.

\subsection{Earth}

 \begin{figure*}
  \centering
  \includegraphics[width = \textwidth]{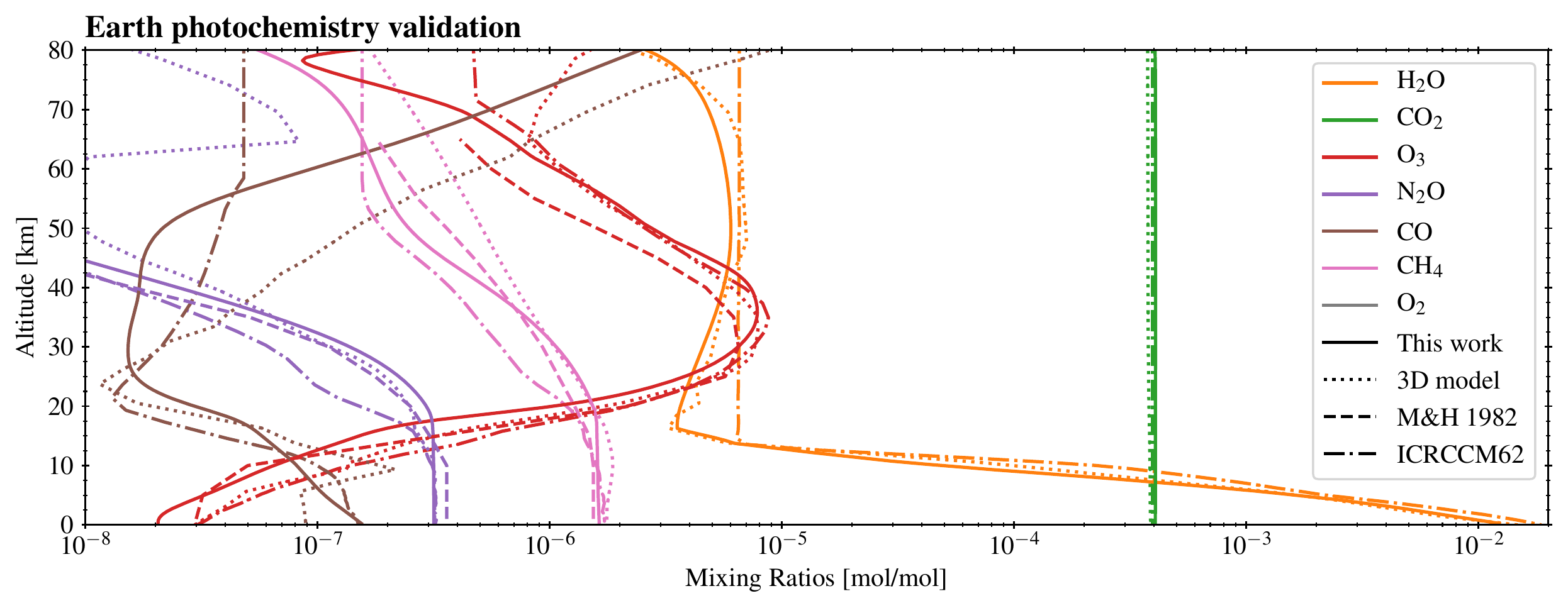}
  \caption{Photochemical equilibrium validation for Earth. Mixing ratio profiles of key species are shown from this work (solid lines), an average derived from the 3D Spectral Earth model \citep[][dotted lines]{Robinson:2011}, measurements from \citet{Massie:1981} (dashed lines), and the ICRCCM Earth mid-latitude summer sounding case 62 (dash-dotted lines). 
   \label{fig:earth_valid}} 
\end{figure*}

In addition to the updates presented in the preceding Appendix, we validated Earth using the boundary conditions given in Table~\ref{table:bc_earth}. In particular, this includes fluxes (not fixed mixing ratios) of the important Earth constituents CO, \ce{CH4}, NO, \ce{H2S}, \ce{H2SO4}, OCS, \ce{N2O}, \ce{CS2}, and dimethyl sulfide (\ce{C2H6S}), along with vertically distributed fluxes of \ce{SO2}. Oxygen is fixed at the surface at 21\% and nitrogen at 78\%. Carbon dioxide is fixed at the surface to the current value of 400~ppm. The flux of NO is calibrated to best reproduce the effects of \ce{NO_x} chemistry, as we do not include chlorine species or other pollutants not listed here. The water vapor profile is from \citet{Manabe:1967} using a fixed surface humidity of 70\%, assuming the tropopause is at 14~km, and that the minimum saturation value in the troposphere is 20\% of the surface value. Vertical profiles of important species are shown in Figure~\ref{fig:earth_valid} along with several sources for comparison.

\begin{table}[]
\begin{center}
\caption{Earth Validation Lower Boundary Conditions.}
\label{table:bc_earth}
{\iftwocol
    \small\selectfont
\else
    \footnotesize\selectfont
\fi
\begin{tabular}[t]{ll}
\hline
\textbf{Species} & \textbf{Boundary Condition} \\ \hline
O       &   $v_\text{dep}=1.$ \\
\ce{O2} &   $r_0=0.21$ \\
\ce{H2O} &   $rh_0=0.70$ \\
\ce{H} &   $v_\text{dep}=1.$ \\
\ce{OH} &   $v_\text{dep}=1.$ \\
\ce{HO2} &   $v_\text{dep}=1.$ \\
\ce{H2O2} &   $v_\text{dep}=0.2$ \\
\ce{H2} &   $v_\text{dep}=2.4\times10^{-4}$ \\
\ce{CO} &   $v_\text{dep}=0.03$, $F=3.7\times10^{11}$ \\
\ce{HCO} &   $v_\text{dep}=1.$ \\
\ce{H2CO} &   $v_\text{dep}=0.2$ \\
\ce{CH4} &   $F=1.6\times10^{11}$ \\
\ce{NO} &   $v_\text{dep}=1.6\times10^{-2}$, $F=1.0\times10^{9}$ \\
\ce{NO2} &   $v_\text{dep}=3.0\times10^{-3}$ \\ \hline
\end{tabular}
\begin{tabular}[t]{ll}
\hline
\textbf{Species} & \textbf{Boundary Condition} \\ \hline
\ce{HNO} &   $v_\text{dep}=1.$ \\
\ce{H2S} &   $v_\text{dep}=0.02$, $F=2.0\times10^{8}$ \\
\ce{SO2} &   $v_\text{dep}=1.$, $F=9.0\times10^{9}$ dist. 14~km \\
\ce{H2SO4} &   $v_\text{dep}=1.$, $F=7.0\times10^{8}$ \\
\ce{HSO} &   $v_\text{dep}=1.$ \\
\ce{OCS} &   $v_\text{dep}=0.01$, $F=1.5\times10^{7}$ \\
\ce{HNO3} &   $v_\text{dep}=0.2$ \\
\ce{N2O} &   $F=1.53\times10^{9}$ \\
\ce{HO2NO2} &   $v_\text{dep}=0.2$ \\
\ce{CO2} &   $r_0=4.0\times10^{-4}$ \\
\ce{CS2} &   $F=2.0\times10^{7}$ \\
\ce{C2H6S}(DMS) &   $F=3.3\times10^{9}$ \\
\ce{N2} &   $r_0=0.78$ \\ \hline
\end{tabular}
}
\end{center}

Note: All unlisted species have a lower boundary condition $v_\text{dep}=0$. Deposition velocities ($v_\text{dep}$) are given in cm~s$^{-1}$, fluxes ($F$) are given in molecules cm$^{-2}$~s$^{-1}$, $r_0$ is a fixed surface mixing ratio, and $rh_0$ is the surface relative humidity.
\end{table}

\subsection{Venus}

 \begin{figure}
  \centering
  \includegraphics[width = 0.5\textwidth]{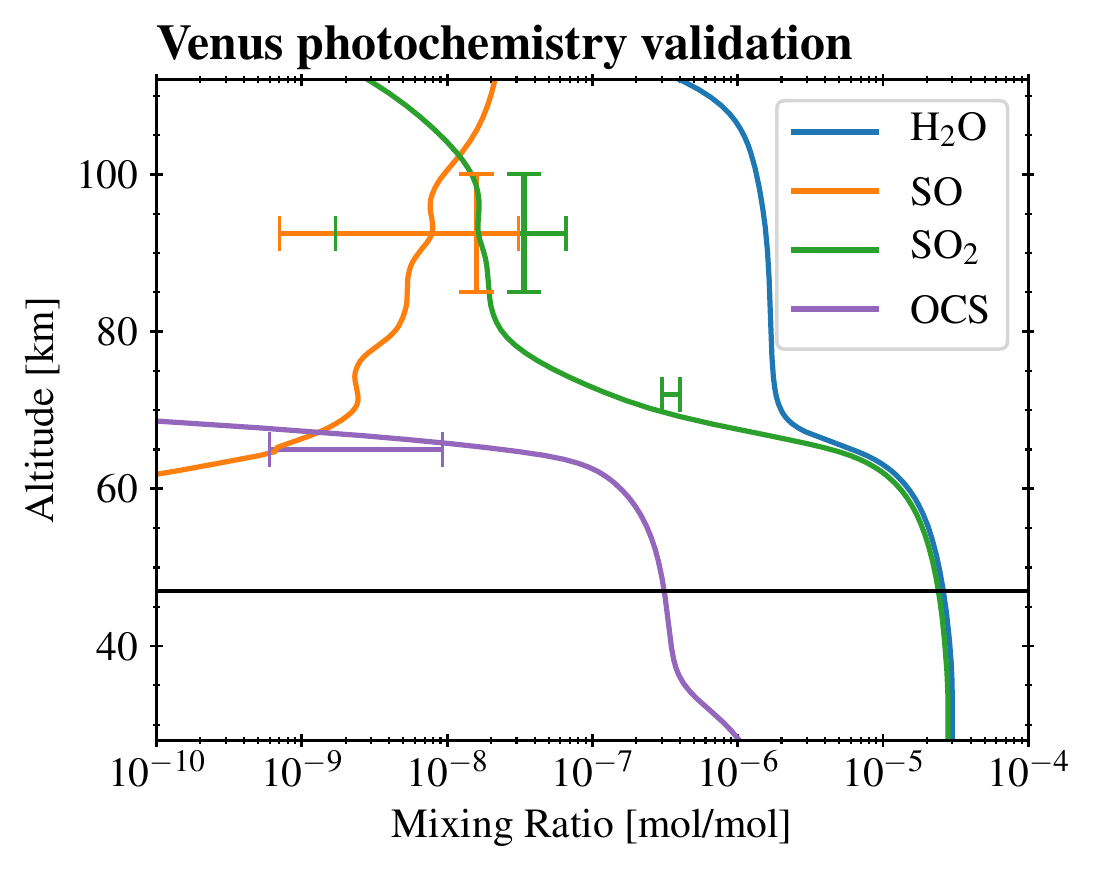}
  \caption{Photochemical equilibrium validation for Venus. Mixing ratios from our model for select species are shown, along with a selection of published measurements. Note that there is considerable variation in these abundances as a function of day vs night and latitute \citep[e.g.][]{Ignatiev:1999,Arney:2014}. 
   \label{fig:venus_valid}} 
\end{figure}

\begin{table}[]
\begin{center}
\caption{Venus Validation Lower Boundary Conditions.}
\label{table:bc_venus}
{\small\selectfont
\begin{tabular}[t]{ll} 
\hline
\textbf{Species} & \textbf{Boundary Condition} \\ \hline
\ce{O2} &   $v_\text{dep}=2v$ \\
\ce{H2O} &   $r_0=3.0\times10^{-5}$ \\
\ce{H2} &   $r_0=4.5\times10^{-9}$ \\
\ce{NO} &   $r_0=5.5\times10^{-9}$ \\
\ce{SO2} &   $r_0=2.8\times10^{-5}$ \\
\ce{OCS} &   $r_0=1.0\times10^{-6}$ \\
\ce{HCl} &   $r_0=4.0\times10^{-7}$ \\
\ce{CO2} &   $r_0=0.965$ \\
\ce{N2} &   $r_0=0.035$ \\ \hline
\end{tabular}
}
\end{center}
Note: All unlisted species have a lower boundary condition $v_\text{dep}=v=K_m/2H$. Boundary conditions are either deposition velocities ($v_\text{dep}$), given in cm~s$^{-1}$, or a fixed surface mixing ratio $r_0$. The Venus lower boundary is at 28~km ($\sim11.5$~bar).
\end{table}

This is the first time our photochemical model has been used for a Venus-like planet. This has required careful selection of boundary conditions, reactions, and of the eddy diffusion profile. Specifically for Venus, few additional changes were required to the model itself. In addition to the model updates listed in Appendix~\ref{app:atmos}, we updated saturation vapor pressure data, added an additional lower boundary condition, compiled a new Venus input template, and conducted a validation against recent Venus literature.

We updated the saturation vapor pressure and temperature data for sulfuric acid solutions using finer intervals of 1\% by fitting splines to published refraction and absorption data, which are available at sulfuric acid concentrations of 25, 38, 50, 75, 84.5, and 95.6\%  \citep{Palmer:1975}.

We drew heavily on recent Venus literature \citep[primarily][]{Krasnopolsky:2012,Zhang:2012} to construct our Venus template, consisting of important species, their boundary conditions, and relevant reactions. The boundary conditions are listed in Table~\ref{table:bc_venus}. As in previous work \citep[e.g.][]{Krasnopolsky:2012,Zhang:2012}, we do not extend our photochemical template for Venus to the surface. Instead, we set the lower boundary at 28~km ($\sim11.5$~bar), as we do not include thermochemical reactions, which are typically treated as a separate regime from the photochemical region \citep[e.g. ][]{Krasnopolsky:2013}. Because the lower boundary is not the surface, the ``deposition velocity'' $v$ depends on the scale height and eddy diffusion coefficient \citep{Krasnopolsky:2012}:
\begin{equation}
    v = \frac{K_m}{2H},
\end{equation}
so the model can calculate the associated downward flux. The reactions in our Venus template are primarily reproduced from those given in \citet{Krasnopolsky:2012}.

The validation for Venus was primarily for major climate and observable species, particularly those involving Venus' sulfuric acid clouds (e.g. \ce{H2O} and \ce{SO2}). We find good agreement with recent modeling by \citet{Krasnopolsky:2012}, on which our reaction and boundary condition template is based. However, there are still disagreements between Venus photochemical modeling in the literature among the abundances retrieved from different observations, particularly observed oxygen abundances and in the features of the \ce{SO2} profile. Future improvements to our Venus reactions can be made as those researchers who focus on Venus solve these outstanding issues.

Our photochemical model includes simplistic aerosol formation, including sulfuric acid aerosols, based on production/loss lifetimes \citep{Pavlov:2001}. We found this model fails to produce the specific distribution of Venus' multiple aerosol models, and is unable to generate aerosols as large as the mode 3 (3.85~\um{}) particles. Nonetheless, it does self-consistently generate aerosols at altitudes consistent with Venus' cloud deck. We find that the transmission spectrum for our model Venus is roughly consistent with the cloud parameterzation from \citet{Crisp:1986}. Therefore, this provides an acceptable starting point for self-consistent climate-chemistry modeling of Venus-like exoplanets with aerosol formation.

\listofchanges 
\end{document}